\documentclass[a4paper,twocolumn,10pt,accepted=2026-08-31]{quantumarticle}
\pdfoutput=1
\usepackage{xpatch}
\makeatletter
\patchcmd{\@ssect@ltx}
    {\addcontentsline{toc}{#1}{\protect\numberline{}#8}}
    {}
    {}
    {}
\makeatother
\pdfoutput=1
\usepackage{color}
\usepackage{enumitem}
\usepackage{mathtools}
\usepackage[dvipsnames]{xcolor}

\usepackage{hyperref}
\definecolor{amaranth}{rgb}{0.9, 0.17, 0.31}
\definecolor{forestForestGreen(web)}{rgb}{0.13, 0.55, 0.13}
\definecolor{blue(munsell)}{HTML}{005567}
\definecolor{bblue}{rgb}{0.0, 0.58, 0.71}
\hypersetup{
pdfstartview={FitH}, 
pdftitle={}, 
colorlinks=true, 
linkcolor=blue(munsell), 
citecolor=blue(munsell), 
filecolor=magenta, 
urlcolor=blue(munsell)
}
\usepackage[numbers]{natbib}
\usepackage{pgfplots}
\pgfplotsset{compat=1.17}
\usepackage{amssymb}
\usepackage{amsfonts}
\usepackage{graphicx}
\usepackage{epstopdf}
\usepackage{dcolumn}
\usepackage{amsmath}
\usepackage{latexsym,bm}
\usepackage{amsthm}
\usepackage{slashed}
\usepackage{float}
\usepackage{color}
\usepackage{url}
\usepackage{rotating}
\usepackage[normalem]{ulem}
\usepackage{faktor}
\usepackage{tikz-cd}
\usepackage{tensor}
\usepackage{array}

\usepackage{tabularx}
\usepackage{makecell}
\usepackage[normalem]{ulem}
\usepackage{changepage}
\numberwithin{equation}{section}

\usepackage{tikz}
\usepackage{diagbox}
\usetikzlibrary{positioning}
\usetikzlibrary{calc}
\usetikzlibrary{decorations.pathreplacing,calligraphy}

\usepackage{xstring}
\usetikzlibrary{decorations.pathmorphing} 
\usetikzlibrary{decorations.markings} 
\usetikzlibrary{arrows} 
\usetikzlibrary{shapes} 
\usetikzlibrary{matrix} 
\usetikzlibrary{positioning} 
\usepackage[english]{babel} 
\usepackage[autostyle]{csquotes}
\usepackage{pifont}
\usetikzlibrary{shapes.multipart}

\tikzset{->-/.style={decoration={
  markings,
  mark=at position .5 with {\arrow{>}}},postaction={decorate}}}

\usepackage{quantikz}

\newcommand{\bea}{\begin{eqnarray}}
\newcommand{\eea}{\end{eqnarray}}
\newcommand{\be}{\begin{equation}}
\newcommand{\ee}{\end{equation}}
\newcommand{\ba}{\begin{aligned}}
\newcommand{\ea}{\end{aligned}}
\newcommand{\bit}{\begin{itemize}}
\newcommand{\eit}{\end{itemize}}
\newcommand{\ben}{\begin{enumerate}}
\newcommand{\een}{\end{enumerate}}
\newcommand{\nn}{\nonumber}

\newcommand{\id}{\text{id}}

\newcommand{\rough}{\text{rough}}
\newcommand{\smooth}{\text{smooth}}
\newcommand{\SPT}{\text{SPT}}

\newcommand{\lb}{\left(}
\newcommand{\rb}{\right)}

\newcommand{\wt}{\widetilde}

\newcommand{\Z}{{\mathbb Z}}

\newcommand{\bC}{{\mathbb C}}

\newcommand{\LL}{{(1)}}
\newcommand{\RR}{{(2)}}

\newcommand{\cA}{\mathcal{A}}

\newcommand{\cC}{\mathcal{C}}

\newcommand{\cH}{\mathcal{H}}

\newcommand{\cL}{\mathcal{L}}

\newcommand{\cX}{\mathcal{X}}

\newcommand{\cZ}{\mathcal{Z}}

\newcommand{\Rep}{\mathsf{Rep}}

\renewcommand{\dim}{\text{dim}}

\def\BlueColor{MidnightBlue!60}
\def\RedColor{Maroon!60}

\makeatother

\usepackage{physics}
\newcommand{\ol}{\overline}

\newcommand{\bbI}{\mathbb I}

\newcommand{\diag}{\text{diag}}
\newcommand{\folded}{\text{folded}}

\usepackage{float}
\newcommand{\vp}{\varphi}

\usepackage{makecell}

\makeatletter
\newcommand\xlabel[2][]{\phantomsection\def\@currentlabelname{#1}\label{#2}}
\makeatother

\makeatletter
\def\l@subsubsection#1#2{}
\makeatother

\begin{document}

\title{Hybrid Lattice Surgery: Non-Clifford Gates via Non-Abelian Surface Codes}

\author{Sheng-Jie Huang}
\affiliation{Mathematical Institute, University of Oxford, Woodstock Road, Oxford, OX2 6GG, United Kingdom}
\author{Alison Warman}
\affiliation{Mathematical Institute, University of Oxford, Woodstock Road, Oxford, OX2 6GG, United Kingdom}
\author{Sakura Sch\"afer-Nameki}
\affiliation{Mathematical Institute, University of Oxford, Woodstock Road, Oxford, OX2 6GG, United Kingdom}
\author{Yanzhu Chen}
\affiliation{Department of Physics, Florida State University, Tallahassee, Florida 32306, USA}
\email{yanzhu.chen@fsu.edu}

\maketitle

\begin{abstract} 
In universal fault-tolerant quantum computing, implementing logical non-Clifford gates often demands substantial spacetime resources for many error-correcting codes, including the high-threshold surface code. A critical mission for realizing large-scale quantum computing is to develop simple and resource-efficient implementations of logical non-Clifford gates.  
We propose a novel way of implementing non-Clifford operations in the standard surface code based on hybrid lattice surgery. First we generalize the standard lattice surgery to hybrid lattice surgery, where operations of rough merge and rough split happen across different topological codes. Then we apply such procedures between Abelian and non-Abelian codes and show that this can provide non-Clifford operations in the standard surface code, in the form of a magic state or a non-Clifford gate teleportation. 
Complementing this, we provide a continuum topological field theory description of this hybrid lattice surgery utilizing interfaces between (2+1)d topological orders.
From these considerations, we can generalize our protocol to non-Clifford gates and magic states at all finite levels of the Clifford hierarchy, as well as gates beyond the hierarchy. 
We also discuss protocols extending this framework to qutrits.
\end{abstract}

\tableofcontents


\section{Introduction}

Fault-tolerant quantum computing relies on quantum error correcting (QEC) codes to protect information against inevitable errors that occur during computation~\cite{Shor1995, Kitaev1997, Cochrane1999, Gottesman2001, Gottesman2009}. The logical information that is encoded on a code patch can have a desired level of errors if the physical errors are adequately detected and corrected.  
For certain logical gates, local errors do not propagate across a code patch during implementation, because the subsystems within the patch are not entangled with each other. Such gates are called transversal gates and are inherently fault-tolerant~\cite{Gottesman2009}. 
Unfortunately, for a given QEC code, the transversal gates do not form a universal set of quantum gates, as famously stated in the Eastin-Knill theorem~\cite{Eastin:2009tem}. To achieve universal fault-tolerant quantum computing, a key challenge is supplementing the transversal gates with non-Clifford operations for the chosen QEC code. 
Beyond the most strictly defined transversal gates, there are other gate operations that, while entangling different subsystems of one code patch, have a limited range of error propagation~\cite{Yoder:2016bys}. 

For the Calderbank–Shor–Steane (CSS) codes~\cite{Steane:1996ghp,Calderbank:1995dw}---a popular family of QEC codes including the two-dimensional (2D) surface code (in 2+1 spacetime dimensions), logical Clifford gates either are strictly transversal or can be implemented with a constant-depth circuit and are therefore fault-tolerant~\cite{Webster:2017odb}. In comparison, the fault-tolerant implementation of logical non-Clifford gates is much more resource-intensive. One approach is magic state injection, where a non-Clifford gate is realized through gate teleportation and the logical error is governed by the fidelity of the injected magic state. The process of improving the magic state fidelity, called magic state distillation~\cite{Bravyi2005magicstate,Fowler2012,OGorman:2016rzu,Gidney2019efficientmagicstate}, requires substantial spacetime resources since the prepared states are discarded when errors are detected. This cost is determined by the error-detecting code chosen for the distillation process and can be improved if a better choice is constructed~\cite{Fowler2012bridge,Litinski2019magicstate,2025lee}. Another approach is transforming the protected logical information temporarily to another QEC code, in which a non-Clifford gate can be implemented transversally~\cite{Anderson2014,Kubica2015,Bombin2015,Beverland2021,Barkeshli:2023bta,Daguerre2024}. Large spacetime resource overheads also incur in this case due to the error correction procedure at intermediate stages.

Topological QEC codes encode logical qubits or qudits in the degenerate ground space of a system with a nontrivial topological order (TO) and are robust against local errors~\cite{Bravyi2010,Cui:2019lvb,Qiu:2020sea}. The locality of the syndrome check operators, the relative ease of scaling up a code patch, and their often high error thresholds make topological codes promising candidates for realizing large-scale fault-tolerant quantum computing. Among these, codes with an Abelian TO are favorable because of the more straightforward decoding procedures~\cite{Duclos-Cianci:2009gwa,Fowler2015,Delfosse:2017npj}. The surface code is one such example with a high error threshold~\cite{Bravyi:1998sy,Dennis2001}. Moreover, on the information protected by topological codes, some gates can be realized by deforming the lattice of the physical qubits or qudits with relatively low resource overheads~\cite{Brown2017}. For example, through merging and splitting surface code patches, collectively called lattice surgery, one can implement 2-qubit Clifford gates~\cite{Horsman:2011hyt,Fowler:2018efb,Litinski:2018lim}. Compared to the transversal implementation, this bypasses the need for entangling operations between all the qubits in the code patches and lessens the demand on hardware connectivity. 

Driven by the need of a resource-efficient implementation of non-Clifford operations, we construct a novel protocol based on lattice surgery in this work. We generalize the standard surface code lattice surgery operations to merge and split operations between two different topological codes. Through the interplay between Abelian TO and non-Abelian TO, our protocol either generates a logical magic state or realizes a non-Clifford gate teleportation in the surface code. Specifically, we consider 2D topological codes given by quantum double models. This method can also be understood in terms of the effective logical entangling operations between different code patches through lattice surgery.

Using the same Abelian and non-Abelian TOs as in this work, we have previously designed a protocol of generating logical magic states that relies on the transformation of a code patch through an intermediate non-Abelian code~\cite{Huang:2025cvt}. Non-Abelian TO has also been exploited in other methods of realizing non-Clifford operations in an Abelian code~\cite{Laubscher:2019rss,Brown2020,Hsin:2024pdi,Davydova:2025ylx,Bauer:2025qly,Sajith:2025rvy}. Compared to the other methods, the lattice surgery-based protocol in this work is simple to implement since it only involves local measurements along boundaries of code patches in addition to the usual error correction procedure. While error correction is required at different stages, this protocol can be applied in parallel with the main computation process and is more time-efficient than code switching.
An upcoming work will appear in \cite{Ellisonetal} also using surface codes with non-Abelian TO for universal quantum computation.

\begin{figure}
\centering
\resizebox{\columnwidth}{!}{
\begin{tikzpicture}
\begin{scope}{shift= (0,0)}
\node[black, left] at (-1,2.5) {$(a)$} ;
\begin{scope}[shift={(-0.3,0)}]
\draw [thick, fill= \BlueColor, opacity=0.5]   (0,0) -- (0,2) -- (2,2) -- (2,0) --(0,0); 
\draw [thick]  (0,0) -- (0,2) -- (2,2) -- (2,0) --(0,0); 
\node at (1,1) {$D(G)$}; 
\node[right] at (2,1) {\text{rough}};
\node[left] at (0,1) {\text{rough}};
\node[above] at (1,2) {\text{smooth}};
\node[below] at (1,0) {\text{smooth}};
\end{scope}
\begin{scope}[shift={(4,0)}]
\draw[thick, fill= \RedColor,opacity=0.5]  (0,0) -- (0,2) -- (2,2) -- (2,0) --(0,0); 
\draw [thick]  (0,0) -- (0,2) -- (2,2) -- (2,0) --(0,0); 
\node at (1,1) {$D(G')$}; 
\node[right] at (2,1) {\text{rough}};
\node[left] at (0,1) {\text{rough}};
\node[above] at (1,2) {\text{smooth}};
\node[below] at (1,0) {\text{smooth}};
\end{scope}
\end{scope}
\node[black, left] at (-1,-0.5) {$(b)$} ;
\begin{scope}[shift={(1,-3)}]
\draw[thick, fill= \BlueColor, opacity=0.5]   (0,0) -- (0,2) -- (2,2) -- (2,0) --(0,0); 
\draw[thick, fill= \RedColor,opacity=0.5]  (2,0) -- (2,2) -- (4,2) -- (4,0) --(2,0); 
\draw [thick]  (0,0) -- (0,2) -- (4,2) -- (4,0) --(0,0); 
\draw[ultra thick, Violet] (2,0) -- (2,2);
\node[above, Violet] at (2,2) {$\mathcal{A}$};
\node at (1,1) {$D(G)$}; 
\node at (3,1) {$D(G')$};
\node[left] at (0,1) {\text{rough}};
\node[above] at (1,2) {\text{smooth}};
\node[below] at (1,0) {\text{smooth}};
\node[right] at (4,1) {\text{rough}};
\node[above] at (3,2) {\text{smooth}};
\node[below] at (3,0) {\text{smooth}};
\end{scope}
\end{tikzpicture}
}
\caption{(a) Code patches for two surface codes realizing the quantum doubles with groups $G$ and $G'$. In both surface codes, we choose the boundary conditions to be rough on the vertical edges and smooth on the horizontal. 
(b) A hybrid code patch after a rough merge between them, resulting in an interface specified by $\cA$ (a particular algebra in the quantum double). }
\label{TwoPatch}
\end{figure}
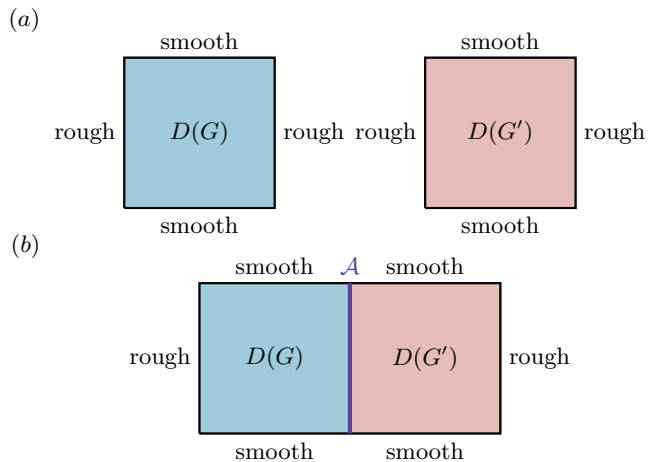

\noindent
{\bf Summary of Protocol.} 
Now we summarize the lattice surgery across different codes, or on hybrid codes, and how it is applied to realizing non-Clifford operations. The general setup that we consider is as follows: we start with two code patches that realize quantum doubles for groups $G$ and $G'$~\cite{Kitaev:1997wr}, which crucially need not be the same and can be non-Abelian, as in Fig.~\ref{TwoPatch}(a). 
The vertical boundaries are chosen to be rough (Dirichlet), the horizontal smooth (Neumann). A rough merge is achieved by performing a particular gauging operation on the rough boundaries, as in Fig.~\ref{TwoPatch}(b). The resulting configuration is a single code patch, with the interface between the two code patches providing an identification between the logical degrees of freedom on each side. In the topological
quantum field theory (TQFT) description, we will see that these are particular topological interfaces between the two topological orders $D(G)$ and $D(G')$. A rough split along the interface of the hybrid code patch in turn reverses this.

\begin{figure}
\includegraphics[width=7cm]{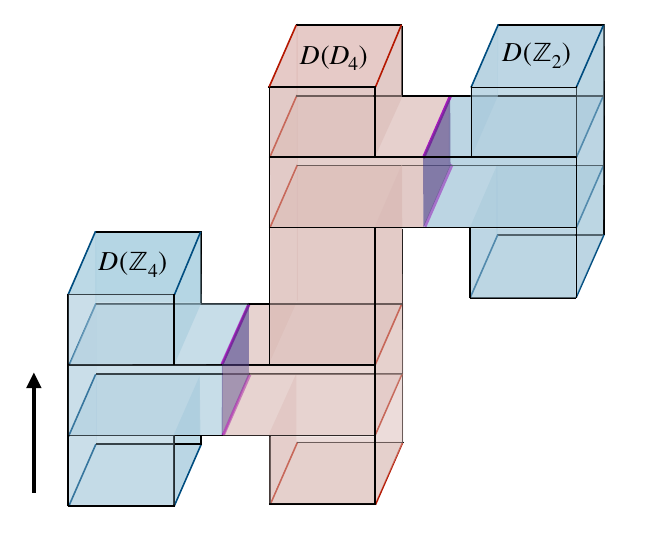}
\caption{Illustration of the lattice surgery operations required for implementing a non-Clifford operation (either $T$ magic state or gate teleportation) on the standard surface code, with time flowing from bottom to top. In practice, the time represents the QEC rounds. The three code patches involved are encoded in the quantum double surface codes associated with groups $\Z_4, D_4, \Z_2$, respectively, where the one with $\Z_2$ is the standard surface code. The protocol starts by initializing a stabilizer state in the $D(\mathbb{Z}_4)$ code patch and performing logical measurements via a sequence of rough merge and split operations, followed by additional measurements in the $D(D_4)$ code patch. The two merge-split processes on the left and right can happen simultaneously. We omit the other measurements or transversal gates in the protocol in this diagram.
\label{fig:Lego} }
\end{figure}

We use this procedure to generate non-Clifford operations in the standard surface code, which is a quantum double with group $\Z_2$. In our method, we first prepare three code patches of quantum doubles with groups $\Z_4$, $D_4$, and $\Z_2\times\Z_2$ (or $\Z_2$), respectively, in some easy-to-prepare states. Then we carry out a rough merge and a rough split between the code patches with $\Z_4$ and $D_4$, followed by a rough merge and a rough split between the code patches with $D_4$ and $\Z_2\times\Z_2$ (or $\Z_2$). Together with some measurements of logical operators, this procedure can either provide a magic state or teleport a non-Clifford gate. The two rough merge-split processes on the two interfaces commute with each other and can be carried out simultaneously. In contrast, one protocol in Ref.~\cite{Davydova:2025ylx} implements a logical non-Clifford entangling gate using three surface code patches by gauging a symmetry of the middle patch together with two interfaces, where operations in the bulk and at the interfaces are treated as a whole. Fig.~\ref{fig:Lego} illustrates the lattice surgery operations in this method. Fault tolerance of our protocols can be achieved by using matching decoders to perform error correction in the Abelian code patches and by using just-in-time decoders~\cite{Bombin:2018wjx,Brown2020,Scruby:2020pvw,Davydova:2025ylx} in the $D(D_4)$ code as well as the hybrid code patches.

\noindent{\bf Continuum TQFT Perspective.}
The proposed protocol relies strongly on our understanding of topological order---in the current paper in the form of quantum doubles $D(G)$ of possibly non-Abelian finite groups $G$, which underlie the surface code. These TQFTs provide a continuum description complementing the lattice approach. 
Central to determining  the results of the lattice surgery (and proposing new hybrid lattice surgeries) is the knowledge of interfaces between topological orders---Abelian or non-Abelian. This provides a gateway to interesting quantum information theoretical applications, such as magic states and non-Clifford gate implementations. 
In the context of lattice surgery, the actual lattice realization is of course absolutely crucial and guides potential future implementations. However, to systematically search for viable interfaces, it is also useful to develop a continuum TQFT picture, which provides a more conceptual perspective and enables a comprehensive exploration of potential lattice surgery setups.

In the second half of the paper we develop such a TQFT picture: 
the rough merge has an interpretation as gauging a non-factored subgroup on the rough boundaries between two 2D (i.e. 2+1 spacetime dimensional) topological orders $D(G)$ and $D(G')$. Such interfaces, or related ones, have been recently studied in the very different context of phases of matter with generalized symmetries, in particular, so-called non-invertible or categorical symmetries (see~\cite{Schafer-Nameki:2023jdn, Shao:2023gho} for recent reviews of this field). 
The anyons, which are the topological line operators in the quantum doubles can either end (condense), pass through, or create anyon excitations (confine) in the presence of an interface. This is depicted in Fig.~\ref{fig:Condensation}.

\begin{figure}
$$
\begin{tikzpicture}
\begin{scope}[shift={(0,0)}]
\draw [fill= \BlueColor, opacity = 0.5, draw=none]   (0,0) -- (0,3) -- (3,3) -- (3,0) --(0,0);
\node at (1.5,2.5) {$D(G)$}; 
\end{scope}
\begin{scope}[shift={(3,0)}]
\draw[fill= \RedColor, opacity = 0.5, draw=none]   (0,0) -- (0,3) -- (3,3) -- (3,0) --(0,0);
\node at (1.5,2.5) {$D(G')$}; 
\node[above,Violet] at (0,3) {$\cA$};
\draw[ultra thick, Violet] (0,3) -- (0,0); 
\end{scope}
\draw[very thick, blue] (0,2) -- (3,2);
\node[red, above] at (1.5, 1.5) {$a$};
\node[red, above] at (4.5, 1.5) {$a'$};
\draw[very thick, red] (0,1.5) -- (6,1.5);
\draw[very thick, green] (0,1) -- (2.95,1) -- (2.95,0);
\end{tikzpicture}
$$
\caption{Anyons (topological line operators) in $D(G)$ can either condense, i.e. terminate, (blue) on the interface $\cA$, pass through it (red), or confine (green). \label{fig:Condensation}}
\end{figure}

In this case the various boundary conditions can incorporate generalized symmetries: the rough boundary of $D(G)$ carries the group symmetry $G$, whereas the smooth boundary the non-invertible (or categorical) symmetry generated by the irreducible representations (irreps) of $G$, $\Rep(G)$. Using the technology developed in this context~\cite{Bhardwaj:2023idu,Bhardwaj:2023bbf, Bhardwaj:2024qrf} we can recast the question of lattice surgery in terms of interfaces between (2+1)d topological orders and apply it to search for interesting settings. 

This approach would be more generally applicable to any (2+1)d (or even higher dimensional) topological field theory, but for the quantum doubles of groups, the situation is even more favorable, as a full classification of interfaces can be obtained using  the mathematical classification of Refs.~\cite{Ostrikmodule,Davydov2009ModularIF,Natale2017,davydov2017lagrangian} whose physical interpretation is discussed in e.g. Refs.~\cite{Beigi:2010htr,delaFuente:2023whm,GaiSchaferNamekiWarman}. This allows us, for example, to find the groups that can realize various non-Clifford magic states and gates such as $T^{1/n}$, systematically. The TQFT analysis can then be readily implemented on the lattice.

This connection between continuum TQFT description and quantum information applications has been very fruitful in various areas, e.g.~\cite{Huang:2025cvt, Davydova:2025ylx}.  In these applications, the interfaces were specified by particular condensation patterns of anyons. For lattice surgery the TQFT picture has to be interpreted in a slightly different fashion. The logical operations are given in terms of the anyon lines in a given patch that end on opposite boundaries. In particular, after the merge, the patch is a combination of $D(G)$ and $D(G')$ modulo an identification of anyons between the two patches. This is distinct from say anyon condensation, where we would map anyons across the interface, from a larger topological order to a smaller one. Here we simply provide an identification whose equivalence classes realize the computational basis in the merged patch. 

There are two extensions which we do not discuss in this paper in detail: in standard lattice surgery, the topological orders involved are Abelian and one can easily study also merge and split along a smooth boundary. As noted above, for non-Abelian quantum doubles $D(G)$, the smooth boundaries correspond to the non-invertible symmetry $\Rep (G)$. Gauging a common sub-symmetry to realize a smooth merge along this boundary is possible, but generically requires gauging a non-invertible symmetry. This could be interesting, but we only discuss some general points at the end of Sec.~\ref{sec:lattice_surgery_interpretation} and leave thorough explorations to the future. 

The other extension is to consider the rough merge with additional cocycle data: already for Abelian groups, one can ask whether before gauging a common subgroup it is possible to first stack with an SPT for the subgroup and then gauge. This is entirely feasible both from the TQFT and lattice model perspective. 
 The quantum information theoretic interpretation remains however more elusive and is another area for future exploration. 
The interfaces that arise in our merge and split operations do not require such SPT stacking.

\noindent{\bf Plan.} 
The plan of the paper is as follows: We start in Sec.~\ref{sec:lattice_surgery} with a general discussion of lattice surgery on QEC code patches protected by quantum double models with arbitrary finite groups $G$. This in particular generalizes the standard lattice surgery between the same code to cases between an Abelian code and a non-Abelian code. We give a description of the rough merge and the rough split, followed by a brief discussion of the merged code patch. 
As an application of the lattice surgery with non-Abelian TO, we construct a protocol in Sec.~\ref{sec:magic} to implement non-Clifford operations in the surface code -- a quantum double model with group $\Z_2$. By making use of a quantum double model with group $D_4$, this produces a logical $T$ magic state or teleports a logical gate equivalent to the $T$ gate up to Clifford gates. In Sec.~\ref{sec:qec}, we discuss the error correction procedure which provides fault tolerance for our protocol.

This lattice part is complemented by a TQFT analysis, where again we first discuss the general setup of non-Abelian lattice surgery from a continuum TO point of view in Sec.~\ref{sec:cont}. We determine what the merge and split correspond to in terms of anyons in the TQFT description. An important input into this analysis is the characterization of interfaces between TOs, which we discuss in detail--both for rough and smooth merge and split. 
This is then applied in Sec.~\ref{sec:ContNonCliff} to universal quantum computation through magic state preparation or non-Clifford gate implementation.

\section{Lattice Surgery with Quantum Double Models} \label{sec:lattice_surgery}

In this section we present the lattice surgery operations between two quantum double model code patches associated with two finite groups $G$ and $G^\prime$. This is a generalization of standard lattice surgery setups in that we do not require the groups to be Abelian or the groups to be the same for both patches. With a suitable choice of boundary conditions, we will discuss the logical operators for a code patch with a non-Abelian $G$, and understand the effect of lattice surgery in terms of the logical operators. In particular, we will focus on the combined effect of a rough merge followed by a rough split on the two code patches. In the special case where $G^\prime \cong G$ for a general finite group $G$, the standard lattice surgery between two code patches protected by the same surface code is recovered~\cite{Cowtan2023algebraic, Cowtan:2025vok}. 

\subsection{$D(G)$ Quantum Doubles as Code Patches} 
\label{subsec:DG}

In this section, we review Kitaev's quantum double model $D(G)$~\cite{Kitaev:1997wr, Beigi:2010htr, Li2025_QDboundary} and place it on a square lattice, with certain boundary conditions, to form a code patch.  In general, a code patch contains a collection of physical degrees of freedom that are prepared in some entangled states. 
Only a smaller number of logical degrees of freedom are encoded in a code subspace of the entire Hilbert space of the code patch and are protected against certain errors.
The quantum double model is defined for a finite group $G$ on an oriented lattice, where on each edge there is a local Hilbert space $\cH=\bC[G]$ equipped with an orthonormal basis labeled by the group elements $\{\ket{h}\;:\;h\in G\}$. We will refer to these local degrees of freedom as $G$-qudits, and the group element basis as the computational basis. The basis states are acted upon by left and right multiplication operators for each $g\in G$ 
\be\ba \label{eq:LgRg}
    L^g\ket{h}&=\ket{gh}, &  R^g\ket{h}&=\ket{hg^{-1}}.\\
\ea\ee
The diagonal projection operators are
\be\ba \label{eq:Tops}
    T_+^g\ket{h}&=\delta_{g,h}\ket{h}, &  T_-^g\ket{h}&=\delta_{g^{-1},h}\ket{h}.
\ea\ee
For a vertex $v$, we define an operator,
\be
    A_v^{(g)}=\prod_{v\rightarrow l} \prod_{l^\prime\rightarrow v} L^g_l R^g_{l^\prime}
\ee
for a group element $g\in G$, where $v\rightarrow l$ and $l^\prime\rightarrow v$ are shorthand notations for the edges starting from and ending at $v$, respectively. For a plaquette $p$, we define
\be
    B_p^{(g)}=\sum_{g_1\in G, g_2\in G, \dots} \delta_{g, \text{P}(g_1, g_2, \dots)} \ket{g_1, g_2, \dots}\bra{g_1, g_2, \dots},
\ee
where $1, 2, \dots$ label the edges around $p$ in a clockwise order. $\text{P}(g_1, g_2, \dots)$ is the product of the group element (or its inverse) at the edge if the orientation of the edge is the same as (or opposite to) the clockwise orientation of the plaquette. 
Placed on a two-dimensional square lattice, where the orientations are chosen to be upward for all vertical edges and rightward for all horizontal edges, these operators are
\be\ba
    A_v^{(g)}= \, & \begin{tikzpicture}[baseline]
\begin{scope}[shift={(0,0)}]
\draw[thick, ->-] (-1,0) to (0,0);
\draw[thick, ->-] (0,0) to (1,0);
\draw[thick, ->-] (0,-1) to (0,0);
\draw[thick, ->-] (0,0) to (0,1);
\node[above] at (0.2, 0) {$v$};
\node[above] at (-0.7,0.05) {$R^g$};
\node[above] at (0.7,0.05) {$L^g$};
\node[right] at (0.05,0.7) {$L^g$};
\node[right] at (0.05,-0.7) {$R^g$};
\draw[fill=black] (0,0) ellipse (0.05 and 0.05);
\end{scope}
\end{tikzpicture},\\
    B_p^{(g)}= & \sum\limits_{g_1,g_2,g_3,g_4}\delta_{g,g_1g_2g_3^{-1}g_4^{-1}} \times \cr 
&
\qquad 
\times  \begin{tikzpicture}[baseline]
\begin{scope}[shift={(0,0)}]
    \draw[->-] (-1,-0.5) to (-1,0.5);
    \node at (-1.5,0) {$\Bigg|$};
    \node at (-1.3,0) {$_{g_1}$};
    \draw[->-] (-1,0.5) to (0,0.5);
    \node at (-0.5,0.8) {$_{g_2}$};
    \draw[->-] (0,-0.5) to (0,0.5);
    \node at (0.3,0) {$_{g_3}$};
    \draw[->-] (-1,-0.5) to (0,-0.5);
    \node at (-0.5,-0.8) {$_{g_4}$};
    \node at (0.5,0) {$\Bigg>$};
    \node at (-0.5,0) {$_p$};
    \begin{scope}[shift={(2.2,0)}]
    \draw[->-] (-1,-0.5) to (-1,0.5);
    \node at (-1.5,0) {$\Bigg<$};
    \node at (-1.3,0) {$_{g_1}$};
    \draw[->-] (-1,0.5) to (0,0.5);
    \node at (-0.5,0.8) {$_{g_2}$};
    \draw[->-] (0,-0.5) to (0,0.5);
    \node at (0.3,0) {$_{g_3}$};
    \draw[->-] (-1,-0.5) to (0,-0.5);
    \node at (-0.5,-0.8) {$_{g_4}$};
    \node at (0.5,0) {$\Bigg|$};
    \node at (-0.5,0) {$_p$};        
    \end{scope}
\end{scope}    
\end{tikzpicture}.
\ea\ee
The quantum double $D(G)$ Hamiltonian is the sum of projectors at all the vertices and plaquettes in the lattice  
\be\ba
    H_{G} = -\sum_v A_v - \sum_p B_p^{(\id)},
\ea\ee
where
\be
    A_v=\frac{1}{\vert G\vert}\sum_{g\in G} A_v^{(g)}.
\ee

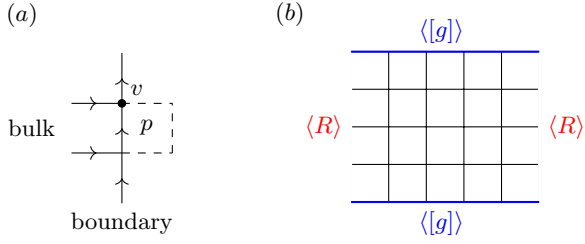
\begin{figure}
\begin{minipage}{0.4\linewidth}
\begin{tikzpicture}
\node[black, left] at (-1,2.5) {$(a)$} ;
\draw[black, ->-] (0,0) -- (0,0.67);
\draw[black, ->-] (0,0.67) -- (0,1.33);
\draw[black, ->-] (0,1.33) -- (0,2);
\draw[black, ->-] (-0.67,0.67) -- (0,0.67);
\draw[black, ->-] (-0.67,1.33) -- (0,1.33);
\draw[black, dashed] (0,0.67) -- (0.67,0.67);
\draw[black, dashed] (0,1.33) -- (0.67,1.33);
\draw[black, dashed] (0.67,0.67) -- (0.67,1.33);
\node[black, below] at (0,0) {boundary} ;
\node[black] at (-1.2,1) {bulk};
\draw[fill=black] (0,1.33) ellipse (0.05 and 0.05);
\node[black, above] at (0.2,1.33) {$v$} ;
\node[black, above] at (0.33,0.8) {$p$} ;
\end{tikzpicture}
\end{minipage}
\hspace{0.1mm}
\begin{minipage}{0.45\linewidth}
\centering
\begin{tikzpicture}
\node[black, left] at (-0.5,2.5) {$(b)$} ;
\draw[step=0.5,black] (0,0) grid (2.5,2);
\draw[white, thick] (2.5,0) -- (2.5,2);
\draw[white, thick] (0,0) -- (0,2);
\draw[blue, thick] (0,2) -- (2.5,2);
\draw[blue, thick] (0,0) -- (2.5,0);
\node[blue, above] at (1.2,2) {$\langle [g] \rangle$} ;
\node[blue, below] at (1.2,0) {$\langle [g] \rangle$} ;
\node[red, right] at (2.5,1) {$\langle  R  \rangle$} ;
\node[red, left] at (0,1) {$\langle  R  \rangle$} ;
\end{tikzpicture}
\end{minipage}
\caption{(a) A boundary site $(v,p)$. (b) The boundary conditions for a $D(G)$ code patch, given by the labels of the anyons that can end on the boundaries. $R$ is an irreducible representation (irrep) of $G$ and labels an electric anyon while $[g]$ is a conjugacy class and labels a magnetic anyon.}
\label{fig:DG_BC}
\end{figure}

A gapped boundary of this Hamiltonian can be specified by a subgroup and a second cohomology class $(K,\vp)$, where $K\subseteq G$ and $\vp\in H^2(K,U(1))$. In addition to the 2-cocycle condition
\be \label{eq:2coc}
\vp(g_1,g_2)\vp(g_1g_2,g_3)=\vp(g_1,g_2g_3)\vp(g_2,g_3)\,,
\ee
the representative of $\vp$ is taken to satisfy~\cite[Lemma 5.1]{Beigi:2010htr} in order to obtain a well-defined Hamiltonian for the quantum double with boundary~\cite{Beigi:2010htr}:
\begin{enumerate}
    \item $\vp(\id,g)=\vp(g,\id)=1$,
    \item $\vp(g,g^{-1})=1$,
    \item $|\vp(g,h)|=1$,
    \item $\vp(g^{-1},h^{-1})=\vp(h,g)^{-1}$.
\end{enumerate}
In Fig.~\ref{fig:DG_BC}(a), we illustrate a boundary site labeled by $(v,p)$. For $(K,\vp)$, the Hamiltonian terms at $(v,p)$ are 
\be\ba
    & \wt{A}_v^{K,\vp} = \frac{1}{\vert K\vert} \sum_{k\in K} \wt{A}_v^{(k)}, \\
    & B_p^{K} = \sum_{k\in K} B_p^{(k)}.
    \label{eq:boundaryAB}
\ea\ee
The term $B_p^{K}$ projects the $G$-qudit on each vertical edge at the boundary to the subspace spanned by $\{\ket{k},\; k\in K\}$. The term $\wt{A}_v^{(k)}$ acts on the boundary in the following way:
\be 
    \wt{A}_v^{(k)} \begin{tikzpicture}[baseline]
\begin{scope}[shift={(0,0)}]
\draw[thick, ->-] (-1,0) to (0,0);
\draw[thick, ->-] (0,-1) to (0,0);
\draw[thick, ->-] (0,0) to (0,1);
\node[] at (0.2, 0) {$v$};
\node[above] at (-0.7,0.05) {$\ket{z}$};
\node[right] at (0.05,0.7) {$\ket{y}$};
\node[right] at (0.05,-0.7) {$\ket{x}$};
\draw[fill=black] (0,0) ellipse (0.05 and 0.05);
\end{scope}
\end{tikzpicture}
= \vp(x, k^{-1})\vp(k, y) \begin{tikzpicture}[baseline]
\begin{scope}[shift={(0,0)}]
\draw[thick, ->-] (-1,0) to (0,0);
\draw[thick, ->-] (0,-1) to (0,0);
\draw[thick, ->-] (0,0) to (0,1);
\node[] at (0.2, 0) {$v$};
\node[above] at (-0.7,0.05) {$\ket{zk^{-1}}$};
\node[right] at (0.05,0.7) {$\ket{ky}$};
\node[right] at (0.05,-0.7) {$\ket{xk^{-1}}$};
\draw[fill=black] (0,0) ellipse (0.05 and 0.05);
\end{scope}
\end{tikzpicture}.
\ee
These operators satisfy the conditions 
\be\ba
    & \wt{A}_v^{(k)}\wt{A}_v^{(k^\prime)}=\wt{A}_v^{(kk^\prime)}, \\
    & \left(\wt{A}_v^{(k)}\right)^\dagger=\wt{A}_v^{(k^{-1})}.
\ea\ee
The Hamiltonian with a boundary is then
\be\ba
    \tilde{H}_{G} = -\sum_{v} A_v - \sum_{p} B_p^{(\id)} - \sum_{(v,p)} \left(\wt{A}_v^{K,\vp} + B_p^{K}\right),
\ea\ee
where the first two terms are summed over bulk vertices $v$ and plaquettes $p$ and the last term is summed over boundary sites $(v,p)$. Below we specify the boundary choice for a $D(G)$ code patch, and later we will discuss the interface between two code patches $D(G)$ and $D(G^\prime)$ by interpreting it as a boundary of the quantum double $D(G\times G^\prime)$ code patch.

For a {\bf $D(G)$ code patch}, we choose the top and bottom boundaries of the square lattice to be smooth, and the left and right boundaries to be rough, as shown in Fig.~\ref{fig:DG_BC}(b). The terms at the boundary vertices and plaquettes are simply the Hamiltonian terms $A_v$ and $B_p^{(\id)}$ truncated according to the lattice. This is equivalent to choosing $K=G, \vp=1$ for the top and bottom boundaries, and $K=\{\id\}, \vp=1$ for the left and right boundaries. 
With this choice, the electric anyons, labeled by the irreducible representations (irreps) of $G$, condense at the left and right boundaries, and the magnetic anyons, labeled by the conjugacy classes, condense at the top and bottom boundaries. Then we have the identification of code space and logical operators as follows: 
\begin{itemize}
\item The ground space of the Hamiltonian is the {\bf code space}. The code space of a $D(G)$ code patch has dimension $|G|$. 
\item The anyon lines that end on opposite boundaries form {\bf logical operators}.
\end{itemize}

While the anyons for the quantum double with a non-Abelian $G$ are more complicated than the Abelian case\footnote{{Each anyon is labeled as $([g],R)$ where $[g]$ is a conjugacy class of $G$ and $R$ is an irrep of the centralizer group $C_G(g)$. Anyons of the form $([g],1_{\text{trivial}})$ are called ``magnetic'' (or pure fluxes) while the $(\id,R)$ anyons are called ``electric'' (or pure charges). }}, we can move them to the boundaries, where they take simpler forms~\cite{Bhardwaj:2023ayw,Li2025_QDboundary}. The TQFT description of the boundary conditions for the $D(G)$ code patch and logical operators is provided in Sec. \ref{sec:cont_BCs}.

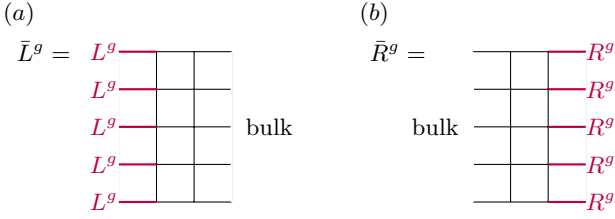
\begin{figure}
\begin{minipage}{0.45\linewidth}
\begin{tikzpicture}
\node[black, left] at (-1,2.5) {$(a)$} ;
\node[black] at (-1,2) {$\bar{L}^g=$};
\draw[step=0.5,black] (0,0) grid (1.5,2);
\draw[white, thick] (1.5,0) -- (1.5,2);
\draw[white, thick] (0,0) -- (0,2);
\draw[purple, thick] (0,0) -- (0.5,0);
\draw[purple, thick] (0,0.5) -- (0.5,0.5);
\draw[purple, thick] (0,1) -- (0.5,1);
\draw[purple, thick] (0,1.5) -- (0.5,1.5);
\draw[purple, thick] (0,2) -- (0.5,2);
\node[purple] at (-0.2,0) {$L^g$};
\node[purple] at (-0.2,0.5) {$L^g$};
\node[purple] at (-0.2,1) {$L^g$};
\node[purple] at (-0.2,1.5) {$L^g$};
\node[purple] at (-0.2,2) {$L^g$};
\node[black] at (2,1) {bulk};
\end{tikzpicture}
\end{minipage}
\hspace{5mm}
\begin{minipage}{0.45\linewidth}
\begin{tikzpicture}
\node[black, left] at (-1,2.5) {$(b)$} ;
\node[black] at (-1,2) {$\bar{R}^g=$};
\draw[step=0.5,black] (0,0) grid (1.5,2);
\draw[white, thick] (1.5,0) -- (1.5,2);
\draw[white, thick] (0,0) -- (0,2);
\draw[purple, thick] (1,0) -- (1.5,0);
\draw[purple, thick] (1,0.5) -- (1.5,0.5);
\draw[purple, thick] (1,1) -- (1.5,1);
\draw[purple, thick] (1,1.5) -- (1.5,1.5);
\draw[purple, thick] (1,2) -- (1.5,2);
\node[purple] at (1.7,0) {$R^g$};
\node[purple] at (1.7,0.5) {$R^g$};
\node[purple] at (1.7,1) {$R^g$};
\node[purple] at (1.7,1.5) {$R^g$};
\node[purple] at (1.7,2) {$R^g$};
\node[black] at (-0.5,1) {bulk};
\end{tikzpicture}
\end{minipage}
\caption{A logical operator labeled by $g$ along the left (a) or the right (b) boundary.}
\label{fig:logical_g}
\end{figure}

On the lattice, anyon lines are realized in terms of ribbon operators, which we briefly summarize in Appendix~\ref{app:ribbon}.
Specifically, we will consider the ribbon operators that make up the magnetic anyon lines ending on the top and bottom boundaries, placed along either the left or the right boundary. At the left (right) boundary, these become a product of left (right) multiplication operators, as illustrated in Fig.~\ref{fig:logical_g}. They are labeled by a group element $g\in G$ and commute with the Hamiltonian. We define them as the logical left and right multiplication operators -- $\{\bar{L}^g\}$ and $\{\bar{R}^g\}$. They satisfy
\be\ba
    \bar{L}^g\bar{L}^h&=\bar{L}^{gh}, & \bar{R}^g\bar{R}^h&=\bar{R}^{gh}, & \bar{L}^g\bar{R}^h&=\bar{R}^{h}\bar{L}^g.
\ea\ee

To form a convenient logical basis, we first define a logical fiducial state as:
\be
    \ket{\Phi_\id} = \vert G\vert^{N_v/2} \left(\prod_{v} A_v\right)\bigotimes_{l} \ket{\id}_l,
\ee
where $N_v$ is the number of vertices and $A_v$'s are truncated on the top and bottom boundaries.  

Then, we act with the logical operators $\{\bar{L}^g\}$ on the fiducial state to obtain the other basis states:
\be
    \ket{\Phi_g} = \bar{L}^g \ket{\Phi_\id}.
\ee
One can check that $\braket{\Phi_g}{\Phi_h}=\delta_{g,h}$. The effect of a logical right multiplication operator is
\be
    \bar{R}^g \ket{\Phi_h} = \ket{\Phi_{hg^{-1}}}.
\ee

We can form another set of logical operators using the horizontal ribbon operators $\{F_{\xi_{H}}^{e,g}\}$, labeled by group elements $\{g\in G\}$, where $\xi_{H}$ is a horizontal ribbon going from left to right:
\be
\begin{tikzpicture}
\fill[gray!20] (0,2) -- (0.25,1.75) -- (2.25,1.75) -- (2.5,2) -- cycle;
\draw[step=0.5,black] (0,0) grid (2.5,2);
\draw[white, thick] (2.5,0) -- (2.5,2);
\draw[white, thick] (0,0) -- (0,2);
\draw[black] (0,0) -- (2.5,0);
\draw[black, dotted] (0,2) -- (0.25,1.75);
\draw[black, dotted] (0.5,2) -- (0.75,1.75);
\draw[black, dotted] (1,2) -- (1.25,1.75);
\draw[black, dotted] (1.5,2) -- (1.75,1.75);
\draw[black, dotted] (2,2) -- (2.25,1.75);
\draw[black, dotted] (0.25,1.75) -- (0.5,2);
\draw[black, dotted] (0.75,1.75) -- (1,2);
\draw[black, dotted] (1.25,1.75) -- (1.5,2);
\draw[black, dotted] (1.75,1.75) -- (2,2);
\draw[black, dotted] (2.25,1.75) -- (2.5,2);
\draw[black, dotted] (0.25,1.75) -- (2.25,1.75);
\end{tikzpicture}
\ee 
We omit the explicit forms of these logical operators, as they are complicated and we do not need to implement them during the lattice surgery protocol. Mathematically, these operators can be used to distinguish the logical basis states
\be
F_{\xi_{H}}^{e,h} \ket{\Phi_g} = \delta_{h,g} \ket{\Phi_g}. 
\ee
The definition of ribbon operators is included in Appendix~\ref{app:ribbon}. These logical operators are related to the ribbon operators of the electric anyon lines $\{F_{\xi_{H}}^{R,ij}\}$ through the Fourier transformation
\be
F_{\xi_{H}}^{R,ij} = \frac{d_{R}}{|G|} \sum_{g \in G} R(g)_{ij} F_{\xi_{H}}^{e,g},
\ee
where $R$ denotes an irrep of $G$ and $d_{R}$ is the dimension of $R$.

\noindent\textbf{Measurements on $G$-qudits.}
In addition to the left and right multiplications $L^g, R^g$, we may also need to measure the $G$-qudits. A measurement of a $G$-qudit in the computational basis projects it to one of the states $\{\ket{g}\;|\; g\in G\}$. When a non-Hermitian (but normal) operator is measured, it means a projective measurement in the eigenbasis of that operator is carried out, where the outcomes are labeled by the eigenvalues. Measurement of an operator supported on multiple $G$-qudits can be realized using entangling gates and ancilla qudits. 
For example, to measure the operator $A_v^{(g)}$ at some vertex $v$ for a group element $g$, we can introduce an ancillary $G$-qudit prepared in the state that is the trivial irrep
\be
    \ket{1_{\rm trivial}} = \frac{1}{\sqrt{\vert{G}\vert}} \sum_{h\in G}\ket{h},
\ee
and apply controlled group multiplication operators targeted on the qudits at the associated edges~\cite{Tantivasadakarn:2022hgp}. The controlled left and right multiplication operators are defined as
\be\ba
    CL_{v-l} = \sum_{h\in G} \ket{h}\bra{h}_v L_l^{h}, \\
    CR_{v-l} = \sum_{h\in G} \ket{h}\bra{h}_v R_l^{h},
\ea\ee
respectively, with the control and the target on qudits at the locations $v$ and $l$. The entangling gates are chosen according to the edge orientations, which are
\be
U_v = \, \begin{tikzpicture}[baseline]
\begin{scope}[shift={(0,0)}]
\draw[thick, -] (-1,0) to (0,0);
\draw[thick, -] (0,0) to (1,0);
\draw[thick, -] (0,-1) to (0,0);
\draw[thick, -] (0,0) to (0,1);
\path[thick, ->] edge[bend right=90] node (0,0) {} (-0.5,0);
\path[thick, ->] edge[bend right=90] node (0,0) {} (0.5,0);
\path[thick, ->] edge[bend right=90] node (0,0) {} (0,-0.5);
\path[thick, ->] edge[bend right=90] node (0,0) {} (0,0.5);
\node[above] at (0.3, 0) {$v$};
\node[above] at (-0.7,0.05) {$CR$};
\node[below] at (0.7,-0.05) {$CL$};
\node[right] at (0.05,0.7) {$CL$};
\node[left] at (-0.05,-0.7) {$CR$};
\draw[fill=black] (0,0) ellipse (0.05 and 0.05);
\end{scope}
\end{tikzpicture},
\ee
on our square lattice. Then the ancillary $G$-qudit is measured in the eigenbasis of $L^g$, where obtaining the eigenvalue $\alpha_g$ corresponds to the measurement outcome of $A_v^{(g)}$ being $\alpha_g^*$. 
To enforce the projector $A_v$ at some vertex $v$, we replace the measurement of the ancillary $G$-qudit with the projection onto the state $\ket{1_{\rm trivial}}$~\cite{Tantivasadakarn:2022hgp,Bravyi:2022zcw}. The projection can be realized by a measurement in the basis given by the irreps of $G$ in the regular representation. When the measurement outcome is the trivial irrep state, no further action is required. If the outcome is one of the nontrivial 1d irrep states, $A_v$ can also be enforced. This can be done by redefining $A_v$ according to feedback from the measurement outcomes
\be
    A_v=\frac{1}{\vert G\vert}\sum_{g\in G} \alpha_g A_v^{(g)},
\ee
when the measurement projects the state to the nontrivial 1d irrep state $\ket{1_\alpha}$ satisfying $L^g\ket{1_\alpha}=\alpha_g\ket{1_\alpha}$. For a solvable group $G$, the projector $A_v$ can be applied via multiple steps~\cite{Tantivasadakarn:2022hgp,Bravyi:2022zcw}.

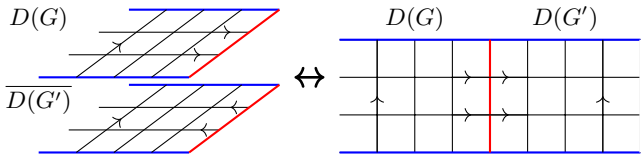
\begin{figure}
    \centering
    \resizebox{\columnwidth}{!}{
    \begin{tikzpicture}
    \draw[blue, thick] (0,0) -- (2,0);
    \draw[black, -] (0.4,0.3) -- (1.9,0.3);
    \draw[black, -] (0.8,0.6) -- (2.3,0.6);
    \draw[blue, thick] (1.2,0.9) -- (3.2,0.9);
    \draw[black, ->-]  (2.4,0.3) -- (1.9,0.3);
    \draw[black, ->-]  (2.8,0.6) -- (2.3,0.6);
    \draw[black, ->-] (0.5,0) -- (1.7,0.9);
    \draw[black, -] (1,0) -- (2.2,0.9);
    \draw[black, -] (1.5,0) -- (2.7,0.9);
    \draw[blue, thick] (0,1) -- (2,1);
    \draw[black, -] (0.4,1.3) -- (1.9,1.3);
    \draw[black, -] (0.8,1.6) -- (2.3,1.6);
    \draw[blue, thick] (1.2,1.9) -- (3.2,1.9);
    \draw[black, ->-] (1.9,1.3) -- (2.4,1.3);
    \draw[black, ->-] (2.3,1.6) -- (2.8,1.6);
    \draw[black, ->-] (0.5,1) -- (1.7,1.9);
    \draw[black, -] (1,1) -- (2.2,1.9);
    \draw[black, -] (1.5,1) -- (2.7,1.9);
    \draw[red, thick] (2,0) -- (3.2,0.9);
    \draw[red, thick] (2,1) -- (3.2,1.9);
    \node[black, above] at (0,0.4) {$\ol{D(G^\prime)}$};
    \node[black, above] at (0,1.5) {${D(G)}$};
    \draw[<->, line width=1pt] (3.4,1) -- (3.8,1);
    \draw[step=0.5,black] (4,0) grid (8,1.5);
    \draw[blue, thick] (4,0) -- (8,0);
    \draw[blue, thick] (4,1.5) -- (8,1.5);
    \draw[red, thick] (6,0) -- (6,1.5);
    \draw[white, thick] (8,0) -- (8,1.5);
    \draw[black, ->-] (5.5,0.5) -- (6,0.5);
    \draw[black, ->-] (5.5,1) -- (6,1);
    \draw[black, ->-] (6,0.5) -- (6.5,0.5);
    \draw[black, ->-] (6,1) -- (6.5,1);
    \draw[black, ->-] (4.5,0) -- (4.5,1.5);
    \draw[black, ->-] (7.5,0) -- (7.5,1.5);
    \node[black, above] at (5,1.5) {$D(G)$};
    \node[black, above] at (7,1.5) {$D(G^\prime)$};
    \end{tikzpicture}
    }
    \caption{The interface between two code patches $D(G)$ and $D(G^\prime)$ viewed as a boundary of a $D(G\times G^\prime)$ patch. In the lower half of the folded patch (the left hand side), the edge orientations are reversed compared to the patch with an interface (the right hand side) and we denote the half patch with $\ol{D(G^\prime)}$.}
    \label{fig:fold}
\end{figure}

\subsection{Rough Merge} \label{sec:Rough_merge}

A rough merge describes a procedure where the rough boundaries of two code patches $D(G)$ and $D(G^\prime)$ are brought together and merged into one hybrid code patch. We begin our discussion by examining the possible interfaces between the two code patches. If we fold the $D(G^\prime)$ code patch to below the $D(G)$ code patch, the interface now becomes a boundary of a $D(G\times G^\prime)$ code patch and as such can be specified by a subgroup $K\subseteq G\times G^\prime$ and a second cohomology class $\vp\in H^2(K,U(1))$. Fig.~\ref{fig:fold} describes this perspective. 
When $\vp=1$, and in case $K$ is a direct product of subgroups of $G$ and $G^\prime$, the merge is trivial and the two code patches remain disjoint. 
 A special case $K=\{\id\}, \vp=1$ is when the two code patches are disjoint {with rough boundary conditions}, which describes the situation before the rough merge.

\begin{figure}
    \centering
    \begin{tikzpicture}
    \draw[step=0.5,black] (4,0) grid (8,1.5);
    \draw[black, thick] (4,0) -- (8,0);
    \draw[black, thick] (4,1.5) -- (8,1.5);
    \draw[white, thick] (8,0) -- (8,1.5);
    \draw[black] (5.5,0.5) -- (6,0.5);
    \draw[black] (5.5,1) -- (6,1);
    \draw[black] (6,0.5) -- (6.5,0.5);
    \draw[black] (6,1) -- (6.5,1);
    \draw[black] (4.5,0) -- (4.5,1.5);
    \draw[black] (7.5,0) -- (7.5,1.5);
    \draw[white, ultra thick] (6,-0.1) -- (6,1.7);
    \node[black, above] at (5,1.5) {$D(G)$};
    \node[black, above] at (7,1.5) {$D(G^\prime)$};
    \draw[fill=black!80, draw=none] (6,0.25) ellipse (0.05 and 0.05);
    \draw[fill=black!80, draw=none] (6,0.75) ellipse (0.05 and 0.05);
    \draw[fill=black!80, draw=none] (6,1.25) ellipse (0.05 and 0.05);
    \end{tikzpicture}
    \caption{A rough merge between $D(G)$ and $D(G')$. Black dots denote the $(G \times G')$-qudits, which are initialized in the state $\ket{\id, \id}$.}
    \label{fig:rough_merge}
\end{figure}
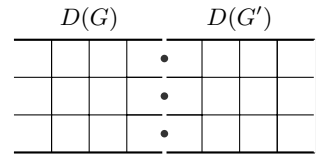

Whenever $K$ is not a direct product of subgroups of $G$ and $G^\prime$, such as when $K$ is a diagonal subgroup, there exists a corresponding merge procedure that results in a non-trivial merged code patch.
Eq.~\eqref{eq:boundaryAB} defines the terms along the folded boundary. To transition from the disjoint case to a nontrivial $K, \vp$, we first introduce a line of vertical edges on the interface, each of which contains a $G$-qudit and a $G^\prime$-qudit, as shown in Fig.~\ref{fig:rough_merge}. 
All the qudits on this line are initialized in the state $\ket{\id,\id}$, which automatically satisfies the $B_p^K=1$ for all the interface plaquettes $\{p\}$. Then the state is projected to the simultaneous $+1$ eigenstate of all the $\{\wt{A}_v^{K,\vp}\}$ terms through measurements at the interface vertices. The measurement(s) to ensure 
\be
\wt{A}_v^{K,\vp}=+1
\ee
may produce an outcome that does not satisfy this condition. For a solvable group $K$, one can redefine the corresponding $\wt{A}_v^{K,\vp}$ in the Hamiltonian or carry out recovery operations similar to error correction. Here we assume $\wt{A}_v^{K,\vp}=+1$ is satisfied for simplicity. In Sec.~\ref{sec:magic}, we will discuss what operator to measure and the possible measurement outcomes for the concrete examples in the magic state generation protocol.

If the states on the disjoint $D(G)$ and $D(G^\prime)$ code patches are $\ket{\Psi_1}_{G}$ and $\ket{\Psi_2}_{G^\prime}$, respectively, the (unnormalized) state on the joint hybrid code patch is 
\be
    \ket{\Psi_{1,2}}_{\mathrm{joint}} = \left( \prod_{v} \wt{A}_v^{K,\vp} \right) \ket{\Psi_1}_{G}\ket{\Psi_2}_{G^\prime} \bigotimes_l \ket{\id,\id}_l,
    \label{eq:joint}
\ee
where $v$ and $l$ label the interface vertices and edges. Being an eigenstate of all the Hamiltonian terms on the left patch, right patch, and the interface, this is a logical state on the hybrid code patch. Since our goal is realizing logical operations on the two separate code patches through rough merge and split, we will refrain from defining a logical basis for the hybrid code at this stage. A discussion can be found in Sec.~\ref{subsec:hybrid}.

Let us now focus on non-trivial merges: these will be labeled by subgroups $K^\diag \subset G\times G'$ which cannot be written as direct products of subgroups of $G$ and of $G'$. As discussed in the TQFT Sec.~\ref{sec:cont}, for any subgroup $K\subseteq G$, we can define a group homomorphism $p:K\rightarrow G^\prime$, where $G^\prime$ contains a subgroup $K/N$, and $N$ is a proper normal subgroup $N\triangleleft K$.

Then
\be
    K^{\diag}=\{(h,p(h))\in G\times G'\;\;\vert\;\; h\in K\}\cong K\,.
\ee
With this choice, the interface vertex term is labeled by $(K^\diag,\vp)$ and is a sum of vertex terms $\wt{A}_v^{(h,p(h))}$ acting simultaneously on both patches:
\be\ba
\wt{A}_v^{K^\diag,\vp}  
& \begin{tikzpicture}[baseline]
\begin{scope}[shift={(0,0)}]
\draw[thick, ->-] (-1,0) to (0,0);
\draw[thick, ->-] (0,-1) to (0,0);
\draw[thick, ->-] (0,0) to (0,1);
\draw[fill=black] (0,0) ellipse (0.05 and 0.05);
\node[] at (0.2, 0) {$v$};
\node[above] at (-0.7,0.05) {$\ket{z}$};
\node[right] at (0.05,0.7) {$\ket{y}$};
\node[right] at (0.05,-0.7) {$\ket{x}$};
\draw[fill=black] (0,0) ellipse (0.05 and 0.05); 
\node[black] at (0.6,0) {$\otimes$};
\draw[thick, ->-] (1,0) to (2,0);
\draw[thick, ->-] (1,-1) to (1,0);
\draw[thick, ->-] (1,0) to (1,1);
\draw[fill=black] (1,0) ellipse (0.05 and 0.05);
\node[above] at (1.8,0.05) {$\ket{p(z)}$};
\node[right] at (1.05,0.7) {$\ket{p(y)}$};
\node[right] at (1.05,-0.7) {$\ket{p(x)}$};
\end{scope}
\end{tikzpicture}\\
& = \frac{1}{\vert{K}\vert}\sum_{h\in K} \vp(x, h^{-1})\vp(h, y) \times\\
& \begin{tikzpicture}[baseline]
\begin{scope}[shift={(0,0)}]
\draw[thick, ->-] (-1,0) to (0,0);
\draw[thick, ->-] (0,-1) to (0,0);
\draw[thick, ->-] (0,0) to (0,1);
\draw[fill=black] (0,0) ellipse (0.05 and 0.05);
\node[] at (0.2, 0) {$v$};
\node[above] at (-0.7,0.05) {$R^h\ket{z}$};
\node[right] at (0.05,1) {$L^h\ket{y}$};
\node[right] at (0.05,-0.7) {$R^h\ket{x}$};
\draw[fill=black] (0,0) ellipse (0.05 and 0.05);
\node[black] at (1,0) {$\otimes$};
\draw[thick, ->-] (2,0) to (3,0);
\draw[thick, ->-] (2,-1) to (2,0);
\draw[thick, ->-] (2,0) to (2,1);
\draw[fill=black] (2,0) ellipse (0.05 and 0.05);
\node[above] at (3.2,0.05) {$L^{p(h)}\ket{p(z)}$};
\node[right] at (2.05,1) {$L^{p(h)}\ket{p(y)}$};
\node[right] at (2.05,-0.7) {$R^{p(h)}\ket{p(x)}$};
\end{scope}
\end{tikzpicture},
\label{eq:Av_interface}
\ea\ee
where the two components in the tensor product act on the $D(G)$ and the $D(G^\prime)$ patches, respectively. The unnormalized state on the hybrid code patch is 
\be\ba
    \ket{\Psi_{1,2}}_{\mathrm{joint}} &= \frac{1}{\vert K\vert^{n_v}}\sum_{\{h_v\}\in K^{n_v}} \left(\prod_v R_{v}^{h_v} \ket{\Psi_1}_{G}\right) \\
    & \otimes \left(\prod_{v^\prime} L_{v^\prime}^{p(h_{v^\prime})} \ket{\Psi_2}_{G^\prime}\right) \\
    & \bigotimes_l \left(\ket{h_{l(1)}(h_{l(2)})^{-1}, p(h_{l(1)})(p(h_{l(2)}))^{-1}} \right. \\
    & \times \left. \vp(h_{l(1)},\id)\vp(h_{l(1)}, (h_{l(2)})^{-1}) \right),
    \label{eq:joint_diag}
\ea\ee
where $l(1),l(2)$ label the starting and ending vertices of an edge $l$, respectively. The indices $v$ and $v^\prime$ run through all the $n_v$ vertices on the interface while the index $l$ runs through all the edges on the interface. An operator $R_v^{g}$ ($L_v^{g}$) acts on the horizontal edge immediately to the left (right) of the vertex $v$. The set $\{h_v\}$ denotes a sequence of group elements assigned to the interface vertices. With our chosen convention for $\vp$, $\vp(h,\id)=1$ for any $h$.

\subsection{Rough Split} \label{sec:Rough_split}

A rough split describes the procedure where a hybrid code patch is returned to the two separate code patches $D(G)$ and $D(G^\prime)$ with rough boundaries in between, by measuring all the vertical edges on the interface in the computational basis $\{\ket{g, g^\prime}\vert g\in G, g^\prime\in G^\prime\}$. Through this procedure, the disjoint interface described by $K=\{\id\}, \vp=1$ is recovered. While each non-disjoint interface corresponds to a unique merge procedure, after a subsequent rough split, the corresponding states are not necessarily all distinct. In the case of the interface defined by a diagonal subgroup, denoting the measurement outcomes by the set of group elements $\{(h_l, p(h_l))\in K^\diag\}$, the unnormalized state for the three parts -- $D(G)$ on the left, $D(G^\prime)$ on the right, and the interface line of vertical edges -- becomes
\be
    \ket{\Psi}_{\mathrm{3parts}} = \left(\bigotimes_l\ket{h_l, p(h_l)}\bra{{h_l, p(h_l)}}\right) \ket{\Psi_{1,2}}_{\mathrm{joint}},
\ee
where $l$ labels the interface edges. The state on the disjoint $D(G)$ and $D(G^\prime)$ code patches is 
\be\ba
    \ket{\Psi_{1,2}(\{h_l\})}_{\mathrm{disjoint}} &= \frac{1}{\vert K\vert^{n_v}}\sum_{k\in K} \left(\prod_{v=1}^{n_v-1}\vp(f_v^{(k)}, (f_{v+1}^{(k)})^{-1})\right) \\
    & \times \left(\prod_{v^\prime=1}^{n_v} R_{v^\prime}^{f_{v^\prime}^{(k)}} \ket{\Psi_1}_{G}\right) \\
    & \otimes \left(\prod_{v''=1}^{n_v} L_{v''}^{p(f_{v''}^{(k)})} \ket{\Psi_2}_{G^\prime}\right),
\ea\ee
where $f^{(k)}_v$ is defined as follows: for the vertex at the bottom $v=1$, $f^{(k)}_1=k$, and for $v=l(2)$, $f^{(k)}_v = h_l^{-1}\cdots h_2^{-1}h_1^{-1}k$. 

When the measurement outcomes are $h_l=\id$ for all edges $\{l\}$, the product of the cocycle factors in the equation above is $1$ as a consequence of the condition $\vp(g^{-1},h^{-1})=\vp(h,g)^{-1}$. The state becomes
\be\ba
    \ket{\Psi_{1,2}(\id,\dots)}_{\mathrm{disjoint}} &= \frac{1}{\vert K\vert^{n_v}}\sum_{k\in K} \left(\prod_{v^\prime=1}^{n_v} R_{v^\prime}^{k} \ket{\Psi_1}_{G}\right) \\
    & \otimes \left(\prod_{v''=1}^{n_v} L_{v''}^{p(k)} \ket{\Psi_2}_{G^\prime}\right).
    \label{eq:disjoint_diag}
\ea\ee
Since $\prod_{v=1}^{n_v} R_{v}^{k}$ and $\prod_{v=1}^{n_v} L_{v}^{p(k)}$ are the logical operators $\bar{R}^{k}$ and $\bar{L}^{p(k)}$ for the code patches $D(G)$ and $D(G^\prime)$, respectively, the combined effect of the rough merge and rough split is projecting the state into a logical subspace stabilized by the diagonal subgroup $K^{\diag}\subset G\times G'$. Here the logical operators in $K^{\diag}$ are $\bar{R}^{k}\otimes\bar{L}^{p(k)}$ for $k\in K\subseteq G$. 

We now restrict to the case $\vp=1$. If the measurements on some edges yield outcomes different from $\id$, the truncated plaquette terms to the left and right of those edges are violated after the split. One can update those terms in the Hamiltonian or return to the standard quantum double Hamiltonian by pairing up the violated terms along the boundary. 

\subsection{Hybrid Code Patch} \label{subsec:hybrid}

Here we briefly discuss the logical basis states of the hybrid code after the merge. Suppose the $D(G)$ and $D(G')$ patches are in the states $\ket{\Phi_{g}}$ and $\ket{\Phi_{g'}}$, respectively. After merging, we label the logical states of the hybrid code as $\ket{\Phi_{(g, g')}}$. This labeling is overcomplete since not all states $\ket{\Phi_{(g, g')}}$ are distinct. To distinguish inequivalent logical states, we use the horizontal logical operators
\begin{equation}
    T^{(h,h')}_{\xi_{H}} = \sum_{(l,l') \in K^{\text{diag}}} F_{\xi_{H}}^{(\id,\id),(h l^{-1}, l' h'^{-1})},
\end{equation}
where $h \in G$ and $h' \in G'$. For the logical state $\ket{\Phi_{(g, g')}}$, the eigenvalues of $T^{(h,h')}_{\xi_{H}}$ follow from
\begin{equation}
    T^{(h,h')}_{\xi_{H}} \ket{\Phi_{(g, g')}} = \sum_{(l,l') \in K^{\text{diag}}} \delta_{h l^{-1},g}  \delta_{l'h'^{-1},g'}  \ket{\Phi_{(g, g')}}.
\end{equation}
One can verify that the logical state $\ket{\Phi_{(g k^{-1}, k' g')}}$ has the same eigenvalues as $\ket{\Phi_{(g, g')}}$, where $(k,k') \in K^{\text{diag}}$. Thus, the logical states satisfy the equivalence relation:
\begin{equation} \label{eq:lattice_equivalence}
    \ket{\Phi_{(g, g')}} \sim \ket{\Phi_{(g k^{-1}, k' g')}} \,.
\end{equation}

\section{Non-Clifford Operations from Lattice Surgery with $D(D_4)$} \label{sec:magic}
\subsection{Overview} \label{subsec:D4}

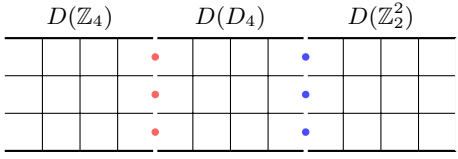
\begin{figure}
    \centering
    \begin{tikzpicture}
    \draw[step=0.5,black] (2,0) grid (8,1.5);
    \draw[black, thick] (2,0) -- (8,0);
    \draw[black, thick] (2,1.5) -- (8,1.5);
    \draw[white, thick] (8,0) -- (8,1.5);
    \draw[white, thick] (4,0) -- (4,1.5);
    \draw[white, ultra thick] (6,-0.1) -- (6,1.7);
    \draw[white, ultra thick] (4,-0.1) -- (4,1.7);
    \node[black, above] at (3,1.5) {$D(\mathbb{Z}_{4})$};
    \node[black, above] at (5,1.5) {$D(D_{4})$};
    \node[black, above] at (7,1.5) {$D(\mathbb{Z}_{2}^{2})$};
    \draw[fill=blue!70, draw=none] (6,0.25) ellipse (0.05 and 0.05);
    \draw[fill=blue!70, draw=none] (6,0.75) ellipse (0.05 and 0.05);
    \draw[fill=blue!70, draw=none] (6,1.25) ellipse (0.05 and 0.05);
    \draw[fill=red!60, draw=none] (4,0.25) ellipse (0.05 and 0.05);
    \draw[fill=red!60, draw=none] (4,0.75) ellipse (0.05 and 0.05);
    \draw[fill=red!60, draw=none] (4,1.25) ellipse (0.05 and 0.05);
    \end{tikzpicture}
    \caption{The three quantum double patches involved in the lattice surgery protocols for the magic state generation and gate teleportation. The red dots denote the $\left(\mathbb{Z}_{4} \times D_{4}\right)$-qudits. They are initialized in the state $\ket{\mathbf{0}}_{\mathbb{Z}_{4}}\ket{\mathbf{0},0}_{D_{4}}$, where the $D_4$-qudit is realized by a 4-dimensional qudit and a 2-dimensional qubit. Blue dots are the $\left(D_{4} \times \mathbb{Z}_{2}^{2}\right)$-qudits, which are initialized in the state $\ket{\mathbf{0},0}_{D_{4}}\ket{0_{\LL},0_{\RR}}_{\mathbb{Z}_{2}^{2}}$. The rightmost $D(\mathbb{Z}_{2}^2)$ patch can be replaced by a $D(\mathbb{Z}_{2})$ patch.}
    \label{fig:Z4D4Z22_magic_surgery}
\end{figure}

To generate a magic state or implement a non-Clifford gate on a standard surface code patch, i.e. a $D(\Z_2)$ code patch, we prepare a $D(\Z_4)$ code patch, a $D(D_4)$ code patch, and a $D(\Z_2\times\Z_2)$ code patch (equivalent to two $D(\Z_2)$ patches on top of each other), all with smooth top and bottom boundaries and rough left and right boundaries, as shown in Fig.~\ref{fig:Z4D4Z22_magic_surgery}. The $D(\Z_2\times\Z_2)$ patch will be the final target code patch on which either a magic state is generated or a non-Clifford gate is implemented. After we describe the entire procedure, we will show that a $D(\Z_2)$ patch instead of a $D(\Z_2\times\Z_2)$ patch can be used as the target code patch in a similar way. 

In Sec.~\ref{sec:ContNonCliff} we provide several generalizations of this method from the TQFT perspective, including gates and magic states in the whole Clifford hierarchy, while we focus on the more experimentally relevant case here.

For the non-Abelian group $D_4$, we label the elements as 
\be
D_4 = \langle r^js^b \;|\; r^4=s^2=\id \,, \ srs=r^3 \rangle \,.
\ee  
We provide a detailed discussion of the quantum double in the TQFT Sec.~\ref{sec:D4_details}. Physically, each $D_4$-qudit can be realized by a 4-dimensional qudit and a 2-dimensional qubit. The physical state $\ket{r^js^b}$ then means that the 4-dimensional qudit is in the state $\ket{\mathbf{j}}$ and the qubit is in the state $\ket{b}$.
Although the 4-dimensional qudit can be decomposed as two qubits in practice~\cite{Moussa2016fusion}, here we retain the qudit expression for simplicity.\footnote{Our qudit-qubit setup is similar to~\cite[App. B]{Albert:2021vts}. An alternative realization of $D_4$-qudits based instead on 3-qubits can be found in~\cite{Warman:2024lir}.} We denote the physical Pauli operators for a qubit by $X, Z$, the generalized Pauli operators for a 4-dimensional qudit by $\mathcal{X}, \mathcal{Z}$, and the charge conjugation operator exchanging $\ket{\mathbf{1}}\leftrightarrow\ket{\mathbf{3}}$ by $\cC$, such that $\cC\cX\cC^\dagger=\cX^3$, $\cC\cZ\cC^\dagger=\cZ^3$. 
The generators of the group multiplication operators are realized in this Hilbert space as
\be\label{LROps}
\ba
 L^{r} &= \mathcal{X} \otimes \bbI\,, \quad & L^{s} &= \mathcal{C} \otimes X\,,
 \\
 R^{r} &= \mathcal{X}^{-Z}\,, \quad & R^{s} &= \bbI \otimes X\,.
\ea
\ee
where the operator $\mathcal{X}^{-Z}$ acts on the basis states as $\mathcal{X}^{-Z}\ket{\mathbf{j}}\ket{b}=\ket{\mathbf{j-(-1)^b}}\ket{b}$.

On a $D(\Z_2)$ or $D(\Z_4)$ patch (more generally, for any Abelian group), the logical operators $\bar{L}^g$ and $\bar{R}^{g^{-1}}$ are equivalent, and can be moved into the bulk.  

Our notation for the surface codes associated to Abelian groups is as follows. For a $D(\Z_2)$ patch we denote the logical operators by $\bar{X},\bar{Z}$, and the logical basis is $\ket{\bar{0}}$ and $\ket{\bar{1}}$. For $D(\Z_2 \times \Z_2)$ we in turn distinguish the two copies by labels $\LL$ and $\RR$, i.e.  $\ket{\bar{0}}_{\LL/\RR}$ and $\ket{\bar{1}}_{\LL/\RR}$, which are the eigenstates of the logical $\bar{Z}_{\LL/\RR}$ operator. 
For the $D(\Z_4)$ patch we denote the logical operators by $\bar{\mathcal{X}},\bar{\mathcal{Z}}$ and the logical basis states are the eigenstates of the logical $\bar{\mathcal{Z}}$ operator and will be denoted by $\ket{\bar{\mathbf{j}}}$, where $\mathbf{j}\in\{0,1,2,3\}$.

Finally, on the non-Abelian surface code patch $D(D_4)$, we define the logical basis states $\ket{\Phi_g}$ as described in Sec.~\ref{subsec:DG}. Writing $g=r^js^b$, we can also interpret the 8-dimensional logical space as a 4-dimensional logical qudit and a logical qubit, whose states are labeled by $j$ and $b$, respectively.

On the $D(\Z_4)$ patch, we generate a logical state
\be\label{SZ4}
    \ket{S}_{\Z_4} = \frac{1}{2}\left( \ket{\bar{\mathbf{0}}} + e^{i\pi/4}\ket{\bar{\mathbf{1}}} - \ket{\bar{\mathbf{2}}} + e^{i\pi/4}\ket{\bar{\mathbf{3}}} \right) 
\ee
using the 4-dimensional qudit $S$ gate in $D(\Z_4)$
\be
    \mathcal{S} \ket{\bar{\mathbf{j}}} = e^{\frac{i \pi j^{2}}{4}} \ket{\bar{\mathbf{j}}}\,,
\label{eq:z4_S}
\ee
which is fold-transversal~\cite{Moussa2016Fold}. The $D(D_4)$ patch is initialized in its fiducial state $\ket{\Phi_\id}_{D_4}$. To generate a magic state on the $D(\Z_2\times\Z_2)$ patch, the $D(\Z_2\times\Z_2)$ patch should be prepared in the state $\ket{\bar{0}}_{\LL}\ket{\bar{0}}_{\RR}$. For the purpose of gate teleportation, the $D(\Z_2\times\Z_2)$ patch is in an arbitrary initial logical state. 

The entire procedure is essentially using lattice surgery to pass the information from the $D(\Z_4)$ patch to the $D(\Z_2\times\Z_2)$ patch via the $D(D_4)$ patch. Between a pair of patches, we first implement a specific rough merge and then the rough split. In the following we describe the process between the $D(\Z_4)$ and $D(D_4)$ patches and the process between the $D(D_4)$ and $D(\Z_2\times\Z_2)$ (or $D(\Z_2)$) patches separately. Later, we will argue that it is possible to simultaneously implement the two processes on the two boundaries of the middle $D(D_4)$ patch.

\subsection{Merge and Split of $D(\Z_4)$ and $D(D_4)$} \label{subsec:Z4_D4}

We consider a rough merge and split between a $D(\Z_4)$ code patch on the left and a $D(D_4)$ code patch on the right. Compared to the general discussion in Sec.~\ref{sec:lattice_surgery}, this exchanges the left and right positions of $D(G)$ and $D(G^\prime)$ patches. For this interface, we choose the trivial $\vp=1$ and the diagonal subgroup of $\Z_4 \times D_4$ generated by
\be \label{Z4diagDef}
\Z_4^\diag=\langle (m, r)\rangle \cong\Z_4 \,.
\ee 
Here, $m$ denotes the $\Z_4$ generator in the $\Z_4$ patch while $r=rs^0$ is the $\Z_4$ generator in $D_4$. This corresponds to introducing a line of vertical edges at the interface, each of which contains two 4-dimensional qudits and one qubit prepared in the state $\ket{\mathbf{0}}\ket{\mathbf{0},0}$, and then applying a projector 
\be\ba
    \wt{A}_v^{\Z_4^\diag,1} 
&= \frac{1}{4} \sum_{j=0}^3
\begin{tikzpicture}[baseline]
\begin{scope}[shift={(0,0)}]
\draw[thick, ->-] (-1,0) to (0,0);
\draw[thick, ->-] (0,-1) to (0,0);
\draw[thick, ->-] (0,0) to (0,1);
\draw[fill=black] (0,0) ellipse (0.05 and 0.05);
\node[] at (0.2, 0) {$v$};
\node[above] at (-0.7,0.05) {$R^{m^j}$};
\node[right] at (0.05,0.7) {$L^{m^j}$};
\node[right] at (0.05,-0.7) {$R^{m^j}$};
\draw[fill=black] (0,0) ellipse (0.05 and 0.05);
\node[black] at (0.6,0) {$\otimes$};
\draw[thick, ->-] (1,0) to (2,0);
\draw[thick, ->-] (1,-1) to (1,0);
\draw[thick, ->-] (1,0) to (1,1);
\draw[fill=black] (1,0) ellipse (0.05 and 0.05);
\node[above] at (1.8,0.05) {$L^{r^j}$};
\node[right] at (1.05,0.7) {$L^{r^j}$};
\node[right] at (1.05,-0.7) {$R^{r^j}$};
\end{scope}
\end{tikzpicture}=\\
&= \frac{1}{4} \sum_{j=0}^3
\begin{tikzpicture}[baseline]
\begin{scope}[shift={(0,0)}]
\draw[thick, ->-] (-1,0) to (0,0);
\draw[thick, ->-] (0,-1) to (0,0);
\draw[thick, ->-] (0,0) to (0,1);
\draw[fill=black] (0,0) ellipse (0.05 and 0.05);
\node[] at (0.2, 0) {$v$};
\node[above] at (-0.7,0.05) {$\mathcal{X}^{-j}$};
\node[right] at (0.05,0.7) {$\mathcal{X}^{j}$};
\node[right] at (0.05,-0.7) {$\mathcal{X}^{-j}$};
\draw[fill=black] (0,0) ellipse (0.05 and 0.05);
\node[black] at (0.6,0) {$\otimes$};
\draw[thick, ->-] (1,0) to (2,0);
\draw[thick, ->-] (1,-1) to (1,0);
\draw[thick, ->-] (1,0) to (1,1);
\draw[fill=black] (1,0) ellipse (0.05 and 0.05);
\node[above] at (1.8,0.05) {$\mathcal{X}^{j}$};
\node[right] at (1.05,0.7) {$\mathcal{X}^{j}$};
\node[right] at (1.05,-0.7) {$\mathcal{X}^{-jZ}$};
\end{scope}
\end{tikzpicture}
\label{eq:Av_z4}
\ea\ee
at each interface vertex. To facilitate this, we can measure the operator 
\be\ba
    \wt{A}_v^{(m, r)} &= 
    \begin{tikzpicture}[baseline]
\begin{scope}[shift={(0,0)}]
\draw[thick, ->-] (-1,0) to (0,0);
\draw[thick, ->-] (0,-1) to (0,0);
\draw[thick, ->-] (0,0) to (0,1);
\draw[fill=black] (0,0) ellipse (0.05 and 0.05);
\node[] at (0.2, 0) {$v$};
\node[above] at (-0.7,0.05) {$R^{m}$};
\node[right] at (0.05,0.7) {$L^{m}$};
\node[right] at (0.05,-0.7) {$R^{m}$};
\draw[fill=black] (0,0) ellipse (0.05 and 0.05);
\node[black] at (0.6,0) {$\otimes$};
\draw[thick, ->-] (1,0) to (2,0);
\draw[thick, ->-] (1,-1) to (1,0);
\draw[thick, ->-] (1,0) to (1,1);
\draw[fill=black] (1,0) ellipse (0.05 and 0.05);
\node[above] at (1.8,0.05) {$L^{r}$};
\node[right] at (1.05,0.7) {$L^{r}$};
\node[right] at (1.05,-0.7) {$R^{r}$};
\end{scope}
\end{tikzpicture}=\\
&=\begin{tikzpicture}[baseline]
\begin{scope}[shift={(0,0)}]
\draw[thick, ->-] (-1,0) to (0,0);
\draw[thick, ->-] (0,-1) to (0,0);
\draw[thick, ->-] (0,0) to (0,1);
\draw[fill=black] (0,0) ellipse (0.05 and 0.05);
\node[] at (0.2, 0) {$v$};
\node[above] at (-0.7,0.05) {$\mathcal{X}^{-1}$};
\node[right] at (0.05,0.7) {$\mathcal{X}^{}$};
\node[right] at (0.05,-0.7) {$\mathcal{X}^{-1}$};
\draw[fill=black] (0,0) ellipse (0.05 and 0.05);
\node[black] at (0.6,0) {$\otimes$};
\draw[thick, ->-] (1,0) to (2,0);
\draw[thick, ->-] (1,-1) to (1,0);
\draw[thick, ->-] (1,0) to (1,1);
\draw[fill=black] (1,0) ellipse (0.05 and 0.05);
\node[above] at (1.8,0.05) {$\mathcal{X}^{}$};
\node[right] at (1.05,0.7) {$\mathcal{X}^{}$};
\node[right] at (1.05,-0.7) {$\mathcal{X}^{-Z}$};
\end{scope}
\end{tikzpicture},
\ea\ee
which yields possible outcomes $i^{m_v}\in\{\pm1,\pm i\}$. The measurement outcomes can be absorbed into the stabilizers $\{\wt{A}_v^{\Z_4^\diag,1}\}$ by changing $\wt{A}_v^{(m, r)}$ to $i^{-m_v}\wt{A}_v^{(m, r)}$ in the definition of the stabilizers. In practice, this means that $\{m_v\}$ must be carried along in subsequent error correction processes. Multiplying the operators $\{\wt{A}_v^{(m, r)}\}$ along the interface, we see that the product of outcomes at all vertices $\prod_v i^{m_v}$ is equal to the measurement of the logical operator $\bar{R}^{m}\otimes\bar{L}^{r}$. Here $\bar{R}^{m}$ and $\bar{L}^{r}$ act on the logical states of the $D(\Z_4)$ and $D(D_4)$ patches, respectively. Depending on the initial states on the two patches, $\prod_v i^{m_v}$ may take different values. In the following discussion, we will simply use Eq.~\eqref{eq:Av_z4}, which corresponds to the situation where $m_v=0$ for any $v$, and discuss the operation associated with $\prod_v i^{m_v}\neq1$ separately.

During the rough split, we assume the measurements on the interface edges project the state on each edge to $\ket{\id,\id}=\ket{\mathbf{0}}\ket{\mathbf{0},0}$ without loss of generality. Any other outcomes would correspond to violation of some truncated plaquette terms on the separated boundaries after the split. For example, obtaining $\ket{\mathbf{j}}\ket{\mathbf{j},0}$ on an interface edge means that for the two plaquettes (each with three edges) to the left and right of that edge, $B_p^{(\id)}=1$ is replaced with $B_p^{(m^j)}=1$ and $B_p^{(r^{-j})}=1$, respectively. We can either correct these terms by pairing them up along the boundary or update the affected $B_p^{(\id)}$ terms. Thus, the logical information is independent of the outcomes. 

If the initial state for the two code patches is $\ket{\Psi_1}_{\Z_4}\ket{\Psi_2}_{D_4}$, using Eq.~\eqref{eq:disjoint_diag}, the state after the rough merge and split is 
\be\ba
    \ket{\Psi_{1,2}}_{\Z_4, D_4} &= \frac{1}{4}\sum_{k=0}^3 \left(\prod_{v=1}^{n_v} R_v^{m^{k}} \ket{\Psi_1}_{\Z_4}\right) \\
    & \qquad\qquad  \otimes \left(\prod_{v'=1}^{n_v} L_{v'}^{r^{k}} \ket{\Psi_2}_{D_4}\right),
    \label{eq:Z4_D4}
\ea\ee
subject to post-measurement operations that enforce $B_p^{(\id)}=1$ for the truncated plaquettes on the two separated boundaries. We note that $\prod_{v=1}^{n_v} R_v^{m^{}}$ is the logical operator $\bar{R}^{m^{}}$ on the $D(\Z_4)$ patch and $\prod_{v'=1}^{n_v} L_{v'}^{r}$ is the logical operator $\bar{L}^{r}$ on the $D(D_4)$ patch. The combined procedure of the rough merge and the rough split effectively measures the logical operator 
\be
  \bar{R}^{m^{}}\otimes\bar{L}^{r}\,.
\ee
Equivalently, this logical operator can be written as $\bar{\mathcal{X}}^{-1}_{\Z_4}\otimes\bar{\mathcal{X}}_{D_4}$. In Eq.~\eqref{eq:Z4_D4}, the state is projected to the $+1$ eigenstate, as a consequence of applying the projectors $\{\wt{A}_v^{\Z_4^\diag,1}\}$ without adapting to the measurement outcomes $\{m_v\}$. Setting
\be 
i^{m_{\mathcal{XX}}}:= \prod_vi^{m_v}\,,
\ee 
Eq.~\eqref{eq:Z4_D4} should be updated to
\be
    \ket{\Psi_{1,2}}_{\Z_4, D_4} = \frac{1}{4}\sum_{k=0}^3 \left( i^{-m_{\mathcal{XX}}} \bar{\mathcal{X}}^{-1}_{\Z_4}\bar{\mathcal{X}}_{D_4} \right)^k \ket{\Psi_1}_{\Z_4}\ket{\Psi_2}_{D_4}.
\ee
When $\ket{\Psi_2}_{D_4}=\ket{\Phi_\id}_{D_4}$, the resulting state can be returned to Eq.~\eqref{eq:Z4_D4} by a logical $\bar{\mathcal{Z}}^{m_{\mathcal{XX}}}$.

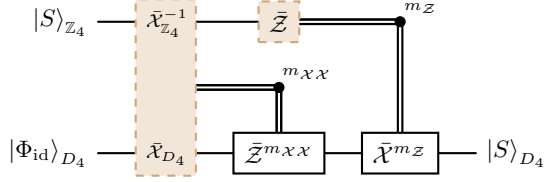
\begin{figure}
\label{fig:Z4-D4_circuit}
\begin{quantikz}
\lstick{$\ket{S}_{\Z_{4}}$} &  \gate[3, style={fill=brown!20, draw=brown!70, dashed}][0.8cm][0.6cm]{}\gateinput[1]{$\bar{\mathcal{X}}_{\Z_{4}}^{-1}$} & \gate[1, style={fill=brown!20, draw=brown!70, dashed}][0.8mm]{\bar{\mathcal{Z}}} & \ctrl[vertical
wire=c]{2} \setwiretype{c} \wire[l][1]["m_{\mathcal{Z}}"{above,pos=-0.25}]{a}  \\
&\setwiretype{n}  & \ctrl[vertical
wire=c]{1} \setwiretype{c} \wire[l][1]["m_{\mathcal{XX}}"{above,pos=-0.4}]{a} \\
\lstick{$\ket{\Phi_{\id}}_{D_{4}}$} &\gateinput[1]{$\bar{\mathcal{X}}_{D_{4}}$}  & \gate[1][0.8mm]{\bar{\mathcal{Z}}^{m_{\mathcal{XX}}}}\wire[u]{c} & \gate[1][0.8mm]{\bar{\mathcal{X}}^{m_{\mathcal{Z}}}} & \rstick{$\ket{S}_{D_{4}}$} 
\end{quantikz}
\caption{The logical-level circuit diagram showing teleportation of $\ket{S}$ via lattice surgery from the $D(\Z_{4})$ code patch to the $D(D_{4})$ code patch. The measurements (brown dashed boxes) of $\bar{\mathcal{X}}^{-1}_{\Z_4}\bar{\mathcal{X}}_{D_4}$ and $\bar{\mathcal{Z}}_{\Z_4}$ yield outcomes $i^{m_{\mathcal{XX}}}$ and $i^{m_{\mathcal{Z}}}$, respectively.}
\label{fig:circ_Z4D4}
\end{figure}

Applying this procedure to the initial logical state $\ket{S}_{\Z_4}\ket{\Phi_\id}_{D_4}$, we obtain the unnormalized state
\be\ba
    \frac{1}{4}\sum_{k=0}^3 \left( \bar{\mathcal{X}}^{-1}_{\Z_4}\bar{\mathcal{X}}_{D_4} \right)^k \ket{S}_{\Z_4}\ket{\Phi_\id}_{D_4} = \frac{1}{4}\sum_{k=0}^3 \ket{\mathbf{k}}_{\Z_4} \bar{\mathcal{X}}_{D_4}^{-k}\ket{S}_{D_4},
\ea\ee
where $\ket{S}_{D_4}$ is defined as
\be
    \ket{S}_{D_4} = \ket{\Phi_{\id}}+e^{i\pi/4}\ket{\Phi_{r}}-\ket{\Phi_{r^2}}+e^{i\pi/4}\ket{\Phi_{r^3}}.
    \label{eq:D4output}
\ee
Therefore, measuring the $D(\Z_4)$ patch in the logical computational basis and applying a logical $\bar{\mathcal{X}}$ to the $D(D_4)$ patch conditioned on the measurement outcome leads to an intermediate logical state $\ket{S}_{D_4}$. We show the effective operations at the logical level with a circuit diagram in Fig.~\ref{fig:circ_Z4D4}. The $D(D_4)$ patch in the state $\ket{S}_{D_4}$ is subsequently used for the lattice surgery with the $D(\Z_2\times\Z_2)$ patch.

\subsection{Merge and Split of $D(D_4)$ and $D(\Z_2\times\Z_2)$}
 \label{subsec:D4_Z2^2}

Next, we consider a rough merge and split between a $D(D_4)$ code patch on the left and a $D(\Z_2\times\Z_2)$ code patch on the right. The interface is chosen to be the one associated with  trivial $\vp=1$ and the diagonal subgroup 
\be \label{Z2Z2DiagDef}
(\Z_2\times \Z_2)^{\diag}= \langle (r^2,m_{\LL}), (r^3s,m_{\RR})\rangle \cong\Z_2\times \Z_2\,,
\ee
where $m_{\LL}$ and $m_{\RR}$ denote the generators of the two $\Z_2$ groups. Again, we introduce a line of vertical edges between the code patches. Each edge comes with a 4-dimensional qudit and three qubits, prepared in the state $\ket{\mathbf{0},0}\ket{0_{\LL},0_{\RR}}$. A projector 
\be
    \wt{A}_v^{\Z_2^{2,\diag},1} = \frac{1}{2}\left(1+\wt{A}_v^{(r^3s,m_{\RR})}\right) \frac{1}{2}\left(1+\wt{A}_v^{(r^2,m_{\LL})}\right)
\label{eq:Av_z2^2}
\ee
is applied at each interface vertex. This can then be done in two consecutive steps by first measuring 
\be\ba
    \wt{A}_v^{(r^2,m_{\LL})} &=
    \begin{tikzpicture}[baseline]
\begin{scope}[shift={(0,0)}]
\draw[thick, ->-] (-1,0) to (0,0);
\draw[thick, ->-] (0,-1) to (0,0);
\draw[thick, ->-] (0,0) to (0,1);
\draw[fill=black] (0,0) ellipse (0.05 and 0.05);
\node[] at (0.2, 0) {$v$};
\node[above] at (-0.7,0.05) {$R^{r^2}$};
\node[right] at (0.05,0.7) {$L^{r^2}$};
\node[right] at (0.05,-0.7) {$R^{r^2}$};
\draw[fill=black] (0,0) ellipse (0.05 and 0.05);
\node[black] at (0.6,0) {$\otimes$};
\draw[thick, ->-] (1,0) to (2,0);
\draw[thick, ->-] (1,-1) to (1,0);
\draw[thick, ->-] (1,0) to (1,1);
\draw[fill=black] (1,0) ellipse (0.05 and 0.05);
\node[above] at (1.8,0.05) {$L^{m_{\LL}}$};
\node[right] at (1.05,0.7) {$L^{m_{\LL}}$};
\node[right] at (1.05,-0.7) {$R^{m_{\LL}}$};
\end{scope}
\end{tikzpicture}=\\
&=\begin{tikzpicture}[baseline]
\begin{scope}[shift={(0,0)}]
\draw[thick, ->-] (-1,0) to (0,0);
\draw[thick, ->-] (0,-1) to (0,0);
\draw[thick, ->-] (0,0) to (0,1);
\draw[fill=black] (0,0) ellipse (0.05 and 0.05);
\node[] at (0.2, 0) {$v$};
\node[above] at (-0.7,0.05) {$\mathcal{X}^{2}$};
\node[right] at (0.05,0.7) {$\mathcal{X}^{2}$};
\node[right] at (0.05,-0.7) {$\mathcal{X}^{2}$};
\draw[fill=black] (0,0) ellipse (0.05 and 0.05);
\node[black] at (0.6,0) {$\otimes$};
\draw[thick, ->-] (1,0) to (2,0);
\draw[thick, ->-] (1,-1) to (1,0);
\draw[thick, ->-] (1,0) to (1,1);
\draw[fill=black] (1,0) ellipse (0.05 and 0.05);
\node[above] at (1.8,0.05) {$X_{\LL}$};
\node[right] at (1.05,0.7) {$X_{\LL}$};
\node[right] at (1.05,-0.7) {$X_{\LL}$};
\end{scope}
\end{tikzpicture},
\ea\ee
followed by the measurement of
\be\ba
    \wt{A}_v^{(r^3s,m_{\RR})} &= 
    \begin{tikzpicture}[baseline]
\begin{scope}[shift={(0,0)}]
\draw[thick, ->-] (-1,0) to (0,0);
\draw[thick, ->-] (0,-1) to (0,0);
\draw[thick, ->-] (0,0) to (0,1);
\draw[fill=black] (0,0) ellipse (0.05 and 0.05);
\node[] at (0.2, 0) {$v$};
\node[above] at (-0.7,0.05) {$R^{r^3s}$};
\node[right] at (0.05,0.7) {$L^{r^3s}$};
\node[right] at (0.05,-0.7) {$R^{r^3s}$};
\draw[fill=black] (0,0) ellipse (0.05 and 0.05);
\node[black] at (0.8,0) {$\otimes$};
\draw[thick, ->-] (1.5,0) to (2.5,0);
\draw[thick, ->-] (1.5,-1) to (1.5,0);
\draw[thick, ->-] (1.5,0) to (1.5,1);
\draw[fill=black] (1.5,0) ellipse (0.05 and 0.05);
\node[above] at (2.3,0.05) {$L^{m_{\RR}}$};
\node[right] at (1.55,0.7) {$L^{m_{\RR}}$};
\node[right] at (1.55,-0.7) {$R^{m_{\RR}}$};
\end{scope}
\end{tikzpicture}=\\
&=\begin{tikzpicture}[baseline]
\begin{scope}[shift={(0,0)}]
\draw[thick, ->-] (-1,0) to (0,0);
\draw[thick, ->-] (0,-1) to (0,0);
\draw[thick, ->-] (0,0) to (0,1);
\draw[fill=black] (0,0) ellipse (0.05 and 0.05);
\node[] at (0.2, 0) {$v$};
\node[above] at (-0.7,0.05) {$\mathcal{X}^{Z}X$};
\node[right] at (0.05,0.7) {$\mathcal{X}^{3}\mathcal{C}X$};
\node[right] at (0.05,-0.7) {$\mathcal{X}^{Z}X$};
\draw[fill=black] (0,0) ellipse (0.05 and 0.05);
\node[black] at (0.8,0) {$\otimes$};
\draw[thick, ->-] (1.5,0) to (2.5,0);
\draw[thick, ->-] (1.5,-1) to (1.5,0);
\draw[thick, ->-] (1.5,0) to (1.5,1);
\draw[fill=black] (1.5,0) ellipse (0.05 and 0.05);
\node[above] at (2.3,0.05) {$X_{\RR}$};
\node[right] at (1.55,0.7) {$X_{\RR}$};
\node[right] at (1.55,-0.7) {$X_{\RR}$};
\end{scope}
\end{tikzpicture}.
\ea\ee
At each vertex, these measurements can yield outcomes $(-1)^{p_v}, (-1)^{q_v}\in\{\pm1\}^2$. Consequently, we should update the stabilizers $\{\wt{A}_v^{\Z_2^{2,\diag},1}\}$ to
\be\ba
    \wt{A}_v^{\Z_2^{2,\diag},1} = &\frac{1}{2}\left(1+(-1)^{q_v}\wt{A}_v^{(r^3s,m_{\RR})}\right)\times \\
    &\frac{1}{2}\left(1+(-1)^{p_v}\wt{A}_v^{(r^2,m_{\LL})}\right)\,.
\ea\ee
Similar to the previous case, we will use Eq.~\eqref{eq:Av_z2^2} for simplicity. 

During the rough split, the qudits and qubits on each interface edge are projected to the state $\ket{\id,\id}=\ket{\mathbf{0},0}\ket{0,0}$. When any other outcomes are obtained, we either correct or update the affected truncated plaquette terms, in the same way discussed in Sec.~\ref{subsec:Z4_D4}. Starting from the initial state $\ket{\Psi_1}_{D_4}\ket{\Psi_2}_{\Z_2^2}$, using Eq.~\eqref{eq:disjoint_diag}, the state after the rough merge and split is 
\be\ba
    \ket{\Psi_{1,2}}_{D_4,\Z_2^2} &= \frac{1}{4}\sum_{p=0}^1\sum_{q=0}^1 \left(\prod_{v=1}^{n_v} (R_v^{r^{3}s})^q \circ (R_v^{r^{2}})^p \ket{\Psi_1}_{D_4}\right) \\
    &\quad  \otimes \left(\prod_{v'=1}^{n_v} (L_{v'}^{m_{\RR}})^q \circ (L_{v'}^{m_{\LL}})^p \ket{\Psi_2}_{\Z_2^2}\right),
    \label{eq:D4_Z2^2}
\ea\ee
subject to post-measurement operations. This is the state after measuring the logical operators
\be
\bar{R}^{r^{2}}\bar{L}^{m_{\LL}} = \prod_{v=1}^{n_v} R_v^{r^{2}} \prod_{v'=1}^{n_v} L_{v'}^{m_{\LL}}
\ee
and
\be
\bar{R}^{r^{3}s}\bar{L}^{m_{\RR}} = \prod_{v=1}^{n_v} R_v^{r^{3}s} \prod_{v'=1}^{n_v} L_{v'}^{m_{\RR}},
\ee
obtaining $+1$ for both outcomes. The two logical operators can also be written as $\bar{\mathcal{X}}^{2}_{D_4}\bar{X}_{\LL}$ and $(\bar{\mathcal{X}}^{\bar{Z}}\bar{X})_{D_4}\bar{X}_{\RR}$. 

\subsection{Magic State and Gate Teleportation} \label{sec:lattice_D4_Z2Z2_magic}

We can use the setup in the last two subsections in two applications: first the magic state generation and then non-Clifford gate teleportation.

\subsubsection{\textbf{\textit{Magic State Generation}}}

For the purpose of generating a magic state, the input state is $\ket{\Psi_2}_{\Z_2^2}=\ket{\bar{0}}_{\LL}\ket{\bar{0}}_{\RR}$. If the measurement outcome of $\bar{\mathcal{X}}^{2}_{D_4}\bar{X}_{\LL}$ (or $(\bar{\mathcal{X}}^{\bar{Z}}\bar{X})_{D_4}\bar{X}_{\RR}$) is $-1$ instead of $+1$, the state in Eq.~\eqref{eq:D4_Z2^2} can be recovered by applying a $\bar{Z}_{\LL}$ (or $\bar{Z}_{\RR}$). 
At the logical level, applying the projectors 
associated with the logical measurements to $\ket{S}_{D_4}\ket{\bar{0}}_{\LL}\ket{\bar{0}}_{\RR}$ yields
\be\ba
     \frac{1}{2}(1+&\bar{R}^{r^{3}s}\bar{L}^{m_{\RR}})\frac{1}{2}(1+\bar{R}^{r^{2}}\bar{L}^{m_{\LL}})\ket{S}_{D_4}\ket{\bar{0}}_{\LL}\ket{\bar{0}}_{\RR} \\
  =  & \frac{1}{8} \big( \ket{\Phi_\id}(\ket{\bar{0}}_{\LL}-\ket{\bar{1}}_{\LL})\ket{\bar{0}}_{\RR}  \\
    &+ e^{i\pi/4}\ket{\Phi_{s}}(\ket{\bar{0}}_{\LL}+\ket{\bar{1}}_{\LL})\ket{\bar{1}}_{\RR}  \\
    & + e^{i\pi/4}\ket{\Phi_{r}}(\ket{\bar{0}}_{\LL}+\ket{\bar{1}}_{\LL})\ket{\bar{0}}_{\RR} \\
    &+ \ket{\Phi_{rs}}(-\ket{\bar{0}}_{\LL}+\ket{\bar{1}}_{\LL})\ket{\bar{1}}_{\RR} \\
    & + \ket{\Phi_{r^2}}(-\ket{\bar{0}}_{\LL}+\ket{\bar{1}}_{\LL})\ket{\bar{0}}_{\RR} \\
    &+ e^{i\pi/4}\ket{\Phi_{r^2s}}(\ket{\bar{0}}_{\LL}+\ket{\bar{1}}_{\LL})\ket{\bar{1}}_{\RR} \\
    & + e^{i\pi/4}\ket{\Phi_{r^3}}(\ket{\bar{0}}_{\LL}+\ket{\bar{1}}_{\LL})\ket{\bar{0}}_{\RR} \\
    &  + \ket{\Phi_{r^3s}}(\ket{\bar{0}}_{\LL}-\ket{\bar{1}}_{\LL})\ket{\bar{1}}_{\RR} \big)\,.
\ea\ee

Next we can disentangle the two patches by measuring the $D(D_4)$ patch. Instead of the logical computational basis, we first measure it in the eigenbasis of the logical operator $\bar{R}^s$ (or equivalently, $\bar{X}_{D_4}$) and obtain an outcome $m_{X}$. This can be done by first measuring
\be
    \wt{A}_v^{(s,1)} = \begin{tikzpicture}[baseline]
\begin{scope}[shift={(0,0.1)}]
\draw[thick, ->-] (-1,0) to (0,0);
\draw[thick, ->-] (0,-1) to (0,0);
\draw[thick, ->-] (0,0) to (0,1);
\draw[fill=black] (0,0) ellipse (0.05 and 0.05);
\node[] at (0.2, 0) {$v$};
\node[above] at (-0.7,0.05) {$R^s$};
\node[right] at (0.05,0.7) {$L^s$};
\node[right] at (0.05,-0.7) {$R^s$};
\draw[fill=black] (0,0) ellipse (0.05 and 0.05);
\node[black] at (0.8,0) {$\otimes$};
\node[black] at (1.5,0) {$\bbI$};
\end{scope}
\end{tikzpicture}=\quad
\begin{tikzpicture}[baseline]
\begin{scope}[shift={(0,0.1)}]
\draw[thick, ->-] (-1,0) to (0,0);
\draw[thick, ->-] (0,-1) to (0,0);
\draw[thick, ->-] (0,0) to (0,1);
\draw[fill=black] (0,0) ellipse (0.05 and 0.05);
\node[] at (0.2, 0) {$v$};
\node[above] at (-0.7,0.05) {$X$};
\node[right] at (0.05,0.7) {$\mathcal{C}X$};
\node[right] at (0.05,-0.7) {$X$};
\draw[fill=black] (0,0) ellipse (0.05 and 0.05);
\node[black] at (0.8,0) {$\otimes$};
\node[black] at (1.5,0) {$\bbI$};
\end{scope}
\end{tikzpicture}
\ee
at all the interface vertices, reusing the $D_4$-qudits on the vertical interface edges, followed by measuring these qudits in the computational basis and performing the necessary correction operations to recover the state $\ket{\id,\id}$. The product of the $\{ \wt{A}_v^{(s,1)} \}$ measurement outcomes gives the outcome $m_{X}$.

\begin{figure*}
\centering
\begin{quantikz}
\lstick{$\ket{\bar{0}}_{(1)}$} & \gate[3, style={fill=brown!20, draw=brown!70, dashed}][0.8cm][0.6cm]{}\gateinput[1]{$\bar{X}_{\LL}$} & \gate[1][0.8mm]{\bar{Z}^{m_{X_{\LL}\mathcal{X}}}} &&& \gate[1][0.8mm]{\bar{X}^{m_{\mathcal{Z}_{2}}}}  &&& \rstick[5]{$\ket{M}_{(1)(2)} = CZ\ket{-}_{\LL}\ket{T}_{\RR}$} \\
&\setwiretype{n}  & \ctrl[vertical
wire=c]{-1} \setwiretype{c}  \wire[l][1]["m_{X_{\LL}\mathcal{X}}"{above,pos=-0.5}]{a}  &\setwiretype{n} &\setwiretype{n} & \ctrl[vertical
wire=c]{1}  \setwiretype{n} \wire[u][1]["m_{\mathcal{Z}}=2m_{\mathcal{Z}_{1}}+m_{\mathcal{Z}_{2}}"{right,pos=0.2}]{c} \\
\lstick{$\ket{S}_{D_{4}}$} &\gateinput[1]{$\bar{\mathcal{X}}_{D_{4}}^{2}$} &  \gate[3, style={fill=brown!20, draw=brown!70, dashed}][1.35cm][0.6cm]{} \gateinput[1]{$\bar{\mathcal{X}}_{D_{4}}^{Z}\bar{X}_{D_{4}}$} && \gate[1, style={fill=brown!20, draw=brown!70, dashed}][0.8mm]{\bar{X}_{D_{4}}} & \gate[1, style={fill=brown!20, draw=brown!70, dashed}][0.8mm]{\bar{\mathcal{Z}}_{D_{4}}}    \\
&\setwiretype{n} &\setwiretype{n} & \ctrl[vertical
wire=c]{1} \setwiretype{c}  \wire[l][1]["m_{\mathcal{X}X_{\RR}}"{above,pos=-0.4}]{a} & \ctrl[vertical
wire=c]{1}  \setwiretype{n} \wire[u][1]["m_{X}"{right,pos=0.2}]{c} & \ctrl[vertical
wire=c]{1}  \setwiretype{n} \wire[u][1]["m_{\mathcal{Z}}=2m_{\mathcal{Z}_{1}}+m_{\mathcal{Z}_{2}}"{right,pos=0.2}]{c} \\
\lstick{$\ket{\bar{0}}_{(2)}$} &&\gateinput[1]{$\quad \bar{X}_{\RR}$} & \gate[1][0.8mm]{\bar{Z}^{m_{\mathcal{X}X_{\RR}}}}\wire[u]{c} & \gate[1][0.8mm]{\bar{Z}^{m_{X}}} & \gate[1][0.8mm]{\bar{Z}^{m_{\mathcal{Z}_{1}}}\bar{X}^{m_{\mathcal{Z}_{2}}}} &&& 
\end{quantikz} 
\caption{The logical-level circuit diagram showing the preparation of a magic state $\ket{M}_{(1)(2)}$ via lattice surgery from the $D(D_{4})$ code patch to the $D(\Z_{2}\times\Z_{2})$ code patch. The shaded boxes represent measurements of logical operators.} 
\label{fig:circ_D4Z22}
\end{figure*}
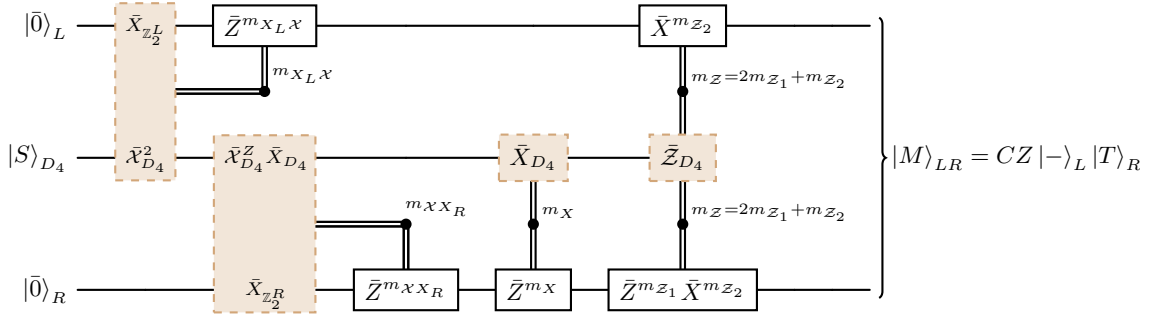

Then, we measure it in the computational basis $\{\ket{\Phi_{r^js^b}}\}$, but only retain the value of $j$. This is equivalent to a measurement of the logical operator $\bar{\mathcal{Z}}_{D_4}$. The outcome $m_{\mathcal{Z}}$ can be written in terms of two bits $m_{\mathcal{Z}}=2m_{\mathcal{Z}1}+m_{\mathcal{Z}2}$ where $m_{\mathcal{Z}1,2}\in\{0,1\}$. At the logical level, the resulting normalized state is $Z_{\RR}^{m_{X}+m_{\mathcal{Z}1}}X_{\RR}^{m_{\mathcal{Z}2}}X_{\LL}^{m_{\mathcal{Z}2}}\ket{M}_{\LL\RR}$, where
\be\ba \label{eq:MLR}
    \ket{M}_{(1)(2)} &= \frac{1}{2}\left(\ket{\bar{0}}_{\LL}\ket{\bar{0}}_{\RR} - \ket{\bar{1}}_{\LL}\ket{\bar{0}}_{\RR} \right. \\
    & \left. + e^{i\pi/4}\ket{\bar{0}}_{\LL}\ket{\bar{1}}_{\RR} + e^{i\pi/4}\ket{\bar{1}}_{\LL}\ket{\bar{1}}_{\RR}\right).
\ea\ee
This is the same as $CZ\ket{-}_{\LL}\ket{T}_{\RR}$ where $\ket{T}=\frac{1}{\sqrt{2}}(\ket{0}+e^{i\pi/4}\ket{1})$. A logical circuit showing the effects of the rough merge and split, together with the measurement procedure afterwards, is in Fig.~\ref{fig:circ_D4Z22}. We can then disentangle the two $D(\Z_2)$ patches using their transversal $CZ$ gate and obtain a copy of the magic state $\ket{T}$.

\subsubsection{\textbf{\textit{Gate Teleportation}}}

\begin{figure*}
\centering
\begin{quantikz}
\lstick[5]{$\ket{\psi}_{(1)(2)} \ket{S}_{D_{4}} $} & \gate[3, style={fill=brown!20, draw=brown!70, dashed}][0.8cm][0.6cm]{}\gateinput[1]{$\bar{X}_{\LL}$} & \gate[1][0.8cm]{\bar{\mathcal{R}}_x(-\pi/2)^{m_{X_{\LL}\mathcal{X}}}} && \gate[1][0.8cm]{\bar{X}^{m_{\mathcal{X}X_{\RR}}}} & \gate[1][0.8cm]{\bar{X}^{m_{X}}} & \gate[1][0.8cm]{\bar{X}^{m_{\mathcal{Z}_{2}}}}  && \\
&\setwiretype{n}  & \ctrl[vertical
wire=c]{3} \setwiretype{c}  \wire[u][1]["m_{X_{\LL}\mathcal{X}}"{right,pos=0.2}]{a}  &\setwiretype{n} &\setwiretype{n} & \ctrl[vertical
wire=c]{1}  \setwiretype{n} \wire[u][1]["m_{X}"{right,pos=0.2}]{c}  & \ctrl[vertical
wire=c]{1}  \setwiretype{n} \wire[u][1]["m_{\mathcal{Z}}=2m_{\mathcal{Z}_{1}}+m_{\mathcal{Z}_{2}}"{right,pos=0.2}]{c} \\
 &\gateinput[1]{$\bar{\mathcal{X}}_{D_{4}}^{2}$} &&  \gate[3, style={fill=brown!20, draw=brown!70, dashed}][1.35cm][0.6cm]{} \gateinput[1]{$\bar{\mathcal{X}}_{D_{4}}^{Z}\bar{X}_{D_{4}}$} && \gate[1, style={fill=brown!20, draw=brown!70, dashed}][0.5cm]{\bar{X}_{D_{4}}} & \gate[1, style={fill=brown!20, draw=brown!70, dashed}][0.5cm]{\bar{\mathcal{Z}}_{D_{4}}}     \\
&\setwiretype{n} &\setwiretype{n} &\setwiretype{n} & \ctrl[vertical
wire=c]{-3} \setwiretype{c}  \wire[u][3]["m_{\mathcal{X}X_{\RR}}"{right,pos=0.2}]{a} &\setwiretype{n}  & \ctrl[vertical
wire=c]{1}  \setwiretype{n} \wire[u][1]["m_{\mathcal{Z}}=2m_{\mathcal{Z}_{1}}+m_{\mathcal{Z}_{2}}"{right,pos=0.2}]{c} \\
 && \gate[1][0.5cm]{\bar{X}^{m_{X_{\LL}\mathcal{X}}}} &\gateinput[1]{$\quad \bar{X}_{\RR}$} &  &  & \gate[1][0.8cm]{\bar{Z}^{m_{\mathcal{Z}_{1}}}\bar{X}^{m_{\mathcal{Z}_{2}}}} && 
\end{quantikz} 
=\begin{quantikz}[align equals at=2.5]
&\setwiretype{n} \\
\lstick[2]{$\ket{\psi}_{(1)(2)}$} & \gate[1][0.5cm]{H} & \ctrl{1} & \gate[1][0.5cm]{T^{\dagger}} &  \gate[1][0.5cm]{H} &
\\
 & \gate[1][0.5cm]{H} & \control{} & \gate[1][0.5cm]{Z} &  \gate[1][0.5cm]{H} &
\end{quantikz}
\caption{The logical-level circuit diagram showing the gate teleportation via lattice surgery between $D(D_{4})$ and $D(\Z_{2} \times \Z_{2})$ code patches. The middle wire represents the $D(D_4)$ patch while the top and bottom wires represent the two $D(\Z_2)$ patches. $\bar{\mathcal{R}}_x(-\pi/2)=e^{-i\bar{X}\pi/4}$ is a 1-qubit rotation. An equivalent logical circuit for the $D(\Z_{2} \times \Z_{2})$ patch is given.} 
\label{fig:gate_teleport_D4Z22}
\end{figure*}
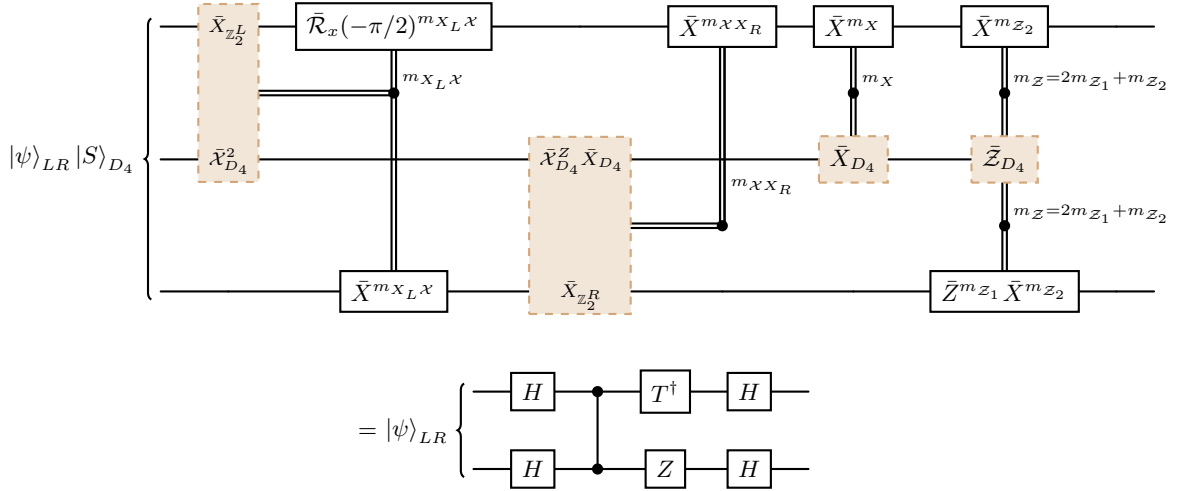

For the application of gate teleportation, with an arbitrary input state $\ket{\Psi_2}_{\Z_2^2}$, we can first focus on the case where the logical measurements of $\bar{\mathcal{X}}^{2}_{D_4}\bar{X}_{\LL}$ and $(\bar{\mathcal{X}}^{\bar{Z}}\bar{X})_{D_4}\bar{X}_{\RR}$ both yield $+1$. The measurement procedure following the rough merge and split is similar to the previous situation. The consecutive logical measurements of $\bar{R}^s=\bar{X}_{D_4}$ and $\bar{\mathcal{Z}}_{D_4}$, with the respective outcomes $m_{X}$ and $m_{\mathcal{Z}}=2m_{\mathcal{Z}1}+m_{\mathcal{Z}2}$, will implement a non-Clifford unitary on the input state $\ket{\Psi_2}_{\Z_2^2}$. The resulting unitary is equivalent to 
\be
    \bar{Z}_{\RR}^{m_{\mathcal{Z}2}}\bar{X}_{\RR}^{m_{\mathcal{Z}2}}\bar{X}_{\LL}^{m_X+{m_{\mathcal{Z}1}}}U_{m_X=0, m_{\mathcal{Z}}=0},
\ee
where
\be
 U_{m_X=0, m_{\mathcal{Z}}=0} = \frac{1}{2}\begin{pmatrix}
1 &  e^{i\pi/4} & -1 &  e^{i\pi/4} \\
 e^{i\pi/4} & 1 &  e^{i\pi/4} & -1 \\
-1 &  e^{i\pi/4} & 1 &  e^{i\pi/4} \\
 e^{i\pi/4} & -1 &  e^{i\pi/4} & 1
\end{pmatrix} \,.
\ee
This can be decomposed as
\be
    U_{m_X=0, m_{\mathcal{Z}}=0} = e^{i\pi/4} H_{\LL}H_{\RR} (CZ) T^\dagger_{\LL} Z_{\RR} H_{\LL}H_{\RR}.
    \label{eq:gate_teleport}
\ee
with 
\be
H=\frac{1}{\sqrt{2}}\begin{pmatrix} 1 & 1\\ 1 & -1 \end{pmatrix} \,.
\ee
Therefore, $U_{m_X=0, m_{\mathcal{Z}}=0}$ can be realized by logical gates conditioned on the measurement outcomes $m_{X}$ and $m_{\mathcal{Z}}$. In general, the measurements of $\bar{\mathcal{X}}^{2}_{D_4}\bar{X}_{\LL}$ and $(\bar{\mathcal{X}}^{\bar{Z}}\bar{X})_{D_4}\bar{X}_{\RR}$ can yield outcomes $+1$ or $-1$, in which case post-measurement operations conditioned on the outcomes will recover the same unitary as in Eq.~\eqref{eq:gate_teleport}. Fig.~\ref{fig:gate_teleport_D4Z22} illustrates the measurements and the corresponding post-measurement operations.

\subsection{Alternative: Merge and Split of $D(D_4)$ and $D(\Z_2)$} \label{subsec:D4_Z2}

Instead of a $D(\Z_2\times\Z_2)$ patch, we can directly carry out a rough merge and split between the $D(D_4)$ patch and a $D(\Z_2)$ patch. We choose the interface in this case to be described by the diagonal subgroup $\Z_2^\diag=\langle(r^3s,m)\rangle \cong\Z_2$ and $\vp=1$ where $m$ generates the $\Z_2$ group. On each edge of the interface, a 4-dimensional qudit and two qubits are initialized in the state $\ket{\mathbf{0},0}\ket{0}$. The operator to be measured at each interface vertex is
\be\ba
    \wt{A}_v^{(r^3s,m)} &=\begin{tikzpicture}[baseline]
\begin{scope}[shift={(0,0)}]
\draw[thick, ->-] (-1,0) to (0,0);
\draw[thick, ->-] (0,-1) to (0,0);
\draw[thick, ->-] (0,0) to (0,1);
\draw[fill=black] (0,0) ellipse (0.05 and 0.05);
\node[] at (0.2, 0) {$v$};
\node[above] at (-0.7,0.05) {$R^{r^3s}$};
\node[right] at (0.05,0.7) {$L^{r^3s}$};
\node[right] at (0.05,-0.7) {$R^{r^3s}$};
\draw[fill=black] (0,0) ellipse (0.05 and 0.05);
\node[black] at (0.8,0) {$\otimes$};
\draw[thick, ->-] (1.5,0) to (2.5,0);
\draw[thick, ->-] (1.5,-1) to (1.5,0);
\draw[thick, ->-] (1.5,0) to (1.5,1);
\draw[fill=black] (1.5,0) ellipse (0.05 and 0.05);
\node[above] at (2.3,0.05) {$L^m$};
\node[right] at (1.55,0.7) {$L^m$};
\node[right] at (1.55,-0.7) {$R^m$};
\end{scope}
\end{tikzpicture}=\\
&=
\begin{tikzpicture}[baseline]
\begin{scope}[shift={(0,0)}]
\draw[thick, ->-] (-1,0) to (0,0);
\draw[thick, ->-] (0,-1) to (0,0);
\draw[thick, ->-] (0,0) to (0,1);
\draw[fill=black] (0,0) ellipse (0.05 and 0.05);
\node[] at (0.2, 0) {$v$};
\node[above] at (-0.7,0.05) {$\mathcal{X}^{Z}X$};
\node[right] at (0.05,0.7) {$\mathcal{X}^{3}\mathcal{C}X$};
\node[right] at (0.05,-0.7) {$\mathcal{X}^{Z}X$};
\draw[fill=black] (0,0) ellipse (0.05 and 0.05);
\node[black] at (0.8,0) {$\otimes$};
\draw[thick, ->-] (1.5,0) to (2.5,0);
\draw[thick, ->-] (1.5,-1) to (1.5,0);
\draw[thick, ->-] (1.5,0) to (1.5,1);
\draw[fill=black] (1.5,0) ellipse (0.05 and 0.05);
\node[above] at (2.3,0.05) {$X$};
\node[right] at (1.55,0.7) {$X$};
\node[right] at (1.55,-0.7) {$X$};
\end{scope}
\end{tikzpicture}.
\label{eq:A^r3s}
\ea\ee
Together with the subsequent rough split, this effectively measures the logical operator $\bar{R}^{r^3s}\otimes\bar{L}^{m}$, or equivalently $(\bar{\mathcal{X}}^{\bar{Z}}\bar{X})_{D_4}\bar{X}_{\Z_2}$, obtaining an outcome $m_{\mathcal{X}X}$. 

\subsubsection{\textbf{\textit{Magic state generation}}}

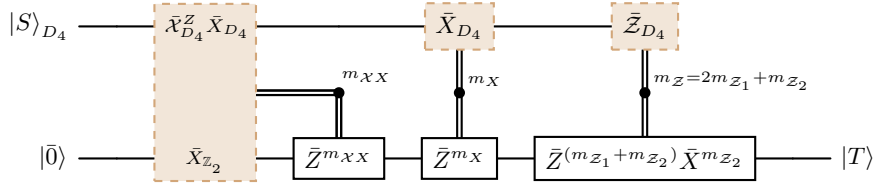
\begin{figure*}
\centering
\begin{quantikz}
\lstick{$\ket{S}_{D_{4}}$} & &  \gate[3, style={fill=brown!20, draw=brown!70, dashed}][1.35cm][0.6cm]{} \gateinput[1]{$\bar{\mathcal{X}}_{D_{4}}^{Z}\bar{X}_{D_{4}}$} && \gate[1, style={fill=brown!20, draw=brown!70, dashed}][0.8mm]{\bar{X}_{D_{4}}} & \gate[1, style={fill=brown!20, draw=brown!70, dashed}][0.8mm]{\bar{\mathcal{Z}}_{D_{4}}}    \\
&\setwiretype{n} &\setwiretype{n} & \ctrl[vertical
wire=c]{1} \setwiretype{c}  \wire[l][1]["m_{\mathcal{X}X_{}}"{above,pos=-0.4}]{a} & \ctrl[vertical
wire=c]{1}  \setwiretype{n} \wire[u][1]["m_{X}"{right,pos=0.2}]{c} & \ctrl[vertical
wire=c]{1}  \setwiretype{n} \wire[u][1]["m_{\mathcal{Z}}=2m_{\mathcal{Z}_{1}}+m_{\mathcal{Z}_{2}}"{right,pos=0.2}]{c} \\
\lstick{$\ket{\bar{0}}_{}$} &&\gateinput[1]{$\quad \bar{X}_{\Z_{2}^{}}$} & \gate[1][0.8mm]{\bar{Z}^{m_{\mathcal{X}X_{}}}}\wire[u]{c} & \gate[1][0.8mm]{\bar{Z}^{m_{X}}} & \gate[1][0.8mm]{\bar{Z}^{(m_{\mathcal{Z}_{1}} + m_{\mathcal{Z}_{2}})  }\bar{X}^{m_{\mathcal{Z}_{2}}}} && \rstick{$\ket{T}_{}$}
\end{quantikz} 
\caption{The logical-level circuit diagram showing the preparation of a magic state $\ket{T}$ via lattice surgery between $D(D_{4})$ and $D(\Z_{2})$ code patches.}
\label{fig:D4-Z2_circuit}
\end{figure*}

To generate a logical magic state, the input state on the $D(\Z_2)$ patch is $\ket{\bar{0}}$. After the rough merge and split, the two operators $\bar{R}^s=\bar{X}_{D_4}$ and $\bar{\mathcal{Z}}_{D_4}$ are measured. Denoting the outcomes with $m_{X}$ and $m_{\mathcal{Z}}=2m_{\mathcal{Z}1}+m_{\mathcal{Z}2}$, the output state is
\be
    \bar{Z}_{}^{m_{\mathcal{X}X}+m_{X}+m_{\mathcal{Z}1}+m_{\mathcal{Z}2}} X_{}^{m_{\mathcal{Z}2}} \ket{T}.
\ee
Conditioned on the outcomes, post-measurement operations can recover the magic state $\ket{T}$. An effective logical circuit diagram is shown in Fig.~\ref{fig:D4-Z2_circuit}.

\subsubsection{\textbf{\textit{Gate Teleportation}}}

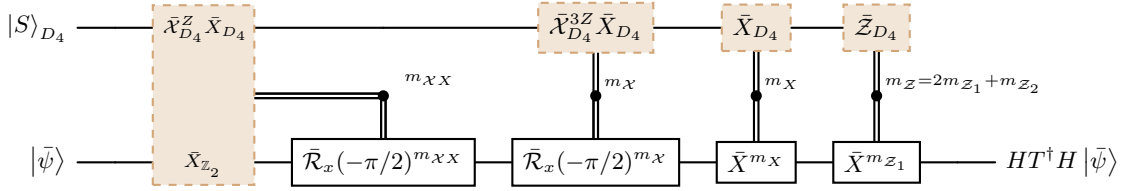
\begin{figure*}
\centering
\begin{quantikz}
\lstick{$\ket{S}_{D_{4}}$} & &  \gate[3, style={fill=brown!20, draw=brown!70, dashed}][1.35cm][0.6cm]{} \gateinput[1]{$\bar{\mathcal{X}}_{D_{4}}^{Z}\bar{X}_{D_{4}}$} && \gate[1, style={fill=brown!20, draw=brown!70, dashed}][0.8mm]{\bar{\mathcal{X}}_{D_{4}}^{3Z}\bar{X}_{D_{4}}} & \gate[1, style={fill=brown!20, draw=brown!70, dashed}][0.8mm]{\bar{X}_{D_{4}}} & \gate[1, style={fill=brown!20, draw=brown!70, dashed}][0.8mm]{\bar{\mathcal{Z}}_{D_{4}}}    \\
&\setwiretype{n} & \setwiretype{n} & \ctrl[vertical
wire=c]{1} \setwiretype{c}  \wire[l][1]["m_{\mathcal{X}X_{}}"{above,pos=-0.4}]{a} & \ctrl[vertical
wire=c]{1}  \setwiretype{n} \wire[u][1]["m_{\mathcal{X}}"{right,pos=0.2}]{c} & \ctrl[vertical
wire=c]{1}  \setwiretype{n} \wire[u][1]["m_{X}"{right,pos=0.2}]{c} & \ctrl[vertical
wire=c]{1}  \setwiretype{n} \wire[u][1]["m_{\mathcal{Z}}=2m_{\mathcal{Z}_{1}}+m_{\mathcal{Z}_{2}}"{right,pos=0.2}]{c} \\
\lstick{$\ket{\bar{\psi}}_{}$} &&\gateinput[1]{$\quad \bar{X}_{\Z_{2}^{}}$} & \gate[1][0.8mm]{\bar{\mathcal{R}}_x(-\pi/2)^{m_{\mathcal{X}X_{}}}}\wire[u]{c} & \gate[1][0.8mm]{\bar{\mathcal{R}}_x(-\pi/2)^{m_{\mathcal{X}}}}\wire[u]{c} & \gate[1][0.8mm]{\bar{X}^{m_{X}}} & \gate[1][0.8mm]{\bar{X}^{m_{\mathcal{Z}_{1}}}} && \rstick{$HT^\dagger H\ket{\bar{\psi}}_{}$}
\end{quantikz} 
\caption{The logical-level circuit diagram showing the gate teleportation via lattice surgery between $D(D_{4})$ and $D(\Z_{2})$ code patches.}
\label{fig:D4-Z2_gate}
\end{figure*}

To implement a gate teleportation, the input state on the $D(\Z_2)$ patch is a general state $\ket{\bar{\psi}}$, and an extra step of measuring the logical operator $\bar{R}^{rs}$ on the $D(D_4)$ patch is required. This corresponds to choosing the interface defined by the subgroup $\langle(r^3s,m), (rs,\id)\rangle \cong\Z_2\times\Z_2$ and $\vp=1$. 
It can be carried out in either of two ways: (A) in addition to the operators in Eq.~\eqref{eq:A^r3s}, measuring the operators 
\be
    \wt{A}_v^{(rs,\id)} =
    \begin{tikzpicture}[baseline]
\begin{scope}[shift={(0,0.1)}]
\draw[thick, ->-] (-1,0) to (0,0);
\draw[thick, ->-] (0,-1) to (0,0);
\draw[thick, ->-] (0,0) to (0,1);
\draw[fill=black] (0,0) ellipse (0.05 and 0.05);
\node[] at (0.2, 0) {$v$};
\node[above] at (-0.7,0.05) {$R^{rs}$};
\node[right] at (0.05,0.7) {$L^{rs}$};
\node[right] at (0.05,-0.7) {$R^{rs}$};
\draw[fill=black] (0,0) ellipse (0.05 and 0.05);
\node[black] at (0.8,0) {$\otimes$};
\node[black] at (1.5,0) {$\bbI$};
\end{scope}
\end{tikzpicture}=\quad
\begin{tikzpicture}[baseline]
\begin{scope}[shift={(0,0.1)}]
\draw[thick, ->-] (-1,0) to (0,0);
\draw[thick, ->-] (0,-1) to (0,0);
\draw[thick, ->-] (0,0) to (0,1);
\draw[fill=black] (0,0) ellipse (0.05 and 0.05);
\node[] at (0.2, 0) {$v$};
\node[above] at (-0.7,0.05) {$\mathcal{X}^{3Z}X$};
\node[right] at (0.05,0.7) {$\mathcal{X}^{}\mathcal{C}X$};
\node[right] at (0.05,-0.7) {$\mathcal{X}^{3Z}X$};
\draw[fill=black] (0,0) ellipse (0.05 and 0.05);
\node[black] at (0.8,0) {$\otimes$};
\node[black] at (1.5,0) {$\bbI$};
\end{scope}
\end{tikzpicture}
\ee
at all the interface vertices during the rough merge; (B) measuring the logical operator $\bar{R}^{rs}$ after the rough split. 
After the rough split, we measure $\bar{R}^s=\bar{X}_{D_4}$ and $\bar{\mathcal{Z}}_{D_4}$ as before, with the respective outcomes $m_X$ and $m_{\mathcal{Z}}$. The overall effect on the state $\ket{\bar{\psi}}$ of the $D(\Z_2)$ patch can be described by a non-Clifford unitary 
\be\ba \label{eq:UTLattice}
    U_{m_X=0, m_{\mathcal{Z}}=0}&=\frac{1}{2}
    \begin{pmatrix}
        1+e^{i\pi/4} & -1+e^{i\pi/4} \\
        -1+e^{i\pi/4} & 1+e^{i\pi/4}
    \end{pmatrix}\\
    &=HX\,T\,XH=e^{i\pi/4}HT^\dagger H\,,
\ea\ee
up to some recovery operations depending on the measurement outcomes. We show a logical circuit diagram in Fig.~\ref{fig:D4-Z2_gate}. A complementary derivation of Eq.~\eqref{eq:UTLattice} from the TQFT is provided in Sec. \ref{sec:T_cont}, which is generalized to gates equivalent to $T^{1/n}$ in Sec. \ref{sec:T^1/n_cont}.

\subsection{Simultaneous Surgery on both Interfaces}
\label{subsec:both}

The two lattice surgery processes, each containing a rough merge and split, on the left and right boundaries of the $D(D_4)$ code patch, respectively, are effectively two logical measurements. These can be expressed as 
\be
\frac{1}{4}\sum_{j=0}^3(\bar{R}^{m}\otimes\bar{L}^{r})^j
\ee
between $D(\Z_4)$ and $D(D_4)$
and 
\be 
\frac{1}{4}(1+\bar{R}^{r^3s}\otimes\bar{L}^{m_{\RR}})(1+\bar{R}^{r^2}\otimes\bar{L}^{m_{\LL}})\ee
[or $\frac{1}{2}(1+\bar{R}^{r^3s}\otimes\bar{L}^{m})$] with the $D(\Z_2\times\Z_2)$ [or $D(\Z_2)$] patch on the right, which commute with each other. With a sufficiently large code distance for the $D(D_4)$ patch, the rough merge and split processes on the two boundaries can be carried out at the same time. Furthermore, the measurement of the $D(\Z_4)$ patch in the logical computational basis commutes with the lattice surgery between the $D(D_4)$ and $D(\Z_2\times\Z_2)$ [or $D(\Z_2)$] patches. Delaying the measurement of the $D(\Z_4)$ patch until after the merge and split processes on both interfaces are completed will not change the result. 

If the two merge-split processes are carried out simultaneously, the hybrid code patch at the intermediate stage will consist of three parts instead of two. Consequently, the error correction procedure for the hybrid patch is more complicated, involving matching of anyons across both interfaces. 

\section{Error Correction and Fault Tolerance}
\label{sec:qec} 

In this section, we discuss the fault tolerance of the merge-split protocol and the associated decoding strategies. The key to achieving fault tolerance lies in the ability to perform error correction within the $D(D_4)$ quantum double model. To set the stage, we first formulate the $D(D_{4})$ quantum double model as a non-commuting stabilizer code~\cite{Davydova:2025ylx}. As in conventional stabilizer codes, the code space consists of the eigenstates with $+1$ eigenvalues of the stabilizers. The difference is that these stabilizers do not commute in the entire Hilbert space.

The generators of the stabilizer group for the $D(D_{4})$ quantum double model are given by 
\begin{align}
A_v^{(r)} &= 
\begin{tikzpicture}[baseline]
\begin{scope}[shift={(0,0.1)}]
\draw[thick, ->-] (-1,0) to (0,0);
\draw[thick, ->-] (0,-1) to (0,0);
\draw[thick, ->-] (0,0) to (0,1);
\draw[thick, ->-] (0,0) to (1,0);
\draw[fill=black] (0,0) ellipse (0.05 and 0.05);
\node[] at (0.15, 0.15) {$v$};
\node[above] at (-0.6,0.05) {$\mathcal{X}^{-Z}$};
\node[right] at (0.04,-0.7) {$\mathcal{X}^{-Z}$};
\node[right] at (0.04,0.8) {$\mathcal{X}$};
\node[above] at (0.75,0.05) {$\mathcal{X}$};
\draw[fill=black] (0,0) ellipse (0.05 and 0.05);
\end{scope}
\end{tikzpicture}, \quad
A_v^{(s)} = 
\begin{tikzpicture}[baseline]
\begin{scope}[shift={(0,0.1)}]
\draw[thick, ->-] (-1,0) to (0,0);
\draw[thick, ->-] (0,-1) to (0,0);
\draw[thick, ->-] (0,0) to (0,1);
\draw[thick, ->-] (0,0) to (1,0);
\draw[fill=black] (0,0) ellipse (0.05 and 0.05);
\node[] at (0.15, 0.15) {$v$};
\node[above] at (-0.7,0.05) {$X$};
\node[right] at (0.04,-0.7) {$X$};
\node[right] at (0.04,0.8) {$\mathcal{C}X$};
\node[above] at (0.75,0.05) {$\mathcal{C}X$};
\draw[fill=black] (0,0) ellipse (0.05 and 0.05);
\end{scope}
\end{tikzpicture} \nonumber
\\
S_{B,p}^{(r)} &=
\begin{tikzpicture}[baseline]
\begin{scope}[shift={(0,-0.4)}]
    \draw[thick, ->-] (-1,0) to (-1,1);
    \node at (-1.4,0.5) {$\mathcal{Z}_{1}$};
    \draw[thick, ->-] (-1,1) to (0,1);
    \node at (-0.5,1.4) {$\mathcal{Z}_{2}^{Z_{1}}$};
    \draw[thick, ->-] (0,0) to (0,1);
    \node at (0.6,0.5) {$\mathcal{Z}_{3}^{-Z_{4}}$};
    \draw[thick, ->-] (-1,0) to (0,0);
    \node at (-0.5,-0.35) {$\mathcal{Z}^{-1}_{4}$};
    \node at (-0.5,0.5) {$_p$};
\end{scope}    
\end{tikzpicture}, \quad
S_{B,p}^{(s)} =
\begin{tikzpicture}[baseline]
\begin{scope}[shift={(0,-0.4)}]
    \draw[thick, ->-] (-1,0) to (-1,1);
    \node at (-1.3,0.5) {$Z$};
    \draw[thick, ->-] (-1,1) to (0,1);
    \node at (-0.5,1.35) {$Z$};
    \draw[thick, ->-] (0,0) to (0,1);
    \node at (0.3,0.5) {$Z$};
    \draw[thick, ->-] (-1,0) to (0,0);
    \node at (-0.5,-0.3) {$Z$};
    \node at (-0.5,0.5) {$_p$};
\end{scope}    
\end{tikzpicture}.
\end{align}
The group commutators between the stabilizer generators always form a subgroup of the stabilizer group:
\begin{align}
    \left[ A_{v}^{(r)}, A_{v}^{(s)} \right] &= A_{v}^{(r^{2})}, \nonumber
    \\ 
    \left[ A_{v}^{(s)}, S_{B,p_{NE}}^{(r)} \right] &= S_{B,p_{NE}}^{(r^{2})}, \nonumber
    \\
    \left[ A_{v}^{(r)}, S_{B,p_{SW}}^{(r)} \right] &= S_{B,p_{SW}}^{(s)},
\end{align}
where $p_{NE}$ and $p_{SW}$ denote the plaquettes located to the northeast and southwest of the vertex $v$, respectively. The commutators for the other stabilizer generators are trivial.

Since the stabilizers do not commute in the full Hilbert space, simultaneous syndrome measurements are not possible. However, within the subspace where the eigenvalues of all $A_{v}^{(r^{2})}$, $S_{B,p}^{(r^{2})}$, and $S_{B,p}^{(s)}$ stabilizers are $+1$, the stabilizer generators do commute. The violation of the $A_{v}^{(r^{2})}$, $S_{B,p}^{(r^{2})}$, and $S_{B,p}^{(s)}$ stabilizers corresponds to some superposition of anyons. We therefore refer to these syndrome defects as non-Abelian syndrome defects. In contrast, errors that violate the remaining stabilizers while keeping all $A_{v}^{(r^{2})}$, $S_{B,p}^{(r^{2})}$, and $S_{B,p}^{(s)}$ stabilizers equal to $+1$ correspond to Abelian anyons. Table~\ref{tb:error_D4} summarizes the correspondence between each syndrome and the associated anyon.

\begin{table}
\begin{center}
\resizebox{\columnwidth}{!}{
 \begin{tabular}{ c | c | c} 
 Errors & Syndromes & Anyons \\
 \hline
 $Z$ & $A_{v}^{(s)} = -1$, $+1$ for other stabilizers & $(1,1_{r})$
 \\
 $\mathcal{Z}^{2}$ & $A_{v}^{(r)} = -1$, $+1$ for other stabilizers & $(1,1_{s})$ 
 \\
 $\mathcal{X}^{2}$ & $S_{B,p}^{(r)} = -1$, $+1$ for other stabilizers & $([r^{2}],1)$
 \\
 \hline
 $\mathcal{Z}$, $\mathcal{Z}^{3}$ & $A_{v}^{(r^{2})} = -1$ & Non-Abelian
 \\
 $\mathcal{X}$, $\mathcal{X}^{3}$ & $S_{B,p}^{r^{2}} = -1$ & Non-Abelian
 \\
 $X$ & $S_{B,p}^{s\phantom{^1}} = -1$ & Non-Abelian
 \end{tabular}
 }
\end{center}
\caption{Errors, syndromes, and the associated anyons in the $D(D_{4})$ quantum double model}
\label{tb:error_D4}
\end{table}

One possible error correction strategy is to first measure the $A_{v}^{(r^{2})}$, $S_{B,p}^{(r^{2})}$, and $S_{B,p}^{(s)}$ stabilizers and immediately correct any detected errors. This step ensures that the system remains in the subspace where all stabilizer generators commute. We can then finish the error correction process by measuring the remaining stabilizers simultaneously in the next step. 

In the presence of measurement errors, the non-Abelian syndrome defects cannot be corrected too quickly, as the correction may be unreliable. One resolution is to apply the just-in-time decoder~\cite{Bombin:2018wjx,Brown2020,Scruby:2020pvw,Davydova:2025ylx}. The key idea is to delay correction until we are sufficiently confident that the observed syndrome is not due to a measurement error. A general rule is that a pair of non-Abelian syndrome defects can be matched after $d_{\text{n}}$ rounds of measurements if the spatial separation of the defects is $d_{\text{n}}$. After the non-Abelian errors are corrected, the remaining errors correspond to Abelian anyons. One can simply store the measurement outcomes of the corresponding stabilizers and correct those errors at the end of the protocol.

The just-in-time decoder has been applied to the $\Z_{2}^{3}$ twisted quantum double (TQD) model, and the existence of a threshold has been proved. The $\Z_{2}^{3}$ TQD model realizes the same $D(D_{4})$ topological order as in the $D_4$ quantum double model. The non-Abelian syndrome defects here play the same role as the ``fluxes" in the TQD model. The proof of the existence of a threshold in \cite{Davydova:2025ylx} can be straightforwardly modified accordingly to show the existence of a threshold for the $D(D_4)$ patch. 

A full fault-tolerant merge-split protocol proceeds as follows. Before merging the Abelian patch with the $D(D_{4})$ patch, we perform $d$ rounds of QEC on both patches, where both patches are prepared to have the same code distance $d$. For the Abelian patch, a QEC round consists of simultaneous stabilizer measurements. For the $D(D_{4})$ patch, we describe $d$ QEC rounds as applying the just-in-time decoder, where the effective ``time” scale is on the order of $d$.

\begin{figure}
\includegraphics[width=8cm]{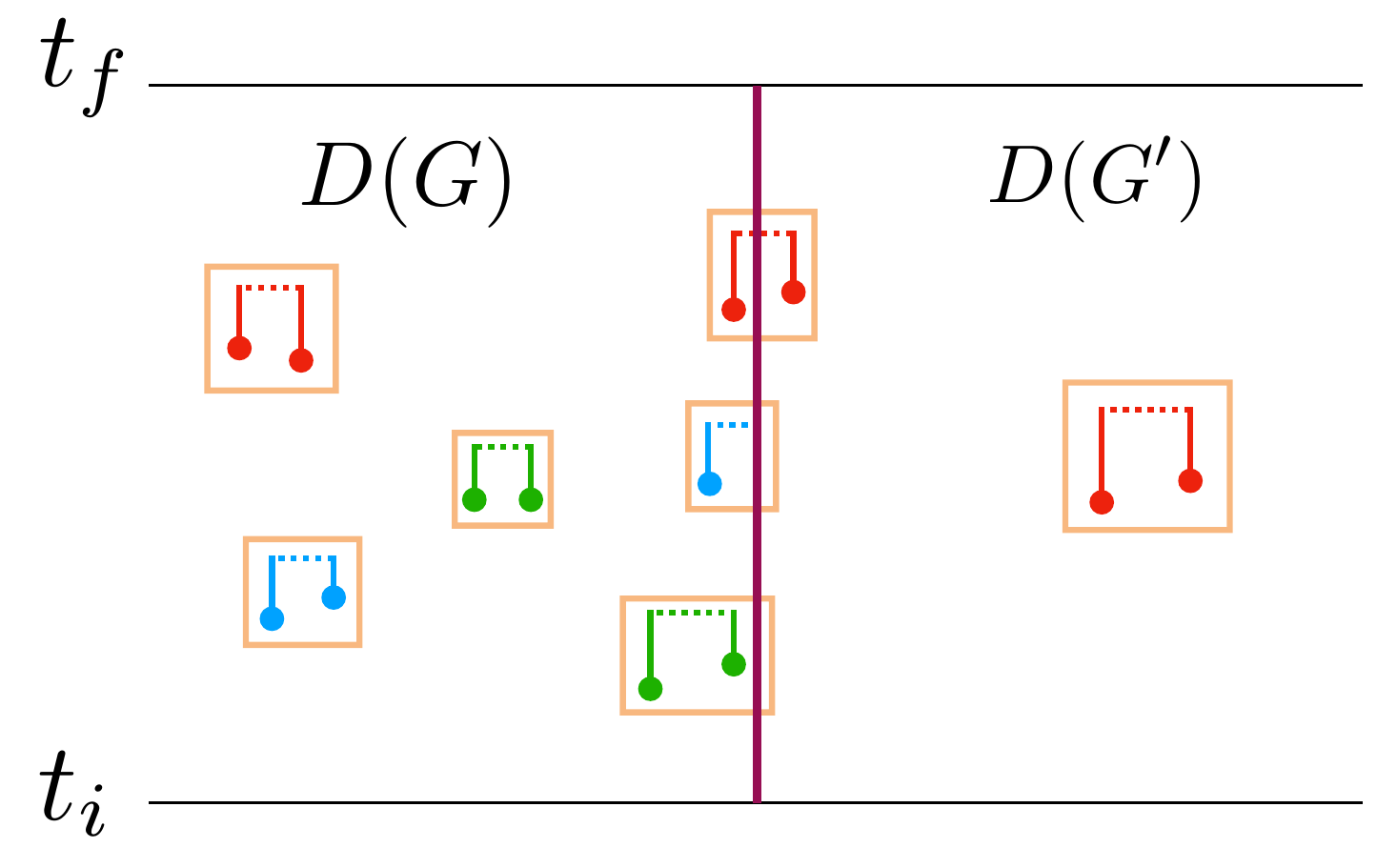}
\caption{Schematic illustration of the just-in-time decoder in the hybrid code patch. The red and blue lines correspond to the worldlines of the syndrome defects that can pass through or end (condense) on the interface, respectively. The green lines correspond to the worldlines of the syndrome defects that are confined on the interface. 
\label{fig:JIT}}
\end{figure}

After the merge, we have a hybrid code with an interface between the Abelian and the non-Abelian regions. In this setting, one can use a modified just-in-time decoder, which takes the presence of the interface into account by appropriately matching the non-Abelian syndrome defects to it~\cite{Davydova:2025ylx}. The essential modifications are as follows. If a non-Abelian syndrome defect can pass through the interface and become an Abelian syndrome defect in the Abelian region, the decoder ignores the interface. If a non-Abelian syndrome defect can end (i.e. condense) on the interface, the decoder treats the interface as a boundary and allows the error to be matched to it, provided the defect is close enough. Finally, if a non-Abelian syndrome defect cannot pass or end on the interface (i.e. it is confined on the interface), the decoder is unchanged and does not match the errors to the interface. This is illustrated in Fig.~\ref{fig:JIT}.

After performing $d$ rounds of QEC for the hybrid code, we split the code back to the original Abelian and $D(D_{4})$ patches, and then carry out $d$ additional QEC rounds on both patches. This completes a fault tolerant merged-split operation. 

\section{TQFT Description of Lattice Surgery} \label{sec:cont}

In this section, we provide a complementary, continuum TQFT point of view on lattice surgery with quantum doubles of finite groups. Here, each code patch will be modeled by a topological field theory, which is the (2+1)d topological order (TO) given by $G$-gauge theory or Dijkgraaf-Witten theory~\cite{Dijkgraaf:1989pz}. Again, $G$ is a finite group  that need not be Abelian.
The lattice merge and split will be characterized in terms of particular---topological or gapped---interfaces between such code patches or TOs. Note that the interfaces can connect two patches that are not necessarily the same. 
The TQFT formulation has several benefits: it gives a general theoretical framework for lattice surgery and, importantly, allows us to systematically search for interesting lattice surgery configurations.

\subsection{Interfaces between Topological Orders}
\label{sec:interf}

The first key input is the concept of a condensation interface between two topological orders, specified in terms of condensable algebras: these are algebras of anyons that can consistently be condensed. From a mathematical viewpoint, they were classified in~\cite{Davydov2009ModularIF,davydov2017lagrangian}. 
If the condensable algebra is maximal, (also called Lagrangian), then it specifies a gapped boundary of $D(G)$~\cite{Beigi:2010htr,delaFuente:2023whm}. 
For lattice surgery we are instead interested in non-maximal condensable algebras $\cA$ that give rise to interfaces between $D(G)$ and $D(G')$. 
\be\label{GLAGR}
\begin{tikzpicture}
\begin{scope}[shift={(0,0)}]
\draw[thick, fill= \BlueColor, opacity=0.5]   (0,0) -- (0,2) -- (2,2) -- (2,0) --(0,0); 
\draw[thick, fill= \RedColor,opacity=0.5]  (2,0) -- (2,2) -- (4,2) -- (4,0) --(2,0); 
\draw [thick]  (0,0) -- (0,2) -- (4,2) -- (4,0) --(0,0); 
\draw[ultra thick, Violet] (2,0) -- (2,2);
\node[above, Violet] at (2,2) {$\mathcal{A}$};
\node at (1,1) {$D(G)$}; 
\node at (3,1) {$D(G')$};
\end{scope}
\end{tikzpicture}
\ee

The representations of a quantum double $D(G)$ form the so-called Drinfeld center $\cZ(G)$, whose simple objects physically correspond to anyons (see e.g.~\cite{Beigi:2010htr}). Each anyon can be labeled by the pair
\be
    ([g],R)
\ee
where $[g]=\{hgh^{-1} : h\in G\}$ is a conjugacy class of $G$
and $R$ is an irreducible representation (irrep) of the centralizer $C_G(g)=\{h\in G: hg=gh\}$ of a representative $g\in[g]$.

A condensable algebra $\cA$ in $\cZ(G)$ is specified
 by the data~\cite{Davydov2009ModularIF}
\begin{equation}
	\cA(K, N, \varphi, \epsilon)\,,
\end{equation}
where
\begin{itemize}
	\item $K\subseteq G$ is a subgroup;
	\item $N \triangleleft K$ is a normal subgroup;
	\item $\varphi\in H^2(N,U(1))$ is a normalized 2-cocycle,
	\item $\epsilon: K\times N \rightarrow U(1)$ satisfies conditions~\cite[(8)-(10)]{Davydov2009ModularIF}.
\end{itemize}
For Lagrangian algebras $K=N$ and $\epsilon$ is completely specified by $\varphi$, so they will be labeled by $(K,\varphi)$. The anyons appearing in the algebra can be computed from character theory~\cite{Beigi:2010htr,davydov2017lagrangian,delaFuente:2023whm}, see~\cite{GaiSchaferNamekiWarman} for a concise formula.

When $N$ is a proper (normal) subgroup of $K$, the algebra $\cA(K,N,\varphi,\epsilon)$ is non-maximal and gives rise to a map of anyons from $D(G)$ to 
$D(G')= D(K/N,\wt{\alpha})$, 
the quantum double of a group
\be
    G'=K/N\,,
\ee
with a possible 3-cocycle twist $\wt{\alpha}$ determined by the algebra data~\cite{davydov2017lagrangian}. 
In this work, we will restrict to condensable algebras for which $\wt{\alpha}$ is trivial, which specify an interface between $D(G)$ and 
\be
D(G')=D(K/N)\,.
\ee

An interface between TOs can also be seen as a gapped boundary condition for the folded TO, in this case
\be
{D(G)} \boxtimes \ol{D(G')} \,.
\ee
The associated gapped boundary condition is encoded in the ``Folded Lagrangian'' (see~\cite{Beigi:2010htr,Chatterjee:2022tyg,Bhardwaj:2023bbf,Bhardwaj:2024qrf,Bhardwaj:2025jtf} for prior work on this topic). It is specified by~\cite{GaiSchaferNamekiWarman}
\be
    \cL^\folded(K^\diag,\vp) \in {\cZ(G)} \boxtimes \ol{\cZ(G')}\,,
\ee
where
\begin{itemize}
    \item $K^\diag \subset G\times G'$ is a subgroup isomorphic to $K\subseteq G$, with elements 
    \be\label{Kdiagdef}
       K^\diag=\{(h,p(h))\;:\;h\in K\}\,.
    \ee
    Here
    \be \label{eq:p_def}
        p:K\to K/N\,,
    \ee
    is the projection from $K$ to $G'=K/N$ whose kernel is $N$,\footnote{$p$ is a group homomorphism, i.e. $p(h_1)p(h_2)=p(h_1h_2)$. Indeed, let $p(h_i)=h_iN$, then $p(h_1)p(h_2)=h_1Nh_2N=h_1h_2N$ where we used that $Nh_2=h_2N$, which follows from the requirement that $N$ is normal in $K$. Therefore $(h_1, p(h_1))(h_2, p(h_2))=(h_1h_2, p(h_1h_2))$ and $K^\diag\cong K$ as a group.}
    \item $\vp$ is a 2-cocycle on $K^\diag$ such that the Folded Lagrangian $\cL^\folded\in {\cZ(G)}\boxtimes \ol{\cZ(G')}$ for $\cA\in \cZ(G)$ contains ${\cA}\otimes1$ as a subalgebra: in the unfolded picture, this means that the anyons in $\cA$ are mapped to the identity anyon $1$ in $\cZ(G')$.
\end{itemize}

\subsection{Merge and Split via Anyon Condensation} 

The merge and split operations in lattice surgery can be realized in terms of the TQFT description and interfaces as follows. This section is general and does not depend on the specific choices of boundary conditions. We will then connect to the lattice surgery rough and smooth merge and split operations in Sec.~\ref{sec:lattice_surgery_interpretation}.

We start with two TOs in two separate patches: this will be depicted in a projection to one dimension lower, as follows, where each boundary is (1+1)d: 
\be
\begin{tikzpicture}
\begin{scope}[shift={(0,0)}]
\draw [thick, fill= \BlueColor, opacity=0.5]   (0,0) -- (0,2) -- (2,2) -- (2,0) --(0,0); 
\draw [thick]  (0,0) -- (0,2) -- (2,2) -- (2,0) --(0,0); 
\node at (1,1) {$D(G)$}; 
\end{scope}
\begin{scope}[shift={(4,0)}]
\draw[thick, fill= \RedColor,opacity=0.5]  (0,0) -- (0,2) -- (2,2) -- (2,0) --(0,0); 
\draw [thick]  (0,0) -- (0,2) -- (2,2) -- (2,0) --(0,0); 
\node at (1,1) {$D(G')$}; 
\end{scope}
\end{tikzpicture}
\ee

This is equivalent to folding the left TO over, and considering 
\be
\begin{tikzpicture} 
 \draw [Violet,  fill=Violet, opacity =0.5]
(0,0) -- (0,2) --(3,2) -- (3,0) -- (0,0) ; 
 \draw [thick]
(0,0) -- (0,2) --(3,2) -- (3,0) -- (0,0) ; 
\draw [very thick] (3,0) -- (3,2)  ;
\node at (1.5,1) {${D(G)}\boxtimes\ol{D(G')}$}; 
\end{tikzpicture}
\ee

To perform the merge 
we consider a condensable algebra $\cA\in\cZ(G)$ such that the reduced TO is 
\be
D(G') = D (K/N) \,.
\ee
This identifies a diagonal subgroup of $G\times G'$ as in (\ref{Kdiagdef})
\be
K^\diag \subset G\times G' \,,
\ee
and a cocycle $\vp$, which in turn specifies a $K^\diag$-SPT in (1+1)d. 
The merge is obtained by choosing the boundary condition on the right to be specified by $\cL^\folded (K^\diag, \vp)$
\be
\begin{tikzpicture} 
 \draw [Violet,  fill=Violet, opacity =0.5]
(0,0) -- (0,2) --(3,2) -- (3,0) -- (0,0) ; 
 \draw [thick]
(0,0) -- (0,2) --(3,2) -- (3,0) -- (0,0) ; 
\draw [ultra thick] (3,0) -- (3,2)  ;
\node at (1.5,1) {${D(G)}\boxtimes\ol{D(G')}$}; 
\node[right] at (3,1) {$\cL^\folded (K^\diag, \vp)$};
\end{tikzpicture}
\ee
Unfolding the TOs, this is precisely the merge of the two topological orders, specified by the algebra $\cA$:
\be \label{eq:setup}
\begin{tikzpicture}
\begin{scope}[shift={(0,0)}]
\draw [thick, fill= \BlueColor, opacity = 0.5]   (0,0) -- (0,2) -- (3,2) -- (3,0) --(0,0); 
\draw [thick]  (0,0) -- (0,2) -- (3,2) -- (3,0) --(0,0); 
\node at (1.5,1) {$D(G)$}; 
\end{scope}
\begin{scope}[shift={(3,0)}]
\draw[thick, fill= \RedColor, opacity = 0.5]   (0,0) -- (0,2) -- (3,2) -- (3,0) --(0,0); 
\draw [thick]  (0,0) -- (0,2) -- (3,2) -- (3,0) --(0,0); 
\node at (1.5,1) {$D(G')$}; 
\node[above,Violet] at (0,2) {$\cA$};
\draw[ultra thick, Violet] (0,2) -- (0,0); 
\end{scope}
\end{tikzpicture}
\ee
This setup describes two TOs with a condensation interface, specified by the algebra $\cA\in \cZ(G)$. It provides in particular an identification between anyons in $\cZ(G)$ and $\cZ(G')$, as it can be written in terms of the anyons as 
\be
\cL^{\folded} (K^\diag, \vp) = \bigoplus_{a \in D (G), b \in D (G')} n_{ab} \;{a} \otimes \ol{b} \,.
\ee
The anyons appearing in $\cL^\folded (K^\diag, \vp)$ can be absorbed by the interface and become trivial after the merge.

The split is now simply the ungauging operation on the interface. If $K^\diag$ is an Abelian group, we simply gauge the Pontryagin dual group $\widehat{K}^\diag$. If $K^\diag$ is non-Abelian, the dual symmetry is $\Rep (K^\diag)$, i.e. a non-invertible symmetry. Ungauging this results back in the two TOs without identification.

\subsection{Lattice Surgery Interpretation}  \label{sec:lattice_surgery_interpretation}

In this Section, we provide the continuum TQFT description of the lattice surgery construction of Sec. \ref{sec:lattice_surgery}.

\subsubsection{\textbf{\textit{Rough and Smooth Boundary Conditions}}} \label{sec:cont_BCs}

Boundary conditions of $D(G)$ are labeled by pairs $(K,\vp)$ for $K\subseteq G$ and $\vp\in H^2(K,U(1))$~\cite{Ostrikmodule,Beigi:2010htr,Natale2017}. $D(G)$ always has two canonical boundary conditions. 
\begin{itemize}
    \item The rough (or Dirichlet) boundary condition corresponds to $K=\{\id\}$ (with necessarily trivial $\vp$) and is specified by the Lagrangian algebra that condenses all irreps $R$ of $G$ each with multiplicity $\dim(R)$: 
    \be 
    \cL_{G}^\rough = \bigoplus_{ R  \in \text{irreps}(G)} \dim(R)\, R \,.
    \ee
    \item Conversely, the smooth (or Neumann) boundary condition corresponds to $K=G$ and trivial $\vp$ and is given by the Lagrangian algebra that condenses all conjugacy classes $[g]$ of $G$:
    \be\label{Lsmooth}
    \cL_{G}^\smooth = \bigoplus_{{[g]}\in \text{ConjClasses}(G)} {[g]} \,.
    \ee
\end{itemize}

Let us choose the boundary conditions of a $D(G)$ patch as follows: 
\begin{itemize}
   \item Top and bottom boundaries are smooth, i.e. all the conjugacy classes $[g]$ are condensed,
   \item Left and right boundaries are rough, i.e. all the irreducible representations (irreps) $ R $ are condensed.
\end{itemize}
This will be depicted in a projection to one dimension lower, as follows, where each boundary is (1+1)d: 
\be\label{eq:tblr}
\begin{tikzpicture}
\begin{scope}[shift={(0,0)}]
\draw [thick, fill= \BlueColor, opacity=0.5]   (0,0) -- (0,2) -- (2,2) -- (2,0) --(0,0); 
\draw [thick]  (0,0) -- (0,2) -- (2,2) -- (2,0) --(0,0); 
\node at (1,1) {$D(G)$}; 
\node[above] at (1,2) {$\langle[g]\rangle$};
\node[below] at (1,0) {$\langle[g]\rangle$};
\node[left] at (0,1) {$\langle R \rangle$};
\node[right] at (2,1) {$\langle R \rangle$};
\end{scope}
\end{tikzpicture}
\ee

\subsubsection{\textbf{\textit{ Bases of logical states}}}

We now turn to the logical bases, whose lattice description we provided in Sec. \ref{subsec:DG}. Note that, in the TQFT, we only discuss \emph{logical} operations, so all states and operators are taken to be logical and we will omit the bar which was used in the lattice section.

\vspace{2mm}
\noindent{\bf Group element basis of logical states.}
Along the rough boundaries, we get local operators at the end of the condensed irrep anyons, from which we can resolve the different group elements in each conjugacy class
\be \label{eq:g_basis}
\begin{tikzpicture}
\begin{scope}[shift={(0,0)}]
\draw [thick, fill= \BlueColor, opacity=0.5]   (0,0) -- (0,2) -- (2,2) -- (2,0) --(0,0); 
\draw [thick]  (0,0) -- (0,2) -- (2,2) -- (2,0) --(0,0); 
\node at (1,1) {$D(G)$}; 
\node[above] at (1,2) {$\langle[g]\rangle$};
\node[below] at (1,0) {$\langle[g]\rangle$};
\node[left] at (0,1) {$\langle R \rangle$};
\node[right] at (2,1) {$\langle R \rangle$};
\node[above,red] at (2,2) {$\bm{g}$};
\draw[very thick, red, ->] (2,0) -- (2,2); 
\node at (4,1) {$\Rightarrow\quad\ket{g}$}; 
\end{scope}
\end{tikzpicture}
\ee
The vertical $([g],1)$ anyons, when pushed to the rough boundaries, thus provide a basis of logical states of order $|G|$
\be \label{eq:ketg_basis}
    \{\ket{g}:g\in G\}\,,
\ee
which is the ``logical computational basis''.

\vspace{2mm}
\noindent{\bf Irrep basis of logical states.}
Along each smooth boundary, every irrep anyon $R$ ends with multiplicity $\dim(R)$,
\be \label{eq:R_basis}
\begin{tikzpicture}
\begin{scope}[shift={(0,0)}]
\draw [thick, fill= \BlueColor, opacity=0.5]   (0,0) -- (0,2) -- (2,2) -- (2,0) --(0,0); 
\draw [thick]  (0,0) -- (0,2) -- (2,2) -- (2,0) --(0,0); 
\node at (1,1) {$D(G)$}; 
\node[above] at (1,2) {$\langle[g]\rangle$};
\node[below] at (1,0) {$\langle[g]\rangle$};
\node[left] at (0,1) {$\langle R \rangle$};
\node[right] at (2,1) {$\langle R \rangle$};
\node[left,blue] at (0,0.55) {$\bm{i}$};
\node[right,blue] at (2,0.55) {$\bm{j}$};
\node[below,blue] at (1,0.55) {$\bm{R}$};
\draw[very thick, blue, ->] (0,0.55) -- (2,0.55); 
\node at (5.2,1) {$\Rightarrow\;\ket{R^{i,j}}\; i,j=1,...,\dim(R)$}; 
\end{scope}
\end{tikzpicture}
\ee
The horizontal anyons therefore provide an alternative basis of logical states labeled by
\be
   \{\ket{R^{i,j}}\;:\; R\in \text{Irreps}(G),\;i,j=1,...,\dim(R)\}\,.
\ee
Indeed, it is of the same order as the basis \ref{eq:ketg_basis}, since
\be
    \sum_R \dim(R)^2=|G|\,,
\ee
and all its states are orthogonal thanks to the Schur orthogonality relations.

\subsubsection{\textbf{\textit{Rough Merge}}}
In this section, we provide the TQFT description of rough merges, complementing the lattice realization provided in Sec. \ref{sec:Rough_merge}.

To perform a rough merge between $D(G)$ and $D(G')$, for $G$ and $G'$ finite groups, we choose the boundary conditions as follows: 
\be\label{eq:rough_BC}
\begin{tikzpicture}
\begin{scope}[shift={(0,0)}]
\draw [thick, fill= \BlueColor, opacity=0.5]   (0,0) -- (0,2) -- (2,2) -- (2,0) --(0,0); 
\draw [thick]  (0,0) -- (0,2) -- (2,2) -- (2,0) --(0,0); 
\node at (1,1) {$D(G)$}; 
\node[above] at (1,2) {$\langle[g]\rangle$};
\node[below] at (1,0) {$\langle[g]\rangle$};
\node[left] at (0,1) {$\langle R \rangle$};
\node[right] at (2,1) {$\langle R \rangle$};
\end{scope}
\begin{scope}[shift={(4,0)}]
\draw[thick, fill= \RedColor,opacity=0.5]  (0,0) -- (0,2) -- (2,2) -- (2,0) --(0,0); 
\draw [thick]  (0,0) -- (0,2) -- (2,2) -- (2,0) --(0,0); 
\node at (1,1) {$D(G')$}; 
\node[above] at (1,2) {$\langle[g']\rangle$};
\node[below] at (1,0) {$\langle[g']\rangle$};
\node[left] at (0,1) {$\langle R' \rangle$};
\node[right] at (2,1) {$\langle R' \rangle$};
\end{scope}
\end{tikzpicture}
\ee

As discussed earlier this is equivalent to folding the left TO over
\be
\begin{tikzpicture} 
 \draw [Violet,  fill=Violet, opacity =0.5]
(0,0) -- (0,2) --(3,2) -- (3,0) -- (0,0) ; 
 \draw [thick]
(0,0) -- (0,2) --(3,2) -- (3,0) -- (0,0) ; 
\draw [very thick] (3,0) -- (3,2)  ;
\node at (1.5,1) {${D(G)}\boxtimes\ol{D(G')}$}; 
\node[above] at (1.5,2) {${\langle[g]\rangle}\otimes \ol{\langle[g']\rangle}$};
\node[below] at (1.5,0) {${\langle[g]\rangle}\otimes \ol{\langle[g']\rangle}$};
\node[left] at (0,1) {${\langle {R} \rangle} \otimes \ol{\langle { R' } \rangle} $};
\node[right] at (3,1) {${\langle {R} \rangle} \otimes\ol{\langle { R' } \rangle} $};
\end{tikzpicture}
\ee
Here ${\langle[g]\rangle}\otimes\ol{\langle[g']\rangle}$ and ${\langle {R} \rangle} \otimes\ol{\langle { R' } \rangle} $ denote the tensor product of Lagrangian algebras for the smooth and rough boundaries. In particular, ${\langle {R} \rangle} \otimes\ol{\langle { R' } \rangle} $ is
    \be 
    \cL_{G}^\rough\otimes \ol{\cL_{G'}^\rough}  = \bigoplus_{ \substack{R  \in \text{Irreps}(G)\\
    R'\in \text{Irreps}(G')}} \dim(R)\dim(R')\, R\otimes R' \,.
    \ee
The merge is obtained by changing the boundary condition on the right to be specified by $\cL^\folded (K^\diag, \vp)$: this is achieved by stacking this rough boundary with an SPT for $K^{\diag}$, specified by $\vp$, and gauging $K^\diag$: 
\be\label{LFolded}
\cL^\folded(K^\diag,\vp) ={ \left( {\cL_{G}^\rough}  \otimes \ol{\cL_{G'}^{\rough}}\right) \otimes \SPT^\vp_{K^\diag}\over K^\diag }\,,
\ee
and gives
\be\label{FoldedNonProduct}
\begin{tikzpicture} 
 \draw [Violet,  fill=Violet, opacity =0.5]
(0,0) -- (0,2) --(3,2) -- (3,0) -- (0,0) ; 
 \draw [thick]
(0,0) -- (0,2) --(3,2) -- (3,0) -- (0,0) ; 
\draw [very thick] (3,0) -- (3,2)  ;
\node at (1.5,1) {${D(G)}\boxtimes\ol{D(G')}$}; 
\node[above] at (1.5,2) {${\langle[g]\rangle}\otimes \ol{\langle[g']\rangle}$};
\node[below] at (1.5,0) {${\langle[g]\rangle}\otimes\ol{\langle[g']\rangle}$};
\node[left] at (0,1) {${\langle {R} \rangle} \otimes\ol{\langle { R' } \rangle} $};
\node[right] at (3,1) {$\cL^\folded (K^\diag, \vp)$};
\end{tikzpicture}
\ee

Unfolding the TOs, this is the rough merge of the two topological orders, specified by the condensable algebra $\cA$ with Folded Lagrangian $\cL^\folded (K^\diag, \vp)$:
\be 
\begin{tikzpicture}
\begin{scope}[shift={(0,0)}]
\draw [thick, fill= \BlueColor, opacity = 0.5]   (0,0) -- (0,2) -- (3,2) -- (3,0) --(0,0); 
\draw [thick]  (0,0) -- (0,2) -- (3,2) -- (3,0) --(0,0); 
\node at (1.5,1) {$D(G)$}; 
\node[above] at (1.5,2) {$\langle[g]\rangle$};
\node[below] at (1.5,0) {$\langle[g]\rangle$};
\node[left] at (0,1) {$\langle R \rangle$};
\end{scope}
\begin{scope}[shift={(3,0)}]
\draw[thick, fill= \RedColor, opacity = 0.5]   (0,0) -- (0,2) -- (3,2) -- (3,0) --(0,0); 
\draw [thick]  (0,0) -- (0,2) -- (3,2) -- (3,0) --(0,0); 
\node at (1.5,1) {$D(G')$}; 
\node[above] at (1.5,2) {$\langle[g']\rangle$};
\node[below] at (1.5,0) {$\langle[g']\rangle$};
\node[right] at (3,1) {$\langle R' \rangle$};
\node[above,Violet] at (0,2) {$\cA$};
\draw[ultra thick, Violet] (0,2) -- (0,0); 
\end{scope}
\end{tikzpicture}
\ee

For a lattice surgery rough merge, the non-trivial logical states in the computational basis correspond to vertical anyons that can end on the top and bottom boundaries: they are labeled by conjugacy classes in the bulk and split into group elements on the rough boundaries, as shown in \eqref{eq:g_basis}. After the merge, the logical states that differ by multiplication by group elements in $K^\diag$ are identified. This follows from the Folded Lagrangian \eqref{LFolded} (see~\cite{GaiSchaferNamekiWarman} for a concise version of the formula from group-theory data to anyons), since the anyons in $\cL^\folded(K^\diag,\vp)$ have conjugacy classes labeled by group elements in $K^\diag$: having these anyons condensed means that they become trivial and hence group elements related by multiplication by elements in $K^\diag$ are equivalent.
We will write the equivalence relation in the unfolded picture as:
\be \label{eq:sim}
   (g, g') \sim (gh^{-1},\, p(h)g')\,.
\ee
for $g\in G,\;g'\in G',(h,p(h))\in K^\diag$, where this convention (right multiplication on the first entry and left on the second entry) is chosen to match the lattice conventions \eqref{eq:lattice_equivalence}.\footnote{Note that the identification \eqref{eq:sim} is compatible with the group multiplication structure since
\be
    (gh_1^{-1}h_2^{-1},\, p(h_2)p(h_1)g')=(g(h_2h_1)^{-1},\, p(h_2h_1)g')\,.
\ee}
Because of the identification \eqref{eq:sim}, we see that operators for group elements in $K^\diag$ do not change the logical state post-merge: the logical space has therefore been made smaller by the merge procedure.
This merged topological order is what we refer to as the {\bf hybrid code-patch} in the lattice section Sec.~\ref{subsec:hybrid}.

\subsubsection{\textbf{\textit{Rough Split}}}
When considering a single topological order, one can define a split in the usual way, introducing a trivial interface and suitably gauging on it. 
More generally, we start with two TOs $D(G)$ and $D(G')$, which are separated by an interface, as in (\ref{eq:setup}). 

In order to split these into two distinct TOs, we reverse the stacking and gauging operation that produced the merge, i.e. in the folded picture, we have (\ref{FoldedNonProduct}). The rough split is then produced in two steps: 
\begin{itemize}
\item Gauging the dual symmetry $\Rep (K^\diag)$ (i.e. the symmetry that we get on the interface after we have gauged $K^\diag$ in the merge)
\item Stacking with the inverse SPT, i.e.  $(\SPT^\varphi_{K^\diag})^{-1}$.
\end{itemize}
That is 
\be
{\cL^\folded (K^\diag, \varphi) \over \Rep (K^\diag)} \boxtimes (\SPT^\varphi_{K^\diag})^{-1} = \cL_{G}^\rough  \otimes \overline{\cL_{G'}^{\rough}} \,.
\ee
This maps the configuration back to (\ref{eq:rough_BC}).

\vspace{2mm}
\noindent{\bf Change of BCs after Rough Merge and Split.}
After having performed a rough merge and split between $D(G)$ and $D(G')$, we will have returned to the configuration (\ref{eq:rough_BC}). The rough boundary of $D(G)$ carries symmetry $G$, as shown in \eqref{eq:g_basis}. We can then gauge any subgroup of $G$ which corresponds to a change of boundary condition to $(K,\vp)$ for $K\subseteq G$ and $\vp\in H^2(K,U(1))$: 
\be\label{changedBC}
\begin{tikzpicture}
\begin{scope}[shift={(0,0)}]
\draw [thick, fill= \BlueColor, opacity=0.5]   (0,0) -- (0,2) -- (2,2) -- (2,0) --(0,0); 
\draw [thick]  (0,0) -- (0,2) -- (2,2) -- (2,0) --(0,0); 
\node at (1,1) {$D(G)$}; 
\node[above] at (1,2) {$\langle[g]\rangle$};
\node[below] at (1,0) {$\langle[g]\rangle$};
\node[left] at (0,1) {$\langle R \rangle$};
\node[right] at (2,1) {$\cL(K,\vp)$};
\end{scope}
\begin{scope}[shift={(4.5,0)}]
\draw[thick, fill= \RedColor,opacity=0.5]  (0,0) -- (0,2) -- (2,2) -- (2,0) --(0,0); 
\draw [thick]  (0,0) -- (0,2) -- (2,2) -- (2,0) --(0,0); 
\node at (1,1) {$D(G')$}; 
\node[above] at (1,2) {$\langle[g']\rangle$};
\node[below] at (1,0) {$\langle[g']\rangle$};
\node[left] at (0,1) {$\langle R' \rangle$};
\node[right] at (2,1) {$\langle R' \rangle$};
\end{scope}
\end{tikzpicture}
\ee 
This specifies a gapped boundary of $D(G)$ which condenses the anyons in a Lagrangian algebra $\cL(K,\vp)$. We will use this step in our protocols to produce magic states and non-Clifford gates.

\subsubsection{\textbf{\textit{Smooth Merge and Split}}}

Let us briefly discuss the other merge and split lattice surgery in the TQFT. Instead of the Dirichlet (rough) boundary, we can merge along the smooth (Neumann) boundary, given by 
(\ref{Lsmooth}), i.e. the anyons that are in the conjugacy classes can end on the boundaries along which we merge. 

The starting configuration is the following setup: 
\be
\begin{tikzpicture}
\begin{scope}[shift={(0,0)}]
\draw [thick, fill= \BlueColor, opacity=0.5]   (0,0) -- (0,2) -- (2,2) -- (2,0) --(0,0); 
\draw [thick]  (0,0) -- (0,2) -- (2,2) -- (2,0) --(0,0); 
\node at (1,1) {$D(G)$}; 
\node[above] at (1,2) {$\langle R \rangle$};
\node[below] at (1,0) {$\langle R \rangle$};
\node[left] at (0,1) {$\langle[g]\rangle$};
\node[right] at (2,1) {$\langle[g]\rangle$};
\node[above,thick,\BlueColor] at (2,2) {$\phantom{p} R \phantom{p}$};
\draw[very thick, \BlueColor, ->] (2,0) -- (2,2); 
\end{scope}
\begin{scope}[shift={(4,0)}]
\draw[thick, fill= \RedColor,opacity=0.5]  (0,0) -- (0,2) -- (2,2) -- (2,0) --(0,0); 
\draw [thick]  (0,0) -- (0,2) -- (2,2) -- (2,0) --(0,0); 
\node at (1,1) {$D(G')$}; 
\node[above] at (1,2) {$\langle R' \rangle$};
\node[below] at (1,0) {$\langle R' \rangle$};
\node[left] at (0,1) {$\langle[g']\rangle$};
\node[right] at (2,1) {$\langle[g']\rangle$};
\node[above,\RedColor] at (0,2) {$\bm{\ol{R'}}$};
\draw[very thick, \RedColor, ->] (0,2) -- (0,0); 
\end{scope}
\end{tikzpicture}
\ee
The symmetry along the interface is $\Rep (G)\boxtimes \Rep (G')$. We can now again gauge any subsymmetry (i.e. pick a module category) to generate different interfaces. For example, if the groups are Abelian, then $\Rep (G)$ is simply ${\widehat{G}}$, i.e. the group of characters, which is isomorphic to $G$. In this case, the TQFT analysis of the merge and split works in the same fashion as for the rough merge and split.

\subsection{Example: $D(\Z_n)$ with $D(\Z_n)$ merge and split}
\vspace{2mm}
\noindent{\bf $D(\Z_2)$ with $D(\Z_2)$ rough merge.}
Let us first review the rough merge of two $D(\Z_2)$ topological orders, first proposed in~\cite{Horsman:2011hyt}.
The starting configuration is the following setup: 
\be
\resizebox{\columnwidth}{!}{
\begin{tikzpicture}
\begin{scope}[shift={(0,0)}]
\draw [thick, fill= \BlueColor, opacity=0.5]   (0,0) -- (0,2) -- (2,2) -- (2,0) --(0,0); 
\draw [thick]  (0,0) -- (0,2) -- (2,2) -- (2,0) --(0,0); 
\node at (1,1) {$D(\Z_{2})$}; 
\node[above] at (1,2) {$\langle e_{\LL} \rangle$};
\node[below] at (1,0) {$\langle e_{\LL} \rangle$};
\node[left] at (0,1) {$\langle m_{\LL}\rangle$};
\node[right] at (2,1) {$\langle m_{\LL} \rangle$};
\node[above,\BlueColor] at (2,2) {$\bm{m_{\LL}}$};
\draw[very thick, \BlueColor, ->] (2,2) -- (2,0); 
\end{scope}
\begin{scope}[shift={(4.5,0)}]
\draw[thick, fill= \RedColor,opacity=0.5]  (0,0) -- (0,2) -- (2,2) -- (2,0) --(0,0); 
\draw [thick]  (0,0) -- (0,2) -- (2,2) -- (2,0) --(0,0); 
\node at (1,1) {$D(\Z_{2})$}; 
\node[above] at (1,2) {$\langle e_{\RR} \rangle$};
\node[below] at (1,0) {$\langle e_{\RR} \rangle$};
\node[left] at (0,1) {$\langle m_{\RR} \rangle$};
\node[right] at (2,1) {$\langle m_{\RR} \rangle$};
\node[above,\RedColor] at (0,2) {$\bm{m_{\RR}}$};
\draw[very thick, \RedColor, ->] (0,0) -- (0,2); 
\end{scope}
\end{tikzpicture}
}
\ee
where we take $G=G'=\Z_2$ and label the two copies by $L$ and $R$. In each patch, we fix smooth ($m$ condensed) boundaries at the top and bottom, and rough ($e$ condensed) boundaries at the left and right.
In this case we take the condensable algebra $\cA\in \cZ(\Z_2)$ to be the trivial algebra
\be
\cA=1\,,
\ee
in which $K=\Z_2$, $N=\{\id\}$ and only the identity anyon $1$ is condensed. The group homomorphism $p$, defined in Eq. \eqref{eq:p_def}, in this case maps the $\Z_2$ generator $m_{\LL}$ on the left to the $\Z_2$ generator $m_{\RR}$ on the right:
\be\ba
    p:\;\;\Z_2&\to\Z_2\\
        m_{\LL}&\mapsto m_{\RR}\,.
\ea\ee
Therefore, in this case, the diagonal subgroup $K^\diag$, defined in \eqref{Kdiagdef}, of $G\times G'=\Z_2\times\Z_2$ is
\be
    K^\diag=\Z_2^\diag=\langle m_{\LL}m_{\RR}\rangle\,,
\ee
with corresponding Folded Lagrangian in $D(\Z_2)\boxtimes\ol{D(\Z_2)}$
\be \label{eq:LFolded_Z2}
    \cL^\folded(\Z_2^\diag)=1\oplus m_{\LL}\ol{m_{\RR}} \oplus e_{\LL}\ol{e_{\RR}} \oplus e_{\LL}m_{\LL}\ol{e_{\RR}m_{\RR}}\,.
\ee

The merge operation can then be depicted as the identification \eqref{eq:sim} on logical states:
\begin{align}\label{eq:Z2_triv}
\begin{tikzpicture}
 \begin{scope}[shift={(0,0)}]
    \draw[-] (-1,-0.5) to (-1,0.5);
    \draw[-] (-1,0.5) to (0,0.5);
    \draw[-] (0,-0.5) to (0,0.5);
    \draw[-] (-1,-0.5) to (0,-0.5);
    \end{scope}
       \begin{scope}[shift={(1,0)}]
    \draw[-] (-1,-0.5) to (-1,0.5);
    \draw[-] (-1,0.5) to (0,0.5);
    \draw[-] (0,-0.5) to (0,0.5);
    \draw[-] (-1,-0.5) to (0,-0.5);
    \end{scope}
    \begin{scope}[shift={(4,0)}]
    \node at (-2,0) {$\sim$};
    \draw[-] (-1,-0.5) to (-1,0.5);
    \draw[-] (-1,0.5) to (0,0.5);
    \node[red] at (-0.5,0.7) {$_{m_{\LL}}$};
    \draw[-] (0,-0.5) to (0,0.5);
    \draw[-] (-1,-0.5) to (0,-0.5);
    \draw[-,thick,red] (-0.5,-0.5) to (-0.5,0.5);
    \end{scope}
       \begin{scope}[shift={(5,0)}]
    \draw[-] (-1,-0.5) to (-1,0.5);
    \draw[-] (-1,0.5) to (0,0.5);
    \node[red] at (-0.5,0.7) {$_{m_{\RR}}$};
    \draw[-] (0,-0.5) to (0,0.5);
    \draw[-] (-1,-0.5) to (0,-0.5);
    \draw[-,thick,red] (-0.5,-0.5) to (-0.5,0.5);
    \end{scope}
\end{tikzpicture}\\
\label{eq:Z2_non-triv}
\begin{tikzpicture}
 \begin{scope}[shift={(0,0)}]
    \draw[-] (-1,-0.5) to (-1,0.5);
    \draw[-] (-1,0.5) to (0,0.5);
    \node[red] at (-0.5,0.7) {$_{m_{\LL}}$};
    \draw[-] (0,-0.5) to (0,0.5);
    \draw[-] (-1,-0.5) to (0,-0.5);
    \draw[-,thick,red] (-0.5,-0.5) to (-0.5,0.5);
    \end{scope}
       \begin{scope}[shift={(1,0)}]
    \draw[-] (-1,-0.5) to (-1,0.5);
    \draw[-] (-1,0.5) to (0,0.5);
    \draw[-] (0,-0.5) to (0,0.5);
    \draw[-] (-1,-0.5) to (0,-0.5);
    \end{scope}
    \begin{scope}[shift={(4,0)}]
    \node at (-2,0) {$\sim$};
    \draw[-] (-1,-0.5) to (-1,0.5);
    \draw[-] (-1,0.5) to (0,0.5);
    \draw[-] (0,-0.5) to (0,0.5);
    \draw[-] (-1,-0.5) to (0,-0.5);
    \end{scope}
       \begin{scope}[shift={(5,0)}]
    \draw[-] (-1,-0.5) to (-1,0.5);
    \draw[-] (-1,0.5) to (0,0.5);
    \node[red] at (-0.5,0.7) {$_{m_{\RR}}$};
    \draw[-] (0,-0.5) to (0,0.5);
    \draw[-] (-1,-0.5) to (0,-0.5);
    \draw[-,thick,red] (-0.5,-0.5) to (-0.5,0.5);
    \end{scope}
\end{tikzpicture}
\end{align}
The states in \eqref{eq:Z2_triv} are trivial while those in \eqref{eq:Z2_non-triv} are non-trivial: this can be detected by linking with the horizontal $e_{\LL}e_{\RR}$ anyon present in \eqref{eq:LFolded_Z2}:
\be
\begin{tabular}{c}
    \hspace{1cm} Anyon configuration \hspace{1cm} Linking phase \\
   \begin{tikzpicture}
 \begin{scope}[shift={(0,0)}]
    \draw[-] (-1,-0.5) to (-1,0.5);
    \draw[-] (-1,0.5) to (0,0.5);
    \node[blue] at (-0.5,0.2) {$_{e_{\LL}}$};
    \draw[-] (0,-0.5) to (0,0.5);
    \draw[-] (-1,-0.5) to (0,-0.5);
    \draw[-,thick,blue] (-1,0) to (0,0);
    \end{scope}
       \begin{scope}[shift={(1,0)}]
    \draw[-] (-1,-0.5) to (-1,0.5);
    \draw[-] (-1,0.5) to (0,0.5);
        \node[blue] at (-0.5,0.2) {$_{e_{\RR}}$};
    \draw[-] (0,-0.5) to (0,0.5);
    \draw[-] (-1,-0.5) to (0,-0.5);
    \draw[-,thick,blue] (-1,0) to (0,0);
    \end{scope}
    \begin{scope}[shift={(3,0)}]
     \node at (-1.5,0) {$\sim$};
    \draw[-] (-1,-0.5) to (-1,0.5);
    \draw[-] (-1,0.5) to (0,0.5);
    \node[red] at (-0.5,0.7) {$_{m_{\LL}}$};
    \node[blue] at (-0.7,0.2) {$_{e_{\LL}}$};
    \draw[-] (0,-0.5) to (0,0.5);
    \draw[-] (-1,-0.5) to (0,-0.5);
    \draw[-,thick,red] (-0.5,-0.5) to (-0.5,0.5);
    \draw[-,thick,blue] (-1,0) to (0,0);
    \end{scope}
       \begin{scope}[shift={(4,0)}]
    \draw[-] (-1,-0.5) to (-1,0.5);
    \draw[-] (-1,0.5) to (0,0.5);
    \node[red] at (-0.5,0.7) {$_{m_{\RR}}$};
    \node[blue] at (-0.7,0.2) {$_{e_{\RR}}$};
    \draw[-] (0,-0.5) to (0,0.5);
    \draw[-] (-1,-0.5) to (0,-0.5);
    \draw[-,thick,red] (-0.5,-0.5) to (-0.5,0.5);
    \draw[-,thick,blue] (-1,0) to (0,0);
         \node at (1.5,0) {$+1$};
    \end{scope}
\end{tikzpicture} 
\\
\begin{tikzpicture}
 \begin{scope}[shift={(0,0)}]
    \draw[-] (-1,-0.5) to (-1,0.5);
    \draw[-] (-1,0.5) to (0,0.5);
  \node[red] at (-0.5,0.7) {$_{m_{\LL}}$};
    \node[blue] at (-0.7,0.2) {$_{e_{\LL}}$};
    \draw[-] (0,-0.5) to (0,0.5);
    \draw[-] (-1,-0.5) to (0,-0.5);
    \draw[-,thick,red] (-0.5,-0.5) to (-0.5,0.5);
    \draw[-,thick,blue] (-1,0) to (0,0);
    \end{scope}
       \begin{scope}[shift={(1,0)}]
    \draw[-] (-1,-0.5) to (-1,0.5);
    \draw[-] (-1,0.5) to (0,0.5);
        \node[blue] at (-0.5,0.2) {$_{e_{\RR}}$};
    \draw[-] (0,-0.5) to (0,0.5);
    \draw[-] (-1,-0.5) to (0,-0.5);
    \draw[-,thick,blue] (-1,0) to (0,0);
    \end{scope}
    \begin{scope}[shift={(3,0)}]
     \node at (-1.5,0) {$\sim$};
    \draw[-] (-1,-0.5) to (-1,0.5);
    \draw[-] (-1,0.5) to (0,0.5);
    \node[blue] at (-0.7,0.2) {$_{e_{\LL}}$};
    \draw[-] (0,-0.5) to (0,0.5);
    \draw[-] (-1,-0.5) to (0,-0.5);
    \draw[-,thick,blue] (-1,0) to (0,0);
    \end{scope}
       \begin{scope}[shift={(4,0)}]
    \draw[-] (-1,-0.5) to (-1,0.5);
    \draw[-] (-1,0.5) to (0,0.5);
    \node[red] at (-0.5,0.7) {$_{m_{\RR}}$};
    \node[blue] at (-0.7,0.2) {$_{e_{\RR}}$};
    \draw[-] (0,-0.5) to (0,0.5);
    \draw[-] (-1,-0.5) to (0,-0.5);
    \draw[-,thick,red] (-0.5,-0.5) to (-0.5,0.5);
    \draw[-,thick,blue] (-1,0) to (0,0);
         \node at (1.5,0) {$-1$};
    \end{scope}
\end{tikzpicture}
\end{tabular}
\ee

\vspace{2mm}
\noindent{\bf $D(\Z_n)$ with $D(\Z_n)$ rough merge.}
We now show how our formalism reproduces the results of~\cite{Cowtan:2022csx,Cowtan:2025vok} for $\Z_n$-qudits. Let us take:
\be
    G=\Z_n\,,\quad G'=\Z_n\,,\quad K^\diag=\Z_n^\diag\,,
\ee
whose generators we denote by $m_{\LL},m_{\RR},m_{\LL}m_{\RR}$ respectively.
The Folded Lagrangian in this case is:
\be \label{eq:Am1m2}
\cL^\folded_{\Z_n^\diag}=\langle m_{\LL}\ol{m}_{\RR},\;e_{\LL}\ol{e}_{\RR}\rangle\,.
\ee
After the merge and unfolding, we have the identification on magnetic anyons given by \eqref{eq:sim}
\be
    m_{\LL}^im_{\RR}^j\sim m_{\LL}^{i-k}m_{\RR}^{j+k}
\ee
therefore only the sum $i+j$ is relevant. We can detect different logical states after the merge by linking with the horizontal electric anyon $e_{\LL}e_{\RR}$ appearing in \eqref{eq:Am1m2}\footnote{When unfolding we remove the bar, i.e. we take the complex conjugate representation.}
\be
\begin{tabular}{c}
   Anyon configuration \hspace{1cm} Linking phase \\
   \begin{tikzpicture}
    \begin{scope}[shift={(3,0)}]
    \draw[-] (-1,-0.5) to (-1,0.5);
    \draw[-] (-1,0.5) to (0,0.5);
    \node[red] at (-0.5,0.7) {$_{m_{\LL}^{i-k}}$};
    \node[blue] at (-0.7,0.2) {$_{e_{\LL}}$};
    \draw[-] (0,-0.5) to (0,0.5);
    \draw[-] (-1,-0.5) to (0,-0.5);
    \draw[-,thick,red] (-0.5,-0.5) to (-0.5,0.5);
    \draw[-,thick,blue] (-1,0) to (0,0);
    \end{scope}
       \begin{scope}[shift={(4,0)}]
    \draw[-] (-1,-0.5) to (-1,0.5);
    \draw[-] (-1,0.5) to (0,0.5);
    \node[red] at (-0.5,0.7) {$_{m_{\RR}^{j+k}}$};
    \node[blue] at (-0.7,0.2) {$_{e_{\RR}}$};
    \draw[-] (0,-0.5) to (0,0.5);
    \draw[-] (-1,-0.5) to (0,-0.5);
    \draw[-,thick,red] (-0.5,-0.5) to (-0.5,0.5);
    \draw[-,thick,blue] (-1,0) to (0,0);
         \node at (2.5,0.2) {$e^{\frac{2\pi i}{n}(i+j)}$};
    \end{scope}
\end{tikzpicture} 
\end{tabular}
\ee
with linking phase independent of the choice of representative of magnetic anyons (which follows from the fact that anyons in a condensable algebra have trivial mutual braiding).

\section{TQFT Description of Non-Clifford Operations}
\label{sec:ContNonCliff}

In this section, we will first provide details on $D(D_4)$, then illustrate the TQFT description of the lattice surgery operations involving $D(D_4)$, whose lattice implementation we provided in Sec.~\ref{sec:magic}. We explain how condensable algebras determine the subgroups $K^\diag$ used in the lattice construction and how we can use the algebra data to determine the logical magic states and non-Clifford gates. This enables us to systematically search for generalizations, which we will describe.

\subsection{Details on $D(D_4)$} 
\label{sec:D4_details}

We denote by $D_4$ the group of symmetries of a square (generated by a $\pi/2$ rotation $r$ and reflection $s$):
\be
    D_4=\langle r,s \;|\; r^4=s^2=\id\,,\quad srs=r^3\rangle\,,
\ee
whose conjugacy classes are
\be\ba
    [\id]&=\{\id\},&[r^2]&=\{r^2\},
    &[r]&=\{r,r^3\},\\
    [s]&=\{s,r^2s\}, &[rs]&=\{rs,r^3s\}\,.
\ea\ee
We denote the $D_4$ irreps by $1,\,1_r,\,1_s,\,1_{rs},\,E$, with character table
\be
\begin{array}{|c|ccccc|}
\hline
    & [\id] & [r^2] & [r] & [s] & [rs] \\
    \hline
    1 &  1 & 1 & 1 & 1 & 1\\
    1_r &  1 & 1 & 1 & -1 & -1\\
    1_s &  1 & 1 & -1 & 1 & -1\\
    1_{rs} &  1 & 1 & -1 & -1 & 1\\
    E &  2 & -2 & 0 & 0 & 0\\
     \hline
\end{array}
\ee
The anyons in $\cZ(D_4)$ are shown in table \ref{tab:D4_anyons}. We will adopt a slightly more concise version of the first column, e.g. by omitting the $\id$ group element and the trivial representation $1$. All Condensable Algebras and Folded Lagrangians in $\cZ(D_4)$ were computed in~\cite{Bhardwaj:2024qrf}.
\begin{table}[H]
    \centering
    \begin{tabular}{|c|c|c|c|}
    \hline
    $([g],R)$ & Color Label & Dim & $T$ \\
    \hline\hline
$([\id],1)$		&	$1$	&	$1$	&	$1$ \\
$([\id],1_r)$	&	$e_{RG}$	&	$1$	&	$1$ \\
$([\id],1_s)$	&	$e_R$	&	$1$	&	$1$ \\
$([\id],1_{rs})$	&	$e_G$	&	$1$	&	$1$ \\
$([\id],E)$		&	$m_B$	&	$2$	&	$1$ \\
\hline
$([r^2],1)$	&	$e_{RGB}$	&	$1$	&	$1$ \\
$([r^2],1_r)$	&	$e_B$	&	$1$	&	$1$ \\
$([r^2],1_s)$	&	$e_{GB}$	&	$1$	&	$1$ \\
$([r^2],1_{rs})$	&	$e_{RB}$	&	$1$	&	$1$ \\
\hline
$([r],1)$		&	$m_{RG}$	&	$2$	&	$1$ \\
$([r],i)$		&	$s_{RGB}$	&	$2$	&	$i$ \\
$([r],-1)$		&	$f_{RG}$	&	$2$	&	$-1$ \\
$([r],-i)$		&	$\bar{s}_{RGB}$	&	$2$	&	$-i$ \\
\hline
$([s],+,+)$	&	$m_{GB}$	&	$2$	&	$1$ \\
$([s],+,-)$	&	$m_G$	&	$2$	&	$1$ \\
$([s],-,-)$	&	$f_G$	&	$2$	&	$-1$ \\
$([s],-,+)$	&	$f_{GB}$	&	$2$	&	$-1$ \\
\hline
$([rs],+,+)$	&	$m_{RB}$	&	$2$	&	$1$ \\
$([rs],+,-)$	&	$m_R$	&	$2$	&	$1$ \\
$([rs],-,-)$	&	$f_R$	&	$2$	&	$-1$ \\
$([rs],-,+)$	&	$f_{RB}$	&	$2$	&	$-1$ \\
    \hline
    \end{tabular}
    \caption{Anyons in $\cZ(D_4)$ can be classified by a choice of conjugacy class $[g]$ and an irreducible representation $R$ of the corresponding centralizer $C_G(g)$ (first column). Equivalently, they can be labeled in terms of three copies of the toric code (second column)~\cite{Iqbal:2023wvm,Bhardwaj:2024qrf}. The quantum dimensions of each anyon and diagonal $T$-matrix elements (which encode the anyon spins) are listed in columns 3 and 4 respectively.}
    \label{tab:D4_anyons}
\end{table}

The logical group multiplication operators  for $D_4$ are generated by
\be\ba
L^r&=\ket{r}\bra{\id}+\ket{r^2}\bra{r}+\ket{r^3}\bra{r^2}+\ket{\id}\bra{r^3}\\
&+\ket{rs}\bra{s}+\ket{r^2s}\bra{rs}+\ket{r^3s}\bra{r^2s}+\ket{s}\bra{r^3s}\,,\\
L^s&=\ket{s}\bra{\id}+\ket{r^3s}\bra{r}+\ket{r^2s}\bra{r^2}+\ket{rs}\bra{r^3}\\
&+\ket{\id}\bra{s}+\ket{r^3}\bra{rs}+\ket{r^2}\bra{r^2s}+\ket{r}\bra{r^3s}\,,\\
R^r&=\ket{r^3}\bra{\id}+\ket{\id}\bra{r}+\ket{r}\bra{r^2}+\ket{r^2}\bra{r^3}\\
&+\ket{rs}\bra{s}+\ket{r^2s}\bra{rs}+\ket{r^3s}\bra{r^2s}+\ket{s}\bra{r^3s}\,,\\
R^s&=\ket{s}\bra{\id}+\ket{rs}\bra{r}+\ket{r^2s}\bra{r^2}+\ket{r^3s}\bra{r^3}\\
&+\ket{\id}\bra{s}+\ket{r}\bra{rs}+\ket{r^2}\bra{r^2s}+\ket{r^3}\bra{r^3s}\,,
\ea\ee
whose realization on a quadrit-qubit local Hilbert space $\{\ket{r^as^b}: a=0,1,2,3,b=0,1\}$ is (\ref{LROps}).
The other group multiplication operators can be obtained using
\be
    L^{gh}=L^{g}L^{h}\,,\quad R^{gh}=R^{g}R^{h}\,.
\ee
The irrep operators are diagonal. For 1d irreps, they just multiply each basis state by its character, while for the 2-dimensional $E$ we have 4 operators $Z_E^{i,j}$ for $i,j=1,2$, each of which multiplies $\ket{g}$ by the $i,j$ component of its matrix representation. Concretely:
\begin{align}
Z_{1_r}&=\ket{\id}\bra{\id}+\ket{r}\bra{r}+\ket{r^2}\bra{r^2}+\ket{r^3}\bra{r^3}+\nn\\
&-\ket{s}\bra{s}-\ket{rs}\bra{rs}-\ket{r^2s}\bra{r^2s}-\ket{r^3s}\bra{r^3s}\,,\nn\\
Z_{1_s}&=\ket{\id}\bra{\id}-\ket{r}\bra{r}+\ket{r^2}\bra{r^2}-\ket{r^3}\bra{r^3}+\nn\\
&+\ket{s}\bra{s}-\ket{rs}\bra{rs}+\ket{r^2s}\bra{r^2s}-\ket{r^3s}\bra{r^3s}\,,\nn\\
Z_{1_{rs}}&=Z_{1_r}Z_{1_s}\,,\\
Z_{E}^{1,1}&=\ket{\id}\bra{\id}+i\ket{r}\bra{r}-\ket{r^2}\bra{r^2}-i\ket{r^3}\bra{r^3}\,,\nn\\
Z_{E}^{1,2}&=\ket{s}\bra{s}+i\ket{rs}\bra{rs}-\ket{r^2s}\bra{r^2s}-i\ket{r^3s}\bra{r^3s}\,,\nn\\
Z_{E}^{2,2}&=(Z_{E}^{1,1})^\dagger\,, \qquad Z_{E}^{2,1}=(Z_{E}^{1,2})^\dagger\,,\nn
\end{align}
from which we have that e.g. $\cZ_{D_4}=Z_E^{1,1}+Z_E^{1,2}$\,.

\subsection{Merge and Split of $D(\Z_4)$ and $D(D_4)$}  \label{sec:Z4_D4_cont}

In the TQFT, we can merge $D(D_4)$ and $D(\Z_4)$ 
by condensing the $\cZ(D_4)$ algebra:
\be
\cA(\Z_4,\{\id\},1,1)=1\oplus 1_r\,,
\ee
that can be completed to a Folded Lagrangian, which corresponds to  gauging the {diagonal group defined in (\ref{Z4diagDef}), i.e. }
\be
    \Z_4^\diag=\langle(m,r)\rangle\,\subset \Z_4\times D_4\,,
\ee
where we label the $\Z_4$ generator as $m$.

The folded Lagrangian is:
\be\ba \label{eq:LfoldedZ4}
\cL^\folded_{\Z_4^\diag}&=1\oplus 1_r\oplus1_s\ol{e}^2\oplus1_{rs}\ol{e}^2\oplus E(\ol{e}\oplus \ol{e}^3)\\[-1.5mm]
&\oplus [r](\ol{m}\oplus \ol{m}^3) \oplus [r]_{-1}(\ol{e}^2\ol{m}\oplus \ol{e}^2\ol{m}^3)\\
&\oplus [r]_i(\ol{e}\ol{m}\oplus\ol{e}^3\ol{m}^3) \oplus [r]_{-i}(\ol{e}^3\ol{m}\oplus \ol{e}\ol{m}^3)\\
&\oplus [r^2]\ol{m}^2\oplus [r^2]1_r\ol{m}^2\oplus[r^2]1_s\ol{e}^2\ol{m}^2\\
&\oplus [r^2]1_{rs}\ol{e}^2\ol{m}^2\oplus [r^2]E(\ol{e}\ol{m}^2\oplus \ol{e}^3\ol{m}^2)\,.
\ea\ee

The equivalence relation on group elements is determined from \eqref{eq:sim}:
\be
    (m^i,r^j)\sim(m^im^{-k},r^kr^j),\quad i,j,k\in\{0,1,2,3\}\,,
\ee
in particular
\be \label{eq:m_sim_r}
 (m,\id)\sim(\id,r)\,.
\ee
Note that the order-2 $D_4$ generator $s$ remains non-trivial. A basis of logical states after the merge is thus given by the $D_4$ group elements.                    

We can detect the post-merge logical states by linking with linear combinations of the horizontal electric anyons appearing in \eqref{eq:LfoldedZ4} and ending on left and right boundaries:
\be \label{eq:D4Z4_irreps}
    1,\;1_r,\;1_{s}e^2,\;1_{rs}e^2,\;2Ee,\;2Ee^3\,.
\ee
where the 2-dim $Ee$ and $Ee^3$ end with multiplicity $\dim(E)=2$ on the $D(D_4)$ rough boundary and $\dim(e)=\dim(e^3)=1$ on the $D(\Z_4)$ rough boundary.
There are 8 independent linear combinations of \eqref{eq:D4Z4_irreps} that can detect the 8 independent logical states labeled by $D_4$ group elements after the merge.

\vspace{2mm}
\noindent{\bf Map of Logical States.}
Because of the identification \eqref{eq:m_sim_r}, the merge and split between $D(\Z_4)$ and $D(D_4)$ maps the logical $D(\Z_4)$ input state (\ref{SZ4}) 
\be
\ba
  \ket{S}_{\Z_4} &= \frac{1}{2}\left( \ket{\bar{\mathbf{0}}} + e^{i\pi/4}\ket{\bar{\mathbf{1}}} - \ket{\bar{\mathbf{2}}} + e^{i\pi/4}\ket{\bar{\mathbf{3}}} \right)  \cr 
  &=  \frac{1}{2}\left( \ket{\id } +e^{i\pi/4} \ket{m } 
  - \ket{m^2 } + e^{i\pi/4}\ket{m^3} \right) \,,
\ea
\ee
to the logical $D(D_4)$ state
\be \label{eq:SD4_cont}
 \ket{S}_{D_4} = \frac{1}{2}\left(\ket{\id} + e^{i\pi/4} \ket{r} - \ket{r^2} +e^{i\pi/4} \ket{r^3} \right)\,.
\ee

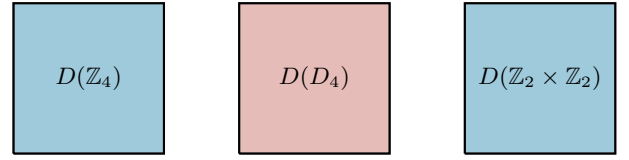
\begin{figure}
$$
\begin{tikzpicture}
\begin{scope}[shift={(0,0)}]
\draw [thick, fill= \BlueColor, opacity=0.5]   (0,0) -- (0,2) -- (2,2) -- (2,0) --(0,0); 
\draw [thick]  (0,0) -- (0,2) -- (2,2) -- (2,0) --(0,0); 
\node at (1,1) {$D(\Z_4)$}; 
\end{scope}
\begin{scope}[shift={(3,0)}]
\draw[thick, fill= \RedColor,opacity=0.5]  (0,0) -- (0,2) -- (2,2) -- (2,0) --(0,0); 
\draw [thick]  (0,0) -- (0,2) -- (2,2) -- (2,0) --(0,0); 
\node at (1,1) {$D(D_4)$}; 
\end{scope}
\begin{scope}[shift={(6,0)}]
\draw [thick, fill= \BlueColor, opacity=0.5]   (0,0) -- (0,2) -- (2,2) -- (2,0) --(0,0); 
\draw [thick]  (0,0) -- (0,2) -- (2,2) -- (2,0) --(0,0); 
\node at (1,1) {$D(\Z_2\times \Z_2)$}; 
\end{scope}
\end{tikzpicture}
$$
\caption{The three TOs used for the magic state preparation and gate teleportation. Each vertical edge is a rough boundary and we perform two rough merge-and-split operations. \label{fig:karaage}}
\end{figure}

\subsection{Merge and Split of $D(D_4)$ and $D(\Z_2\times\Z_2)$} \label{sec:D4_Z2Z2_cont}
After having merged $D(\Z_4)$ and $D(D_4)$ we then merge the latter with $D(\Z_2\times\Z_2)$, as depicted in Figure \ref{fig:karaage}.
This is achieved by condensing the $\cZ(D_4)$ algebra:
\be
\cA(\Z_2^{r^2}\times\Z_2^{r^3s},\{\id\},1,1)=1\oplus 1_{rs}\,,
\ee
that can be completed to a Folded Lagrangian by gauging the diagonal group (\ref{Z2Z2DiagDef}), i.e. 
\be\ba
    &(\Z_2\times\Z_2)^\diag=\langle (r^2,m_{\LL}),\,(r^3s,m_{\RR})\rangle \subset D_4\times (\Z_2\times\Z_2)\,,
\ea\ee
where we label the $\Z_2\times\Z_2$ generators as $m_{\LL},m_{\RR}$.

The Folded Lagrangian is:
\be\ba \label{eq:FoldedLagD4Z2Z2}
\cL^\folded_{(\Z_2\times\Z_2)^\diag}&=1\oplus 1_{rs}\oplus 1_r\ol{e}_{\RR}\oplus 1_s\ol{e}_{\RR}\oplus E(\ol{e}_{\LL}\oplus\ol{e}_{\LL}\ol{e}_{\RR})\\
&\oplus [r^2]\ol{m}_{\LL}\oplus[r^2]1_{rs}\ol{m}_{\LL} \oplus [r^2]1_r\ol{e}_{\RR}\ol{m}_{\LL}\\
&\oplus [r^2]1_s\ol{e}_{\RR}\ol{m}_{\LL}\oplus [r^2]E(\ol{e}_{\LL}\ol{m}_{\LL}\oplus\ol{e}_{\LL}\ol{e}_{\RR}\ol{m}_{\LL})\\
&\oplus[rs]_{++}(\ol{m}_{\LL}\ol{m}_{\RR}\oplus \ol{m}_{\RR}) \\
&\oplus [rs]_{+-}(\ol{e}_{\LL}\ol{e}_{\RR}\ol{m}_{\LL}\ol{m}_{\RR}\oplus \ol{e}_{\LL}\ol{m}_{\RR})\\
&\oplus [rs]_{-+}(\ol{e}_{\RR}\ol{m}_{\LL}\ol{m}_{\RR}\oplus \ol{e}_{\RR}\ol{m}_{\RR})\\
&\oplus[rs]_{--}(\ol{e}_{\LL}\ol{m}_{\LL}\ol{m}_{\RR}\oplus \ol{e}_{\LL}\ol{e}_{\RR}\ol{m}_{\RR})\,. \\
\ea\ee

The identification on group elements implies:
\be
    (r^2,\id)\sim (\id,m_{\LL}),\; (r^3s,\id)\sim (\id,m_{\RR})\,.
\ee
We can detect the 8 independent post-merge logical states by linking with linear combinations of the horizontal electric anyons appearing in \eqref{eq:FoldedLagD4Z2Z2}:
\be
    1,\;1_{rs},\;1_r{e}_{\RR},\;1_s{e}_{\RR},\;2E{e}_{\LL},\;2E{e}_{\LL}{e}_{\RR}\,.
\ee
where, as before, $E$ ends with multiplicity $\dim(E)=2$ on the $D(D_4)$ rough boundary.

\vspace{2mm}
\noindent{\bf Map of Logical States.}
The merge and split between $D(D_4)$ and $D(\Z_2 \times \Z_2)$ with the state \eqref{eq:SD4_cont} and the trivial logical state in $D(\Z_2\times\Z_2)$, by applying
\be
P_{\Z_2^{2,\diag}} = \frac{1}{4}\big( \bbI\otimes\bbI + R^{r^2} \otimes L^{m_{\LL}} + R^{r^3s}\otimes L^{m_{\RR}} + R^{rs}\otimes L^{m_{\LL} m_{\RR}} \big).
\ee
results in the following state  in $D(D_4)\boxtimes D(\Z_2 \times \Z_2)$ 
\be \label{eq:PZ2Z2state}
\ba
&P_{\Z_2^{2,\diag}} \ket{S}_{D_4}\ket{\id}_{\Z_2^2}=\\
&=\frac{1}{8}\big[\ket{\id}(\ket{\id}-\ket{m_{\LL}}) \cr 
&+e^{i\pi/4}\ket{r} (\ket{\id}+\ket{m_{\LL}})\cr 
&-\ket{r^2} (\ket{\id}-\ket{m_{\LL}})\cr 
&+e^{i\pi/4} \ket{r^3} (\ket{\id}+\ket{m_{\LL}}) \cr 
&+e^{i\pi/4}\ket{s} (\ket{m_{\RR}}+\ket{m_{\LL}m_{\RR}})\cr 
&-\ket{rs} (\ket{m_{\RR}}-\ket{m_{\LL}m_{\RR}})\cr 
&+e^{i\pi/4}\ket{r^2s} (\ket{m_{\RR}}+\ket{m_{\LL}m_{\RR}})\cr 
&+\ket{r^3s} (\ket{m_{\RR}}-\ket{m_{\LL}m_{\RR}})\big] \,.
\ea
\ee
 As we did in the lattice construction of Sec. \ref{sec:lattice_D4_Z2Z2_magic}, we now apply a projection which essentially amounts to regrouping the states into eigenstates of $R^s$ and from the TQFT approach corresponds to gauging $\Z_2^s$ on the rough boundary of $D(D_4)$ as discussed around \ref{changedBC}. The Lagrangian algebra $\cL(\Z_2^s,1)\in\cZ(D_4)$ is
\be
\cL(\Z_2^s,1)=1\oplus 1_s\oplus E\oplus [s]_{++}\oplus [s]_{+-}\,.
\ee
On the logical state \eqref{eq:PZ2Z2state}, we apply ${1\over 2} (\bbI + R^s )\otimes\bbI $. This results in the following state:
\be\ba
&\frac{1}{2}\left[(\bbI+R^s)\otimes\bbI\right]P_{\Z_2^{2,\diag}}\ket{S}_{D_4}\ket{\id}_{\Z_2^2}=\cr 
&\frac{1}{16}\big[\ket{\id} ( \ket{\id}-\ket{m_{\LL}}+e^{i\pi/4}\ket{m_{\RR}}+e^{i\pi/4}\ket{m_{\LL}m_{\RR}})\cr
+&\ket{r} ( e^{i\pi/4}\ket{\id}+e^{i\pi/4}\ket{m_{\LL}}-\ket{m_{\RR}}+\ket{m_{\LL}m_{\RR}})\cr
+&\ket{r^2} ( -\ket{\id}+\ket{m_{\LL}}+e^{i\pi/4}\ket{m_{\RR}}+e^{i\pi/4}\ket{m_{\LL}m_{\RR}})\cr
+&\ket{r^3} ( e^{i\pi/4}\ket{\id}+e^{i\pi/4}\ket{m_{\LL}}+\ket{m_{\RR}}-\ket{m_{\LL}m_{\RR}})\cr
+&\ket{s} ( \ket{\id}-\ket{m_{\LL}}+e^{i\pi/4}\ket{m_{\RR}}+e^{i\pi/4}\ket{m_{\LL}m_{\RR}})\cr
+&\ket{rs} ( e^{i\pi/4}\ket{\id}+e^{i\pi/4}\ket{m_{\LL}}-\ket{m_{\RR}}+\ket{m_{\LL}m_{\RR}})\cr
+&\ket{r^2s} ( -\ket{\id}+\ket{m_{\LL}}+e^{i\pi/4}\ket{m_{\RR}}+e^{i\pi/4}\ket{m_{\LL}m_{\RR}})\cr
+&\ket{r^3s} ( e^{i\pi/4}\ket{\id}+e^{i\pi/4}\ket{m_{\LL}}+\ket{m_{\RR}}-\ket{m_{\LL}m_{\RR}})\big]\,.
\ea\ee
As explained in the lattice Sec. \ref{sec:lattice_D4_Z2Z2_magic}, we now make a measurement on the $D(D_4)$ patch in the group element basis. Taking the outcome to be $\ket{\id}_{D_4}$ means that we restrict the state to the terms with $D(D_4)$ component $\ket{\id}$ and thus obtain 
\be
\ket{\psi}_{\Z_2^2}\!\! =\!\frac{1}{2}\!\left[\ket{\id}\!-\!\ket{m_{\LL}}\!+\!e^{i\pi/4}\ket{m_{\RR}}\!+\!e^{i\pi/4}\ket{m_{\LL}m_{\RR}}\right],
\ee
which is the magic state \eqref{eq:MLR} in $D(\Z_2^2)$.

\subsection{Alternative Protocol:\\ $D(\Z_4)|D(D_4)|D(\Z_2)$}

As in Sec.~\ref{subsec:D4_Z2}, we can also consider the third patch to be $D(\Z_2)$, i.e. go directly to magic states for qubits. 
After the $D(\Z_4)$ and $D(D_4)$ merge, and inputting the $\Z_4$ $S$-state, we get $\ket{S}_{D_4} $, Eq. \eqref{eq:SD4_cont}. We then gauge a diagonal $\Z_2$ subgroup of $D_4\times\Z_2$ 
\be \label{eq:Z2diag}
\Z_2^{\diag} = \langle(rs,m)\rangle \,,
\ee
to merge $D(D_4)$ and $D(\Z_2)$. This corresponds to the condensable algebra
\be
\cA(\Z_2^{rs},\{\id\},1,1)=1\oplus 1_{rs}\oplus E\,,
\ee
with Folded Lagrangian
\be\ba
\cL^\folded_{\Z_2^\diag}&=1\oplus 1_{rs}\oplus E \oplus1_r\ol{e}\oplus1_s\ol{e}\oplus E\ol{e}\\[-1.5mm]
&\oplus([rs]_{++}\oplus [rs]_{+-})\ol{m}\oplus([rs]_{-+}\oplus [rs]_{--})\ol{e}\ol{m}
\,. \\
\ea\ee

\vspace{2mm}
\noindent{\bf Map of Logical States.} On the logical states, to merge $D(D_4)$ and $D(\Z_2)$ by gauging \eqref{eq:Z2diag}, we apply 
\be
\frac{1}{2}(\bbI\otimes\bbI+R^{rs}\otimes L^m)
\ee
to $\ket{S_{D_4}}\ket{\id}$:
\be\ba
    &\frac{1}{2}(\bbI\otimes\bbI+R^{rs}\otimes L^m)\ket{S_{D_4}}\ket{\id}=\\
    &\frac{1}{4}\big[(\ket{\id} + e^{i\pi/4} \ket{r} - \ket{r^2} +e^{i\pi/4} \ket{r^3} )\ket{\id}+\\
    &(\ket{rs} + e^{i\pi/4} \ket{r^2s} - \ket{r^3s} +e^{i\pi/4} \ket{s})\ket{m}\big]\,.
\ea\ee
After this merge and split, we apply $\frac{1}{2}(\bbI+R^s)\otimes\bbI$ on the $D(D_4)$ patch
\be\ba
    &\frac{1}{2}\left[(\bbI+R^s)\otimes\bbI\right]\frac{1}{2}(\bbI\otimes\bbI+R^{rs}\otimes L^m)\ket{S_{D_4}}\ket{\id}=\\
    &\frac{1}{8}\big[(\ket{\id}+\ket{s})(\ket{\id}+e^{i\pi/4}\ket{m})\\
    &+(\ket{r}+\ket{rs})(e^{i\pi/4}\ket{\id}+\ket{m})\\
    &-(\ket{r^2}+\ket{r^2s})(\ket{\id}-e^{i\pi/4}\ket{m})\\
    &+(\ket{r^3}+\ket{r^3s})(e^{i\pi/4}\ket{\id}-\ket{m})\big]
\ea\ee
and finally measure $\ket{\id}_{D_4}$. This results in (up to normalization) the $\Z_2$ $T$-magic state 
\be
\ket{T}=\ket{\id} + e^{i \pi /4} \ket{m} \,.
\ee

\subsection{Generalization: $T^{1/n}$ Magic States} \label{sec:qubit_magic_states_cont}

We can construct more generally $n$th roots of the $T$-gate magic states by considering the following three TOs instead, again with vertical rough boundary and horizontal smooth boundaries:
\be \label{Z4D4Z2}
\begin{tikzpicture}
\begin{scope}[shift={(0,0)}]
\draw [thick, fill= \BlueColor, opacity=0.5]   (0,0) -- (0,2) -- (2,2) -- (2,0) --(0,0); 
\draw [thick]  (0,0) -- (0,2) -- (2,2) -- (2,0) --(0,0); 
\node at (1,1) {$D(\Z_{4n})$}; 
\end{scope}
\begin{scope}[shift={(3,0)}]
\draw[thick, fill= \RedColor,opacity=0.5]  (0,0) -- (0,2) -- (2,2) -- (2,0) --(0,0); 
\draw [thick]  (0,0) -- (0,2) -- (2,2) -- (2,0) --(0,0); 
\node at (1,1) {$D(D_{4n})$}; 
\end{scope}
\begin{scope}[shift={(6,0)}]
\draw [thick, fill= \BlueColor, opacity=0.5]   (0,0) -- (0,2) -- (2,2) -- (2,0) --(0,0); 
\draw [thick]  (0,0) -- (0,2) -- (2,2) -- (2,0) --(0,0); 
\node at (1,1) {$D(\Z_2)$}; 
\end{scope}
\end{tikzpicture}
\ee
The input state is the $D(\Z_{4n})$ $S$-state~\cite{Farinholt2014}, which after the first merge of 
$D(\Z_{4n})$ and $D(D_{4n})$
becomes, by similar computations to Sec.~\ref{sec:Z4_D4_cont} the (unnormalized) $D(D_{4n})$ state:
\be
    \ket{S}_{D_{4n}}=\sum_{j=0}^{4n-1} e^{\frac{i\pi}{4n} j^2} \ket{r^j}\,.
\ee
We then gauge a diagonal $\Z_2$ subgroup with $\frac{1}{2}(\bbI\otimes\bbI+R^{rs}\otimes L^m)$ 
\be
\Z_2^{\diag} = \langle(rs,m)\rangle \,.
\ee
and subsequently apply  $\frac{1}{2}(\bbI+R^s)\otimes\bbI$, schematically:
\be\ba
&\ket{S}_{D_{4n}}\ket{\id}_{\Z_2} \xrightarrow{(\bbI\otimes\bbI+R^{rs}\otimes L^m)/2}\\
&\frac{1}{2}(\ket{\id}_{D_{4n}}\ket{\id}_{\Z_2}+e^{\frac{i\pi}{4n}}\ket{s}_{D_{4n}}\ket{m}_{\Z_2}+...)\xrightarrow{(\bbI+R^s)/2\otimes\bbI}\\
&\frac{1}{4}(\ket{\id}_{D_{4n}}\ket{\id}_{\Z_2}+e^{\frac{i\pi}{4n}}\ket{\id}_{D_{4n}}\ket{m}_{\Z_2}+...),
\ea\ee
where the second term in the second line comes from $R^{rs}\otimes L^m(e^{\frac{i\pi}{4n}}\ket{r^{-1}}_{D_{4n}}\ket{\id}_{\Z_2})$.
We finally project onto $\bra{\id}_{D_{4n}}$ and obtain (up to normalization) the $\Z_2$ magic state:
\be
\ket{T^{1/n}}=\ket{\id} + e^{i \pi /4n} \ket{m} \,.
\ee
We can therefore generate all the $\ket{T^{1/n}}$ magic states, and hence all finite levels of the Clifford hierarchy and beyond. We will describe how to produce the corresponding non-Clifford $T^{1/n}$ \emph{gates} in Sec. \ref{sec:T^1/n_cont}.

\subsection{Generalization: Magic States from $D(S_3)$}
In this section we discuss magic state preparation for qubits and for qutrits in which the non-Abelian TO is $D(S_3)$ which has interfaces to $D(\Z_2)$ and $D(\Z_3)$.{\footnote{A proposal for non-Clifford qubit-qutrit gates from $D(S_3)$ was presented in~\cite{Sajith:2025rvy}.}}

\subsubsection{\textbf{\textit{Preliminaries on $D(S_3)$}}}

{We will use the following notation for
\be
    S_3=\langle r,s\;|\; r^3=s^2=\id\,,srs=r^2 \rangle\,,
\ee
and denote its irreps by $1,P,E$ with character table
\be \label{eq:S3_chars}
\begin{array}{|r|ccc|}
    \hline
     & [1]  & [r] & [s] \\
    \hline
    1:& 1 & 1 & 1\\
    P:& 1 & 1 & -1\\
    E:& 2 & -1 & 0\\
     \hline
\end{array}
\ee

The condensable algebras in $\cZ(S_3)$ with corresponding Folded Lagrangians were computed in \cite{Bhardwaj:2023bbf}. We will summarize the two of interest to us here.

\vspace{2mm}
\noindent{\bf Interface from $D(S_3)$ to $D(\Z_3)$.}
The $\cZ(S_3)$ algebra
\be
    \cA(\Z_3,\{\id\},1,1)=1\oplus P\,,
\ee
provides an interface to the double of $G'=\Z_3$, whose generator we denote as $\wt{m}$ and fundamental irrep as $\wt{e}$. The subgroup $K^\diag$ can be taken to be:
\be
    \Z_3^{\diag}=\langle (\wt{m},r)\rangle\subset \Z_3\times S_3
\ee
with corresponding Folded Lagrangian:\footnote{{We have implemented the bar on anyons in $\cZ(\Z_3)$ as complex-conjugation.}}
\be\ba
    \cL(\Z_3^\diag,1)&=1\oplus P \oplus E(\wt{e}\oplus \wt{e}^2) \oplus [r](\wt{m}\oplus \wt{m}^2)\\
    &\oplus [r]_\omega(\wt{e}^2\wt{m}\oplus \wt{e}\wt{m}^2)\oplus [r]_{\omega^2}(\wt{e}\wt{m}\oplus \wt{e}^2\wt{m}^2)\,,
\ea\ee
where $[r]_{\omega^k}$ for $k=0,1,2$ denote the $D(S_3)$ anyons labeled by $[r]$ and an irrep of $C_{S_3}(r)=\Z_3$ in which $r$ has character $\omega^k$ with $\omega=e^{2\pi i/3}$.

\vspace{2mm}
\noindent{\bf Interface from $D(S_3)$ to $D(\Z_2)$.}
The $\cZ(S_3)$ algebra
\be
    \cA(\Z_2,\{\id\},1,1)=1\oplus E\,,
\ee
instead gives rise to an interface to the double of $G'=\Z_2$, whose generator we denote as $m$ and non-trivial irrep as $e$. We can take the $\Z_2\subset S_3$ to be generated by $r^2s$, with corresponding $K^\diag$ given by
\be
    \Z_2^\diag=\langle (r^2s,m) \rangle \subset S_3\times\Z_2
\ee  
and Folded Lagrangian:\footnote{{Note that $[s]=\{s,rs,r^2s\}$, so $r^2s$ is conjugate to $s$. Condensable algebras labeled by conjugate subgroups are equivalent \cite{Ostrikmodule,Natale2017,davydov2017lagrangian,GaiSchaferNamekiWarman}.}}
\be\ba
    \cL(\Z_3^\diag,1)&=1\oplus E(1\oplus \ol{e}) \oplus P\ol{e}\oplus [s]\ol{m}\oplus [s]_-\ol{e}\ol{m}\,,
\ea\ee
where $[s]_-$ ($[s]$) denotes the anyon labeled by the conjugacy class $[s]$ and the non-trivial (trivial) irrep of $C_{S_3}(s)=\Z_2$.
}

\subsubsection{\textbf{\textit{Qubit Magic State from $D(S_3)$}}}

We can prepare qubit magic state in $D(\Z_2)$ starting with a qutrit stabilizer state in $D(\Z_3)$ and performing lattice surgery going through $D(S_3)$.

The TOs are arranged as follows:
\be \label{Z3S3Z2}
\begin{tikzpicture}
\begin{scope}[shift={(0,0)}]
\draw [thick, fill= \BlueColor, opacity=0.5]   (0,0) -- (0,2) -- (2,2) -- (2,0) --(0,0); 
\draw [thick]  (0,0) -- (0,2) -- (2,2) -- (2,0) --(0,0); 
\node at (1,1) {$D(\Z_3)$}; 
\end{scope}
\begin{scope}[shift={(3,0)}]
\draw[thick, fill= \RedColor,opacity=0.5]  (0,0) -- (0,2) -- (2,2) -- (2,0) --(0,0); 
\draw [thick]  (0,0) -- (0,2) -- (2,2) -- (2,0) --(0,0); 
\node at (1,1) {$D(S_3)$}; 
\end{scope}
\begin{scope}[shift={(6,0)}]
\draw [thick, fill= \BlueColor, opacity=0.5]   (0,0) -- (0,2) -- (2,2) -- (2,0) --(0,0); 
\draw [thick]  (0,0) -- (0,2) -- (2,2) -- (2,0) --(0,0); 
\node at (1,1) {$D(\Z_2)$}; 
\end{scope}
\end{tikzpicture}
\ee
 We start by inputting the $\Z_3$ stabilizer state
\be
\ket{\phi}_{\Z_3}=\frac{1}{\sqrt{3}}\lb\ket{\id}+e^{2\pi i/3}\ket{\wt{m}}+e^{4\pi i/3}\ket{\wt{m}^2}\rb\,,
\ee
with $\wt{m}$ the $\Z_3$ generator.
We then gauge 
\be
    \Z_3^{\diag}=\langle (\wt{m},r)\rangle
\ee
along the interface between $D(\Z_3)$ and $D(S_3)$. This identifies $\wt{m}\sim r$ and gives the $D(S_3)$ state
\be
\ket{\phi}_{S_3}=\frac{1}{\sqrt{3}}\lb\ket{\id}+e^{2\pi i/3}\ket{r}+e^{4\pi i/3}\ket{r^2}\rb\,.
\ee

We now turn to the interface between $D(S_3)$ and $D(\Z_2)$ denoting the $\Z_2$ generator as $m$. To merge the TOs, we gauge
\be
    \Z_2^\diag=\langle (r^2s,m) \rangle\subset S_3\times \Z_2
\ee
along the interface between them. Initializing the qubit logical state to $\ket{\id}_{\Z_2}$, we apply the projector
\be
    \frac{1}{2}(\bbI+R^{r^2s}\otimes L^m)\,,
\ee
and obtain:
\be\ba
 &\frac{1}{2}(\bbI+R^{r^2s}\otimes L^m)\ket{\phi}_{S_3}\ket{\id}_{\Z_2}=\\
 &\frac{1}{2\sqrt{3}}\lb\ket{\id}+e^{2\pi i/3}\ket{r}+e^{4\pi i/3}\ket{r^2}\rb\ket{\id}+\\
 &\frac{1}{2\sqrt{3}}\lb\ket{r^2s}+e^{2\pi i/3}\ket{s}+e^{4\pi i/3}\ket{rs}\rb\ket{m}\,.
\ea\ee

We then gauge $\Z_2^s$ on the $D(S_3)$ boundary, which on the logical state means applying $\frac{1}{2}(\bbI+R^s)\otimes\bbI$ and then consider the measurement outcome $\ket{\id}_{S_3}$. This gives the qubit magic state 
\be
    \ket{\psi}_{\Z_2}=\frac{1}{\sqrt{2}}\lb\ket{\id}+e^{2\pi i/3}\ket{m}\rb\,.
\ee

\subsubsection{\textbf{\textit{Qutrit Magic States from $D(S_3)$}}} 

Conversely, we can prepare qutrit magic states in $D(\Z_3)$ starting with a qubit stabilizer state in $D(\Z_2)$ and performing lattice surgery going through $D(S_3)$, with TOs arranged as follows:
\be \label{Z2S3Z3}
\begin{tikzpicture}
\begin{scope}[shift={(0,0)}]
\draw [thick, fill= \BlueColor, opacity=0.5]   (0,0) -- (0,2) -- (2,2) -- (2,0) --(0,0); 
\draw [thick]  (0,0) -- (0,2) -- (2,2) -- (2,0) --(0,0); 
\node at (1,1) {$D(\Z_2)$}; 
\end{scope}
\begin{scope}[shift={(3,0)}]
\draw[thick, fill= \RedColor,opacity=0.5]  (0,0) -- (0,2) -- (2,2) -- (2,0) --(0,0); 
\draw [thick]  (0,0) -- (0,2) -- (2,2) -- (2,0) --(0,0); 
\node at (1,1) {$D(S_3)$}; 
\end{scope}
\begin{scope}[shift={(6,0)}]
\draw [thick, fill= \BlueColor, opacity=0.5]   (0,0) -- (0,2) -- (2,2) -- (2,0) --(0,0); 
\draw [thick]  (0,0) -- (0,2) -- (2,2) -- (2,0) --(0,0); 
\node at (1,1) {$D(\Z_3)$}; 
\end{scope}
\end{tikzpicture}
\ee
We start by inputting a $\Z_2$ stabilizer state
\be
\ket{\phi}_{\Z_2}=\frac{1}{\sqrt{2}}\lb\ket{\id}+\theta\ket{m}\rb\,,\quad \theta\in\{1,-1,i\}\,,
\ee
with the $\Z_2$ generator denoted as $m$ and $\theta$ chosen such that $\ket{\phi}_{\Z_2}$ is a stabilizer state.
We then gauge 
\be
    \Z_2^\diag=\langle (m,r^2s) \rangle\subset \Z_2\times S_3
\ee
along the interface between $D(\Z_2)$ and $D(S_3)$ which identifies $m\sim r^2s$ and produces the $D(S_3)$ state
\be
\ket{\phi}_{S_3}=\frac{1}{\sqrt{2}}\lb\ket{\id}+\theta\ket{r^2s}\rb\,.
\ee
We now turn to the interface between $D(S_3)$ and $D(\Z_3)$ denoting the $\Z_3$ generator as $\wt{m}$. To merge the TOs, we gauge
\be
    \Z_3^{\diag}=\langle (r,\wt{m})\rangle\subset S_3\times\Z_3
\ee
along the interface between them. 

Initializing the qutrit logical state to 
\be
    \ket{\psi}_{\Z_3}=\alpha\ket{\id}+\beta\ket{\wt{m}}+\gamma\ket{\wt{m}^2}\,,
\ee
we apply the projector
\be
    \frac{1}{3}(\bbI+R^r\otimes L^{\wt{m}}+R^{r^2}\otimes L^{\wt{m}^2})\,,
\ee
and obtain:
\be\ba
 &\frac{1}{3}(\bbI+R^r\otimes L^{\wt{m}}+R^{r^2}\otimes L^{\wt{m}^2})\ket{\phi}_{S_3}\ket{\psi}_{\Z_3}=\\
 &=\frac{1}{3\sqrt{2}}\ket{\id}\lb\alpha\ket{\id}+\beta\ket{\wt{m}}+\gamma\ket{\wt{m}^2}\rb\\
 &+\frac{\theta}{3\sqrt{2}}\ket{r^2s}\lb\alpha\ket{\id}+\beta\ket{\wt{m}}+\gamma\ket{\wt{m}^2}\rb\\
 &+\frac{1}{3\sqrt{2}}\ket{r^2}\lb\alpha\ket{\wt{m}}+\beta\ket{\wt{m}^2}+\gamma\ket{\id}\rb\\
 &+\frac{\theta}{3\sqrt{2}}\ket{s}\lb\alpha\ket{\wt{m}}+\beta\ket{\wt{m}^2}+\gamma\ket{\id}\rb\\
  &+\frac{1}{3\sqrt{2}}\ket{r}\lb\alpha\ket{\wt{m}^2}+\beta\ket{\id}+\gamma\ket{\wt{m}}\rb\\
 &+\frac{\theta}{3\sqrt{2}}\ket{rs}\lb\alpha\ket{\wt{m}^2}+\beta\ket{\id}+\gamma\ket{\wt{m}}\rb\,.
\ea\ee
We then gauge $\Z_2^s$ on the $D(S_3)$ boundary, which on the logical state means applying $\frac{1}{2}(\bbI+R^s)\otimes\bbI$ and then consider the measurement outcome $\ket{\id}_{S_3}$. This gives the (unnormalized) qutrit magic state 
\be\ba
    \ket{\psi'}_{\Z_3}&=\alpha\lb\ket{\id}+\theta\ket{\wt{m}}\rb\\
    &+\beta\lb\ket{\wt{m}}+\theta\ket{\wt{m}^2}\rb\\
    &+\gamma\lb\ket{\wt{m}^2}+\theta\ket{\id}\rb\,.  
\ea\ee
For example, setting one of $\alpha,\beta,\gamma$ to $1$ and the other two to $0$ we get:
\be\ba
    \ket{\psi'_\alpha}_{\Z_3}&=\frac{1}{\sqrt{2}}\lb\ket{\id}+\theta\ket{\wt{m}}\rb\,,\\
     \ket{\psi'_\beta}_{\Z_3}&=\frac{1}{\sqrt{2}}\lb\ket{\wt{m}}+\theta\ket{\wt{m}^2}\rb\,,\\
    \ket{\psi'_\gamma}_{\Z_3}&=\frac{1}{\sqrt{2}}\lb\ket{\wt{m}^2}+\theta\ket{\id}\rb\,,
\ea\ee
where $\theta\in\{1,-1,i\}$. These are qutrit magic states~\cite{Wang:2023uog}.

\subsection{$T$ and $T^{1/n}$ Non-Clifford Gates}

Using non-Abelian TOs we can also construct more generally non-Clifford gates as follows using lattice surgery. We will illustrate this with the example of the $T$-gate and the $T^{1/n}$ gates for qubits.

\subsubsection{\textbf{\textit{$T$-gate for $D(\Z_2\times \Z_2)$}}}
We first merge $D(\Z_4)$ and $D(D_4)$ as described in Sec.~\ref{sec:Z4_D4_cont}: this gives the $D(D_4)$ logical state 
\be
\ba
\ket{S}_{D_4}& = \frac{1}{2}\Big( \ket{\id} + e^{i\pi/4} \ket{r}  - \ket{r^2} + e^{i\pi/4} \ket{r^3} \Big) \cr 
\ea
\ee
We then merge $D(D_4)$ and $D(\Z_2\times\Z_2)$ following the same protocol discussed in \ref{sec:D4_Z2Z2_cont}, but now inputting a generic $D(\Z_2\times\Z_2)$ logical state.
We will denote the $D(\Z_2\times\Z_2)$ basis states as follows:
\be\ba \label{eq:Z2Z2_basis_states}
    \ket{\psi_1}&\equiv\ket{\id} & \ket{\psi_2}&\equiv\ket{m_{\LL}}\\
        \ket{\psi_3}&\equiv\ket{m_{\RR}} & \ket{\psi_4}&\equiv\ket{m_{\LL}m_{\RR}}\,,
\ea\ee 
so that we can write a generic
 $D(\Z_2\times\Z_2)$ logical state as
\be\ba
\ket{\psi}_{\Z_2\times \Z_2}&=\sum_{k=1}^{4} c_k\ket{\psi_k}
\ea\ee
in which $c_k$ for $k=1,2,3,4$ are arbitrary complex coefficients and the basis states are defined in \eqref{eq:Z2Z2_basis_states}.

As discussed in Sec.~\ref{sec:D4_Z2Z2_cont}, we merge $D(D_4)$ and $D(\Z_2\times\Z_2)$ by applying the projector
\be
P_{\Z_2^{2,\diag}} = \frac{1}{4}\big( \bbI\otimes\bbI + R^{r^3s}\otimes L^{m_{\RR}} \big)\big( \bbI\otimes\bbI + R^{r^2} \otimes L^{m_{\LL}}\big)\,.
\ee
Subsequently, we change the boundary conditions of the right boundary of $D(D_4)$, as we discussed in generality around \eqref{changedBC}. In this case, we gauge $K=\Z_2^s$: this is implemented on the state by applying the projector 
\be
\frac{1}{2}(\bbI + R^s)\otimes \bbI\,.
\ee
After this, we consider measurement outcome $\ket{\id}_{D_4}$ and obtain:
\be
\ba
\cr 
 \ket{\id}_{D_4} \, U \ket{\psi}_{\Z_2\times \Z_2} \,
\ea\ee
where U is the operator acting on the $D(\Z_2\times\Z_2)$ logical space, whose entries are obtained from the product of the projectors we applied:
\be\ba
    U_{k,l}=\bra{\psi_k}\bra{\id}_{D_4}&(\bbI\otimes\bbI + R^s\otimes \bbI)\times\\
    &\big( \bbI\otimes\bbI + R^{r^3s}\otimes L^{m_{\RR}} \big)\times\\
    &\big( \bbI\otimes\bbI + R^{r^2} \otimes L^{m_{\LL}}\big)\ket{S_{D_4}}\ket{\psi_l}\,,
\ea\ee
explicitly:
\be
U= \frac{1}{2}\begin{pmatrix}
1 &  e^{i\pi/4} & -1 &  e^{i\pi/4} \\
 e^{i\pi/4} & 1 &  e^{i\pi/4} & -1 \\
-1 &  e^{i\pi/4} & 1 &  e^{i\pi/4} \\
 e^{i\pi/4} & -1 &  e^{i\pi/4} & 1
\end{pmatrix}.
\ee

Note that if we denote
\be
H =H_1 \otimes H_2 = \frac{1}{2}\left(\begin{array}{cccc}
1 & 1 & 1 & 1 \\
1 & -1 & 1 & -1 \\
1 & 1 & -1 & -1 \\
1 & -1 & -1 & 1
\end{array}\right)\,,
\ee
we find 
\be
H U H= 
\diag (e^{i\pi/4} , -e^{i\pi/4}, 1 , 1) \,.
\ee
i.e. 
\be
U= H\left(e^{i\pi/4}(T^{\dagger}\otimes Z)\,CZ\right)H \,.
\ee

\subsubsection{\textbf{\textit{$T$-gate for $D(\Z_2)$}}} \label{sec:T_cont}
This protocol describes a merge of $D(D_4)$ and $D(\Z_2)$ that produces a unitary equivalent to the $T$-gate.
We start with the $D(D_4)$ state
\be
\ket{S}_{D_4} =\frac{1}{2}\Big( \ket{\id} + e^{i\pi/4} \ket{r}  - \ket{r^2} + e^{i\pi/4} \ket{r^3} \Big),
\ee
which can be obtained from a merge of $D(\Z_4)$ and $D(D_4)$ as described in Sec.~\ref{sec:Z4_D4_cont}, and a generic qubit state
\be \label{eq:psiZ2}
    \ket{\psi}_{\Z_2}=\alpha\ket{\id}+\beta\ket{m}\,.
\ee
{We now consider the condensable algebra in $\cZ(D_4)$ given by:
\be
    \cA(\Z_2^{rs}\times\Z_2^{r^2},\Z_2^{rs},1,1)=1\oplus1_{rs}\oplus [rs]\,.
\ee
To merge $D(D_4)$ and $D(\Z_2)$ we therefore gauge the following subgroup of $D_4\times\Z_2$ (denoting the $\Z_2$ generator as $m$)
\be \label{eq:lalala}
    \Z_2\times\Z_2^\diag=\{(\id,\,\id),\,(rs,\id),\,(r^2,\,m),\,(r^3s,\,m)\}\,,
\ee
which gives the Folded Lagrangian
\begin{align}
    \cL( \Z_2\times\Z_2^\diag)&=1\oplus 1_{rs} \oplus E\ol{e}\oplus [r^2]\ol{m}\nn\\
    &\oplus [r^2]1_{rs}\ol{m}\oplus[r^2]E\,\ol{em}\\
    &\oplus [rs](1\oplus \ol{m})\oplus [rs]_{+-}\ol{e}\oplus [rs]_{--}\ol{em}\,.\nn
\end{align}
}
On the logical state, we apply the projector corresponding to \eqref{eq:lalala}
\be
P_{\Z_2\times\Z_2^\diag} = \frac{1}{4}\big( \bbI\otimes \bbI + R^{rs}\otimes\bbI + R^{r^2}\otimes L^{m} + R^{r^3s}\otimes L^{m} \big)\,,
\ee
and obtain the state
\be\ba
&P_{\Z_2\times\Z_2^\diag}\ket{S}_{D_4}\ket{\psi}_{\Z_2}=\\
\frac{1}{8}\Big[&(\alpha-\beta)\ket{\id}\ket{\id}-
(\alpha-\beta)\ket{\id}\ket{m}\\+
&(\alpha +\beta) e^{i\pi/4}\ket{r}\ket{\id}+
(\alpha +\beta) e^{i\pi/4}\ket{r}\ket{m}\\-
&(\alpha-\beta)\ket{r^2}\ket{\id}+
(\alpha-\beta)\ket{r^2}\ket{m}\\+
&(\alpha +\beta) e^{i\pi/4}\ket{r^3}\ket{\id}+
(\alpha +\beta) e^{i\pi/4}\ket{r^3}\ket{m}\\+
&(\alpha +\beta) e^{i\pi/4}\ket{s}\ket{\id}+
(\alpha +\beta) e^{i\pi/4}\ket{s}\ket{m}\\+
&(\alpha-\beta)\ket{rs}\ket{\id}-
(\alpha-\beta)\ket{rs}\ket{m}\\+
&(\alpha +\beta) e^{i\pi/4}\ket{r^2s}\ket{\id}+
(\alpha +\beta) e^{i\pi/4}\ket{r^2s}\ket{m}\\-
&(\alpha-\beta)\ket{r^3s}\ket{\id}+
(\alpha-\beta)\ket{r^3s}\ket{m}\,\Big].\\
\ea\ee
As in the previous protocols, we subsequently apply \\$\frac{1}{2}(\bbI+R^s)\otimes\bbI$. This gives:
\begin{align}
\frac{1}{16}\Big[&(\alpha-\beta+\alpha e^{i\pi/4}+\beta e^{i\pi/4})\ket{\id}\ket{\id}\nn\\
+&(-\alpha+\beta+\alpha e^{i\pi/4}+\beta e^{i\pi/4})\ket{\id}\ket{m}\nn\\
+&(\alpha-\beta+\alpha e^{i\pi/4}+\beta e^{i\pi/4})\ket{r}\ket{\id}\nn\\
+&(-\alpha+\beta+\alpha e^{i\pi/4}+\beta e^{i\pi/4})\ket{r}\ket{m}\nn\\
+&(-\alpha+\beta+\alpha e^{i\pi/4}+\beta e^{i\pi/4})\ket{r^2}\ket{\id}\nn\\
+&(\alpha-\beta+\alpha e^{i\pi/4}+\beta e^{i\pi/4})\ket{r^2}\ket{m}\nn\\
+&(-\alpha+\beta+\alpha e^{i\pi/4}+\beta e^{i\pi/4})\ket{r^3}\ket{\id}\nn\\
+&(\alpha-\beta+\alpha e^{i\pi/4}+\beta e^{i\pi/4})\ket{r^3}\ket{m}\nn\\
+&(\alpha-\beta+\alpha e^{i\pi/4}+\beta e^{i\pi/4})\ket{s}\ket{\id}\nn\\
+&(-\alpha+\beta+\alpha e^{i\pi/4}+\beta e^{i\pi/4})\ket{s}\ket{m}\nn\\
+&(\alpha-\beta+\alpha e^{i\pi/4}+\beta e^{i\pi/4})\ket{rs}\ket{\id}\nn\\
+&(-\alpha+\beta+\alpha e^{i\pi/4}+\beta e^{i\pi/4})\ket{rs}\ket{m}\nn\\
+&(-\alpha+\beta+\alpha e^{i\pi/4}+\beta e^{i\pi/4})\ket{r^2s}\ket{\id}\nn\\
+&(\alpha-\beta+\alpha e^{i\pi/4}+\beta e^{i\pi/4})\ket{r^2s}\ket{m}\nn\\
+&(-\alpha+\beta+\alpha e^{i\pi/4}+\beta e^{i\pi/4})\ket{r^3s}\ket{\id}\nn\\
+&(\alpha-\beta+\alpha e^{i\pi/4}+\beta e^{i\pi/4})\ket{r^3s}\ket{m}\Big]\,.
\end{align}

Measuring $\ket{\id}_{D_4}$ we finally get (up to normalization):
\be\ba \label{eq:psipZ2}
\ket{\psi'}_{\Z_2}=&(\alpha-\beta+\alpha e^{i\pi/4}+\beta e^{i\pi/4})\ket{\id}+\\
&(\beta-\alpha+\beta e^{i\pi/4}+\alpha e^{i\pi/4})\ket{m}\,.
\ea\ee
The normalized state corresponding to $\ket{\psi'}_{\Z_2}$ is obtained from the input $\ket{\psi}_{\Z_2}$ by applying the unitary $U$ whose entries can be obtained from the projectors we applied
\begin{align}
     U_{k,l}=\bra{\psi_k}\bra{\id}_{D_4}&(\bbI\otimes\bbI + R^s\otimes \bbI)\times\nn\\
    &\big( \bbI\otimes\bbI + R^{rs}\otimes \bbI \big)\times\\
    &\big( \bbI\otimes\bbI + R^{r^2} \otimes L^{m}\big)\ket{S_{D_4}}\ket{\psi_l}\,,\nn
\end{align}
for $k,l\in\{0,1\}$ with $\ket{\psi_0}=\ket{\id}$ and $\ket{\psi_1}=\ket{m}$. 
Explicitly (as can be checked from Eqs.~\eqref{eq:psiZ2} and \eqref{eq:psipZ2}) the unitary is:
\be\ba \label{eq:UT_continuum}
    U&=\frac{1}{2}
    \begin{pmatrix}
        1+e^{i\pi/4} & -1+e^{i\pi/4} \\
        -1+e^{i\pi/4} & 1+e^{i\pi/4}
    \end{pmatrix}\\
    &=HX\,T\,XH=e^{i\pi/4}HT^\dagger H\,,
\ea\ee
which has the same eigenvalues as the non-Clifford $T$-gate. This is the gate in Eq.~\eqref{eq:UTLattice} of the lattice section.

\subsubsection{\textbf{\textit{$T^{1/n}$-gates for $D(\Z_2)$}}}  
\label{sec:T^1/n_cont}

This protocol generalizes the previous one: by merging $D(D_{4n})$ and $D(\Z_2)$ we will produce a unitary equivalent to the $T^{1/n}$-gate.

We start with the $D(D_{4n})$ state
\be
    \ket{S}_{D_{4n}}=\frac{1}{2\sqrt{n}}\sum_{j=0}^{4n-1} e^{\frac{i\pi}{4n} j^2} \ket{r^j}\,,
\ee
which can be obtained from a merge of $D(\Z_{4n})$ and $D(D_{4n})$ as described in Sec.~\ref{sec:qubit_magic_states_cont}, and a generic qubit state
\be
    \ket{\psi}_{\Z_2}=\alpha\ket{\id}+\beta\ket{m}\,.
\ee
Denoting the $\Z_2$ generator as $m$, we gauge the subgroup of $D_{4n}\times\Z_2$ given by 
\be
    \Z_2\times\Z_2^\diag=\langle (rs,\id),\,(r^{2n}\,,m)\rangle\,.
\ee
Note that both $rs$ and $r^{2n}$ are elements of order-2 in $D_{4n}$ and that $r^{2n}$ is in the center of $D_{4n}$. This is achieved at the logical level by applying the projector
\be
P_{\Z_2\times\Z_2^\diag} = \frac{1}{4}\big( \bbI\otimes \bbI + R^{rs}\otimes\bbI + R^{r^{2n}}\otimes L^{m} + R^{r^{1+2n}s}\otimes L^{m} \big)\,.
\ee
Subsequently, after the rough merge and split, we gauge $\Z_2^s$ by applying  $\frac{1}{2}(\bbI+R^s)\otimes\bbI$. Measuring $\bra{\id}_{D_{4n}}$ gives (up to normalization):
\be\ba 
\ket{\psi'}_{\Z_2}=&(\alpha+(-1)^n\beta+\alpha e^{i\pi/4n}-(-1)^n\beta e^{i\pi/(4n)})\ket{\id}+\\
&(\beta+(-1)^n\alpha+\beta e^{i\pi/4n}-(-1)^n\alpha e^{i\pi/(4n)})\ket{m}\,.
\ea\ee
The normalized state corresponding to $\ket{\psi'}_{\Z_2}$ is obtained from the input $\ket{\psi}_{\Z_2}$ by applying the unitary $U$ whose entries can be obtained from the projectors we applied
\begin{align}
     U_{k,l}=\sqrt{n}\bra{\psi_k}\bra{\id}_{D_{4n}}&(\bbI\otimes\bbI + R^s\otimes \bbI)\times\nn\\
    &\big( \bbI\otimes\bbI + R^{rs}\otimes \bbI \big)\times\\
    &\big( \bbI\otimes\bbI + R^{r^{2n}} \otimes L^{m}\big)\ket{S_{D_{4n}}}\ket{\psi_l}\,.
    \nn
\end{align}
Explicitly (as can be checked from Eq.~\eqref{eq:psiZ2} and the expression above) the unitary is: 
\be\ba
    U&=\frac{1}{2}
    \begin{pmatrix}
        1+e^{i\pi/4n} & (-1)^n(1-e^{i\pi/4n})\\
        (-1)^n(1-e^{i\pi/4n}) & 1+e^{i\pi/4n}
    \end{pmatrix}=\\
    &=HX^n\,T^{1/n} \,X^nH\,,
\ea\ee
which has the same eigenvalues as the non-Clifford gate
\be
    T^{1/n}=\diag(1,e^{i\pi/4n})\,.
\ee
We can therefore generate non-Clifford gates for all finite levels of the Clifford hierarchy, as well as gates outside the hierarchy.\footnote{For $n=2^{k}$, using the classification of~\cite{Cui:2016bxt}, we compute that the $T^{1/n}$ gate is in level $3+k$ of the Clifford hierarchy, where $k$ is a non-negative integer. Otherwise, the gates lie outside any finite Clifford level.}

\section{Discussion}
\label{sec:discussion}

We propose a novel method of implementing non-Clifford operations in the surface code via lattice surgery involving a non-Abelian topological code patch. The topological codes we consider are quantum double models $D(G)$ associated with finite groups $G$, with the case $G=\Z_2$ being the standard surface code. First, we generalize the rough merge and split operations in the standard lattice surgery to the hybrid lattice surgery between different topological codes, particularly involving non-Abelian codes. The rough merge operation is governed by an interface between two quantum double models, which can be understood as gauging a symmetry of the interface. This means that different types of rough merge exist, in contrast to the standard surface code lattice surgery. The rough split is realized by local measurements on the merged interface, similar to the standard lattice surgery, which can be understood as gauging the symmetry dual to the symmetry describing the merged interface.

In our method, a code patch protected by the $D(\Z_4)$ code is prepared in a state through a generalized Clifford gate that can be implemented fold-transversally. It interacts with a $D(D_4)$ code patch---a non-Abelian code---by a specific type of rough merge followed by a rough split. The combined effect of the rough merge and split can be described by the measurement of a logical operator, and as a result the two code patches are entangled. We can either measure out the $D(\Z_4)$ code patch at this stage or delay this measurement until later. Then, another rough merge and split happens between $D(D_4)$ code patch and one or two standard surface code patches, entangling them effectively. With the $D(\Z_4)$ and $D(D_4)$ code patches measured in certain bases, we can either prepare a magic state or implement a non-Clifford gate on the standard surface code patch(es).

The lattice model description of this new method is complemented by a continuum TQFT analysis, based on the topological field theories that underlie the quantum doubles. In this context the interfaces, realized by gauging subgroups of the two adjoining code patches, have a simple interpretation as particular maps between the anyons of the two topological orders. We reproduce the results from the lattice in this language. Moreover the continuum description opens up an avenue to systematically search for interesting lattice surgery configurations: we illustrate this by constructing magic states and non-Clifford $T^{1/n}$ gates for all finite levels of the Clifford hierarchy and beyond using non-Abelian topological orders based on $D(D_{4n})$, where $D_{4n}$ is the dihedral group of order $8n$. Furthermore, we can explore extensions to qudits using other finite groups. The continuum can act in this way as a type of reconnaissance, that allows exploration of interesting code patches for lattice surgery. Writing down the lattice models can be achieved using the general prescription in Sec.~\ref{sec:lattice_surgery}. 

Our protocol offers a new method to implement logical non-Clifford operations through geometrically local operations on 2D surface codes. Based on lattice surgery across different codes, our method is compatible with 2D architectures, where physical entangling operations between the two code patches only happen near the interface, thereby placing less demand on hardware engineering than approaches requiring non-local connectivity.   
The information is protected in topological codes throughout the procedure and logical errors can be systematically reduced by increasing the code distance of the code patches involved. We can correct physical errors when they are detected instead of discarding any prepared states. The resource cost of our method to reach a given fidelity in the implemented logical operations is governed by the error thresholds of the non-Abelian quantum double model involved and of the hybrid code patches. Due to the non-Abelian nature of the code, just-in-time decoding strategies are needed in error correction~\cite{Bombin:2018wjx,Brown2020,Scruby:2020pvw,Davydova:2025ylx}. 
For the surface code on a 3D spatial lattice, \cite{Scruby:2020pvw} analyzed the performance of the just-in-time decoder with numerical simulation. Although the threshold found there is not high ($ \sim 0.1\% $), the threshold for the $D_4$ surface code could be higher due to the intrinsic heralding of the noise~\cite{Jing:2025zuu}. 
We leave the resource analysis for our method to future work. More generally, it is important to develop better decoding strategies for non-Abelian quantum double models, which will shed light on the performance of not only this method, but also other protocols of implementing non-Clifford gates based on the $D(D_4)$ topological order~\cite{Laubscher:2019rss,Brown2020, Huang:2025cvt, Davydova:2025ylx,Bauer:2025qly,Sajith:2025rvy}.

The general continuum analysis opens up a wider exploration of possible magic states and gate teleportation, using the general results on interfaces for quantum doubles of groups (including with anomalies)~\cite{Ostrikmodule,Davydov2009ModularIF,Natale2017,davydov2017lagrangian, Beigi:2010htr,delaFuente:2023whm,GaiSchaferNamekiWarman}. For certain non-Clifford operations, we can systematically determine groups that can realize them via hybrid-surgery, which makes this a very flexible framework, adjustable to specific computational applications.

\subsection*{Acknowledgments}

\noindent
We thank Tyler Ellison, Naren Manjunath, Vieri Mattei, Yuhan Gai, Rajath Radhakrishnan for discussions. In particular we thank the authors of the upcoming work~\cite{Ellisonetal} for discussions on group surface codes. 
We thank the KITP for hospitality, while this work was initiated.  
This material is based upon work supported by the U.S. Department of Energy, Office of Science, Advanced Scientific Computing Research under Award No. DE-SC0026209. 
 The work of S-JH, SSN and AW is supported by the UKRI Frontier Research Grant, underwriting the ERC Advanced Grant ``Generalized Symmetries in Quantum Field Theory and Quantum Gravity''.
This research was supported in part by grant NSF PHY-2309135 to the Kavli Institute for Theoretical Physics. 

\subsection*{Author contributions}

\noindent
All authors contributed extensively to this research and the writing of this manuscript.

\appendix

\section{Ribbon operators}
\label{app:ribbon}
In this appendix, we briefly review the ribbon operators in the quantum double model~\cite{Bombin2008,Beigi:2010htr}. A ribbon $\xi$ consists of a sequence of sites connecting a starting site $s_{0}=(v_{0},p_{0})$ to an ending site $s_{1}=(v_{1},p_{1})$ by adjoining the direct and dual triangles along the path. The ribbon operator $F_{\xi}^{h,g}$ is defined as 
\begin{align}
\label{eq:ribbon_op}
&
\begin{tikzpicture}
\node at (-2.1,0.8) {$F_{\xi}^{h,g}$};
\fill[gray!20] (-0.75,0.75) -- (0,1.5) -- (0.75,0.75) -- cycle;
\fill[gray!20] (0,1.5) -- (0.75,0.75) -- (1.5,1.5) -- cycle;
\fill[gray!20] (0.75,0.75) -- (1.5,1.5) -- (2.25,0.75) -- cycle;
\fill[gray!20] (1.5,1.5) -- (2.25,0.75) -- (3,1.5) -- cycle;
\fill[gray!20] (2.25,0.75) -- (3,1.5) -- (3.75,0.75) -- cycle;
\draw[black, ->-] (0,0) -- (0,1.5);
\draw[black, ->-] (1.5,0) -- (1.5,1.5);
\draw[black, ->-] (3,0) -- (3,1.5);
\draw[black, ->-] (-1.5,1.5) -- (0,1.5);
\draw[black, ->-] (0,1.5) -- (1.5,1.5);
\draw[black, ->-] (1.5,1.5) -- (3,1.5);
\draw[black, ->-] (3,1.5) -- (4.5,1.5);
\draw[black, dotted] (-0.75,0.75) -- (0,1.5);
\draw[black, dotted] (0.75,0.75) -- (0,1.5);
\draw[black, dotted] (0.75,0.75) -- (1.5,1.5);
\draw[black, dotted] (2.25,0.75) -- (1.5,1.5);
\draw[black, dotted] (2.25,0.75) -- (3,1.5);
\draw[black, dotted] (3.75,0.75) -- (3,1.5);
\draw[black, dotted] (-0.75,0.75) -- (3.75,0.75);
\node[above] at (-0.75,0.9) {$s_{0}$};
\node[above] at (3.65,0.9) {$s_{1}$};
\node[above] at (0.75,1.5) {$\ket{y_{1}}$};
\node[above] at (2.25,1.5) {$\ket{y_{2}}$};
\node[above] at (3.75,1.5) {$\ket{y_{3}}$};
\node[below] at (0,0) {$\ket{x_{1}}$};
\node[below] at (1.5,0) {$\ket{x_{2}}$};
\node[below] at (3,0) {$\ket{x_{3}}$};
\end{tikzpicture}
\nonumber
\\
&
\begin{tikzpicture}
\node at (-2.6,0.7) {$=\delta_{g,y_{1}y_{2}y_{3}}$};
\fill[gray!20] (-0.75,0.75) -- (0,1.5) -- (0.75,0.75) -- cycle;
\fill[gray!20] (0,1.5) -- (0.75,0.75) -- (1.5,1.5) -- cycle;
\fill[gray!20] (0.75,0.75) -- (1.5,1.5) -- (2.25,0.75) -- cycle;
\fill[gray!20] (1.5,1.5) -- (2.25,0.75) -- (3,1.5) -- cycle;
\fill[gray!20] (2.25,0.75) -- (3,1.5) -- (3.75,0.75) -- cycle;
\draw[black, ->-] (0,0) -- (0,1.5);
\draw[black, ->-] (1.5,0) -- (1.5,1.5);
\draw[black, ->-] (3,0) -- (3,1.5);
\draw[black, ->-] (-1.5,1.5) -- (0,1.5);
\draw[black, ->-] (0,1.5) -- (1.5,1.5);
\draw[black, ->-] (1.5,1.5) -- (3,1.5);
\draw[black, ->-] (3,1.5) -- (4.5,1.5);
\draw[black, dotted] (-0.75,0.75) -- (0,1.5);
\draw[black, dotted] (0.75,0.75) -- (0,1.5);
\draw[black, dotted] (0.75,0.75) -- (1.5,1.5);
\draw[black, dotted] (2.25,0.75) -- (1.5,1.5);
\draw[black, dotted] (2.25,0.75) -- (3,1.5);
\draw[black, dotted] (3.75,0.75) -- (3,1.5);
\draw[black, dotted] (-0.75,0.75) -- (3.75,0.75);
\node[above] at (-0.75,0.9) {$s_{0}$};
\node[above] at (3.65,0.9) {$s_{1}$};
\node[above] at (0.75,1.5) {$\ket{y_{1}}$};
\node[above] at (2.25,1.5) {$\ket{y_{2}}$};
\node[above] at (3.75,1.5) {$\ket{y_{3}}$};
\node[below] at (0,-0.15) {$\ket{hx_{1}}$};
\node[below] at (1.5,0) {$\ket{(^{\hat{y}_{1}^{-1}}h)x_{2}}$};
\node[below] at (3.4,0) {$\ket{(^{\hat{y}_{2}^{-1}}h)x_{3}}$};
\end{tikzpicture},
\end{align}
where $\hat{y}_{n} = y_{1} \cdots y_{n}$, and $^{y}h = yhy^{-1}$.

The ribbon operators $F_{\xi}^{h,g}$ create a pair of anyonic excitations at the endpoints of the ribbon $\xi$. However, these excitations are a superposition of elementary anyons. The elementary anyons are created by ribbon operators in a new basis labeled by $([g],\pi,\boldsymbol{u},\boldsymbol{v})$, where $\boldsymbol{u}=(i,j)$, $\boldsymbol{v}=(i',j')$ such that $i,i' \in \{1,...,|[g]|\}$ index elements of the conjugacy class $[g]$ and $j,j'$ label matrix entries of the irrep $\pi$. In this basis, we need to define a set $P([g]) = \{ p_{j} \}_{j=1}^{|[g]|}$ of representatives of $G/C_{g}$ such that $c_{j} = p_{j} g p_{j}^{-1}$, where $c_{j}$ enumerate the elements of $[g]$. Every element $g \in G$ can be written in a unique way as $g = p_{j}n$ for some $j \in \{ 1,...,|[g]| \}$ and $n \in C_{g}$. A ribbon operator in the basis $([g],\pi,\boldsymbol{u},\boldsymbol{v})$ is then given by
\begin{equation}
    F_{\xi}^{([g],\pi);(\boldsymbol{u},\boldsymbol{v})} = \frac{\text{dim}(\pi)}{|C_{g}|} \sum_{k \in C_{g}} (\Gamma_{\pi}^{-1}(k))_{j,j'} F_{\xi}^{(c_{i}^{-1},p_{i}kp_{i'}^{-1})}.
\label{eq:ribbon_anyonbasis}
\end{equation}
Here, the pair $([g],\pi)$ labels the anyon type and encodes global degrees of freedom. These labels cannot be changed by local operators at the ends of a ribbon. In contrast, $(\boldsymbol{u},\boldsymbol{v})$ describes local degrees of freedom within each type of anyon and can be changed by applying some local operators at the endpoints.


\bibliographystyle{quantum}
\bibliography{ref_updated}

\end{document}